\documentclass[onecolumn]{aastex62}
\usepackage{amsmath}     
\usepackage{wasysym}           
\usepackage{graphicx}
\usepackage{amssymb}
\usepackage{epstopdf}
\usepackage{mathrsfs}
\usepackage{anyfontsize}
\usepackage{natbib}
\usepackage{color}
\usepackage{lipsum}
\usepackage{diagbox}

\shorttitle{Shock Propagation with Accretion}
\shortauthors{Coughlin, Ro, \& Quataert}
\begin{document}
\title{Weak Shock Propagation with Accretion.~II.~Stability of Self-similar Solutions to Radial Perturbations}
\author[0000-0003-3765-6401]{Eric R.~Coughlin}
\altaffiliation{Einstein Fellow}
\affiliation{Columbia Astrophysics Laboratory, New York, NY 80980}
\affiliation{Astronomy Department and Theoretical Astrophysics Center, University of California, Berkeley, Berkeley, CA 94720}
\author{Stephen Ro}
\affiliation{Astronomy Department and Theoretical Astrophysics Center, University of California, Berkeley, Berkeley, CA 94720}
\author{Eliot Quataert}
\affiliation{Astronomy Department and Theoretical Astrophysics Center, University of California, Berkeley, Berkeley, CA 94720}

\email{eric.r.coughlin@gmail.com}

\begin{abstract}
\citet{coughlin18b} (Paper I) derived and analyzed a new regime of self-similarity that describes weak shocks (Mach number of order unity) in the gravitational field of a point mass. These solutions are relevant to low energy explosions, including failed supernovae. In this paper, we develop a formalism for analyzing the stability of shocks to radial perturbations, and we demonstrate that the self-similar solutions of Paper I are extremely weakly unstable to such radial perturbations. Specifically, we show that perturbations to the shock velocity and post-shock fluid quantities (the velocity, density, and pressure) grow with time as $t^{\alpha}$, where $\alpha \le 0.12$, implying that the ten-folding timescale of such perturbations is roughly ten orders of magnitude in time. We confirm these predictions by performing high-resolution, time-dependent numerical simulations. 
Using the same formalism, we also show that the Sedov-Taylor blastwave is trivially stable to radial perturbations provided that the self-similar, Sedov-Taylor solutions extend to the origin, and we derive simple expressions for the perturbations to the post-shock velocity, density, and pressure. Finally, we show that there is a third, self-similar solution (in addition to the the solutions in Paper I and the Sedov-Taylor solution) to the fluid equations that describes a rarefaction wave, i.e., an outward-propagating sound wave of infinitesimal amplitude. We interpret the stability of shock propagation in light of these three distinct self-similar solutions. 
\end{abstract}


\section{Introduction}
The generation and propagation of a shock wave is central to many otherwise-distinct, astrophysical phenomena, including supernovae, star formation, and galaxy evolution. One of the most powerful methods for describing the time and space-dependent evolution of a fluid in the presence of a shock is that of self-similarity (see \citealt{ostriker88} for a review). This mathematical technique yields exact solutions to the fluid equations -- a set of coupled, nonlinear, partial differential equations -- under a set of basic and non-restrictive assumptions. Moreover, self-similar solutions reproduce the properties of shocks to a high degree of accuracy in a wide range of contexts: the Sedov-Taylor blastwave \citep{sedov59, taylor50}, perhaps one of the best-known examples of a self-similar solution to the fluid equations, can be used to predict the energy released from an atomic bomb and a supernova explosion alike.

Recently, \citet{coughlin18b} (hereafter Paper I) described a new mode of self-similar shock propagation, which, dissimilar from the Sedov-Taylor blastwave, is valid when the shock Mach number is only marginally greater than one and the fluid is situated in the gravitational field of a point mass. These solutions predict not only the outward motion of gas immediately behind the shock, but also the \emph{accretion} of matter onto the point mass. The energy budget of the shocked fluid is also modified by the binding energy of the ambient medium and the accretion of binding energy by the black hole, neither of which is present in the Sedov-Taylor blastwave. These solutions were also shown to accurately reproduce the propagation of a shock in the hydrogen envelope of a massive star following a failed supernova, in which a weak shock is still generated from the mass radiated in neutrinos during core collapse \citep{nadezhin80, lovegrove13, fernandez18, coughlin18a}. 

While the utility of self-similar solutions has been demonstrated both mathematically and phenomenologically, the question of their global stability is of paramount importance for determining the asymptotic, temporal behavior of shock waves. In particular, while a self-similar solution provides an exact solution to the fluid equations and can accurately reproduce the propagation of a shock in certain contexts, the evolution of \emph{deviations} from the self-similar solution must also be understood to determine whether or not the self-similar solution characterizes the state of a shock at late times. If small deviations in the fluid variables from the self-similar prescription are found to grow asymptotically, then the self-similar solution is unstable and cannot describe the long-term evolution of a shock. 

In this paper we analyze the stability of self-similar shocks to radial perturbations, focusing primarily on the solutions presented in \citet{coughlin18b} (though we also analyze the stability of the Sedov-Taylor blastwave in the Appendix). For ease of notation, throughout the remainder of this paper we refer to the self-similar solutions presented in Paper I as the CQR solutions. In Section \ref{sec:equations} we present the basic equations and we develop the formalism for analyzing the evolution of perturbations on top of the CQR solution, and in Section \ref{sec:eigenmodes} we write the solutions to these equations in the form of Eigenmodes. In Section \ref{sec:solutions} we present the Eigenmodes for specific density profiles of the ambient medium, and we show that the CQR solutions are extremely weakly unstable, with perturbations to the shock position growing with time $t$ as $t^{\alpha}$ with $\alpha \lesssim 0.12$. We compare the results of the Eigenmode analysis to simulations in Section \ref{sec:simulations}, showing very good agreement between the two (the simulations will be described in more detail in Paper III of this series). In Section \ref{sec:rarefaction}, we show that there is a second self-similar solution to the fluid equations when the shock Mach number is exactly one; this solution corresponds to a rarefaction wave that travels at the sound speed, and likely represents the asymptotic state of shocks with Mach number less than the CQR value. We discuss the physical origin of the instability in Section \ref{sec:origin} and we summarize and conclude in Section \ref{sec:summary}.

\section{General Considerations and Equations}
\label{sec:equations}
{In this section, we derive the self-similar solutions that describe a shocked fluid as it expands in the gravitational field of a compact object, and we present the equations that govern the evolution of perturbations on top of those self-similar solutions. Our notation will be very similar to that of Paper I, but we will introduce some small differences that are useful for the perturbation analysis of this paper. For ease of reference, we explicitly point out these differences in Table \ref{tab0}. }

\begin{table}[h]
\begin{center}
\begin{tabular}{|c|c|c|}
\hline    
Variable & Paper I & This Paper \\ 
\hline
Time-dependent shock position, velocity & $r_{\rm sh},v_{\rm sh}$ & $R,V$\\
\hline
Ambient length scale, density, pressure & $r_0,\rho_{0},p_0$ & $r_{\rm a},\rho_{\rm a},p_{\rm a}$ \\
\hline
Self-similar shock velocity & $V\sqrt{GM/r_{\rm sh}}$ & $V_{\rm c}\sqrt{GM/R}$ \\
    \hline
    Self-similar post-shock velocity & $v = V\sqrt{GM/r} f(\xi)$ & $v = V(t) f_0(\xi)$ \\
    \hline
    Self-similar post-shock density & $\rho = \rho_{\rm 0}\left(r / r_{0}\right)^{-n}g(\xi)$ & $\rho = \rho_{\rm a}\left(R(t) / {r_{\rm a}}\right)^{-n}g_0(\xi)$  \\ 
    \hline
    Self-similar post-shock pressure & $p = \rho_{0}GM/r\left(r/r_0\right)^{-n}h(\xi)$ & $p = \rho_{\rm a}V(t)^2\left(R(t)/r_{\rm a}\right)^{-n}h_0(\xi)$ \\
    \hline
\end{tabular}
\end{center}
\caption{A summary of the variables defined in Paper I and those introduced here.}
\label{tab0}
\end{table}

We assume that the gas is spherically symmetric, in hydrostatic equilibrium, and the gravitational field is dominated by a compact object of mass $M$. The continuity, radial momentum, and entropy equations that govern the evolution of the fluid velocity, $v$, density, $\rho$, and pressure, $p$, are then

\begin{equation}
\frac{\partial \rho}{\partial t}+\frac{1}{r^2}\frac{\partial}{\partial r}\left[r^2\rho v\right] = 0, \label{cont}
\end{equation}
\begin{equation}
\frac{\partial v}{\partial t}+v\frac{\partial v}{\partial r}+\frac{1}{\rho}\frac{\partial p}{\partial r} = -\frac{GM}{r^2}, \label{rmom}
\end{equation}
\begin{equation}
\frac{\partial}{\partial t}\ln(s)+v\frac{\partial}{\partial r}\ln(s) = 0, \label{gasen}
\end{equation}
where $s = p/\rho^{\gamma}$ is the specific entropy of the gas with adiabatic index $\gamma$. Further assume that there is a shock propagating through the medium with time-dependent position $R(t)$ and velocity $V(t) = dR/dt$. If the pre-shock gas obeys an adiabatic equation of state and we are sufficiently far from any surface, then the density and pressure of the pre-shock gas, $\rho_1$ and $p_1$ respectively, satisfy

\begin{equation}
\rho_1 = \rho_{\rm a}\left(\frac{r}{r_{\rm a}}\right)^{-n}, \label{rhoambient}
\end{equation}
\begin{equation}
p_1 = \frac{1}{n+1}\frac{GM}{r}\rho_{\rm a}\left(\frac{r}{r_{\rm a}}\right)^{-n}, \label{pambient}
\end{equation}
where $\rho_{\rm a}$ is the density of the ambient medium at radius $r_{\rm a}$ and $n$ is a free parameter (note that $n$ has often been denoted by $\omega$ in past works on the Sedov-Taylor blastwave). The shock jump conditions, which guarantee the continuity of energy, momentum, and mass across the shock, further yield a set of boundary conditions that must be satisfied by the fluid at the shock front. If we denote the pre- and post-shock adiabatic indices respectively by $\gamma_1$ and $\gamma_2$, then these boundary conditions are \citep{ostriker88, coughlin18b}:

\begin{multline}
v(R) = \bigg\{1-\frac{\gamma_2}{\gamma_2+1}\left(1+\frac{GM}{\left(n+1\right)RV^2}\right) \\ +\frac{\gamma_2-1}{\gamma_2+1}\sqrt{\left(\frac{\gamma_2}{\gamma_2-1}\right)^2\left(1+\frac{GM}{\left(n+1\right)RV^2}\right)^2-\left(\frac{\gamma_2+1}{\gamma_2-1}\right)\left(1+\frac{2\gamma_1}{\gamma_1-1}\frac{GM}{\left(n+1\right)RV^2}\right)}\bigg\} \times V, \label{bc1}
\end{multline}
\begin{multline}
\rho(R) = \frac{\gamma_2+1}{\gamma_2-1}\bigg\{\frac{\gamma_2}{\gamma_2-1}\left(1+\frac{GM}{\left(n+1\right)RV^2}\right) \\ -\sqrt{\left(\frac{\gamma_2}{\gamma_2-1}\right)^2\left(1+\frac{GM}{\left(n+1\right)RV^2}\right)^2-\left(\frac{\gamma_2+1}{\gamma_2-1}\right)\left(1+\frac{2\gamma_1}{\gamma_1-1}\frac{GM}{\left(n+1\right)RV^2}\right)}\bigg\}^{-1} \times\rho_{\rm a}\left(\frac{R}{r_{\rm a}}\right)^{-n}, \label{bc2}
\end{multline}
\begin{multline}
p(R) = \frac{1}{\gamma_2+1}\bigg\{1+\frac{GM}{\left(n+1\right)RV^2} \\ +\left(\gamma_2-1\right)\sqrt{\left(\frac{\gamma_2}{\gamma_2-1}\right)^2\left(1+\frac{GM}{\left(n+1\right)RV^2}\right)^2-\left(\frac{\gamma_2+1}{\gamma_2-1}\right)\left(1+\frac{2\gamma_1}{\gamma_1-1}\frac{GM}{RV^2}\right)}\bigg\}  
\times \rho_{\rm a}V^2\left(\frac{R}{r_{\rm a}}\right)^{-n}, \label{bc3}
\end{multline}
These expressions are for arbitrary $n$, $\gamma_1$ and $\gamma_2$, though they simplify considerably when $\gamma_1 = \gamma_2$ and when the shock Mach number is much greater than one. Here, however, we will make neither of those assumptions.

Since the shock position changes with time, these boundary conditions take place at a temporally-evolving radius. This observation suggests that we make the change of variables

\begin{equation}
r \rightarrow \xi, \quad \xi = \frac{r}{R(t)},
\end{equation}
as the boundary conditions are then satisfied at a single value of $\xi$ for all $t$. Also, while it is not immediately obvious that it is beneficial to do so, we will make the additional change of variables 

\begin{equation}
    t \rightarrow \tau, \quad \tau = \ln\left(\frac{R}{r_{\rm a}}\right).
\end{equation}
We emphasize that the $R(t)$ appearing in these definitions is the \emph{true} shock position, and we will return to the ramifications of this choice (compared to using the unperturbed, or self-similar, shock position) in later sections. In terms of these variables, Equations \eqref{cont} -- \eqref{gasen} become

\begin{equation}
\frac{\partial \rho}{\partial \tau}-\xi\frac{\partial \rho}{\partial \xi}+\frac{1}{V}\frac{1}{\xi^2}\frac{\partial}{\partial \xi}\left(\xi^2\rho v\right) = 0,
\end{equation}
\begin{equation}
\frac{\partial v}{\partial\tau}-\xi\frac{\partial v}{\partial \xi}+\frac{1}{V}v\frac{\partial v}{\partial \xi}+\frac{1}{V}\frac{1}{\rho}\frac{\partial p}{\partial \xi} = -\frac{GM}{VR}\frac{1}{\xi^2},
\end{equation}
\begin{equation}
\frac{\partial \ln s}{\partial \tau}-\xi\frac{\partial \ln s}{\partial \xi}+\frac{1}{V}v\frac{\partial \ln s}{\partial \xi} = 0,
\end{equation}
where temporal derivatives here are with respect to constant $\xi$. Note that we did not yet assume anything about the temporal dependence of the shock position -- we only used the fact that $V = dR/dt$ to derive the above three equations. 

We now write the fluid quantities in the following forms:

\begin{equation}
v = V\left\{f_0(\xi)+f_1(\xi,\tau)\right\}, \quad \rho = \rho_{\rm a}\left(\frac{R}{r_{\rm a}}\right)^{-n}\left\{g_0(\xi)+g_1(\xi,\tau)\right\}, \quad p = \rho_{\rm a}\left(\frac{R}{r_{\rm a}}\right)^{-n}V^2\left\{h_0(\xi)+h_1(\xi,\tau)\right\}, \label{fluiddefs}
\end{equation}
\begin{equation*}
\Rightarrow \ln s \simeq \ln V^2R^{-n+n\gamma_2}+\ln\left(\frac{h_0}{g_0^{\gamma_2}}\right)+\frac{h_1}{h_0}-\gamma_2\frac{g_1}{g_0} \equiv \ln V^2R^{-n+n\gamma_2} +s_0(\xi) +s_1(\xi,\tau),
\end{equation*}
where in the last line we kept only first-order corrections in the assumed-small ratios $g_1/g_0$ and $h_1/h_0$. There is also an additional constant that appears in the definition of the entropy that is irrelevant to the equations. With these parameterizations, the fluid equations become

\begin{equation}
\frac{\partial g_1}{\partial \tau}-n\left(g_0+g_1\right)-\xi\frac{\partial}{\partial \xi}\left(g_0+g_1\right)+\frac{1}{\xi^2}\frac{\partial}{\partial \xi}\left[\xi^2\left(g_0+g_1\right)\left(f_0+f_1\right)\right] = 0. \label{contfull}
\end{equation}
\begin{equation}
\frac{\partial f_1}{\partial \tau}+\left(\frac{1}{2}\frac{d}{d\tau}\ln V^2\right)\left(f_0+f_1\right)-\xi\frac{\partial}{\partial \xi}\left(f_0+f_1\right)+\left(f_0+f_1\right)\frac{\partial}{\partial \xi}\left(f_0+f_1\right)+\frac{1}{g_0}\left(1+\frac{g_1}{g_0}\right)^{-1}\frac{\partial}{\partial \xi}\left(h_0+h_1\right) = -\frac{GM}{RV^2\xi^2}, \label{rmomfull}
\end{equation}
\begin{equation}
\frac{\partial s_1}{\partial \tau}+
\frac{d}{d\tau}\ln V^2-n+n\gamma_2-\xi\frac{\partial}{\partial \xi}\left[s_0+s_1\right]+\left(f_0+f_1\right)\frac{\partial}{\partial\xi}\left[s_0+s_1\right] = 0. \label{gasenfull}
\end{equation}
These equations can be satisfied exactly with all subscript-1 quantities identically zero if

\begin{equation}
\frac{GM}{RV^2} = \frac{1}{V_{\rm c}^2} \quad \Leftrightarrow \quad V = V_{\rm c}\sqrt{\frac{GM}{R}}, \label{ssexact}
\end{equation}
where $V_{\rm c}$ is an unspecified parameter. If the shock velocity satisfies precisely this scaling, then the solutions to the fluid equations (and the boundary conditions) are self-similar, i.e., the solutions can be written as separable in $t$ and $\xi$, and we recover the following three ordinary differential equations for the unperturbed quantities:

\begin{equation}
-n g_0-\xi g_0'+\frac{1}{\xi^2}\frac{d}{d\xi}\left[\xi^2g_0f_0\right] = 0, \label{ss1}
\end{equation}
\begin{equation}
-\frac{1}{2}f_0-\xi f_0'+f_0f_0'+\frac{1}{g_0}h_0' = -\frac{1}{V_{\rm c}^2\xi^2},
\end{equation}
\begin{equation}
n\gamma_2-n-1+\left(f_0-\xi\right)s_0' = 0. \label{ss3}
\end{equation}
These three equations are the same as those derived in Appendix A of Paper I, and the boundary conditions at $\xi = 1$ can be read off from Equations \eqref{bc1} -- \eqref{bc3}. The parameter $V_{\rm c}$, as shown in Paper I, is determined by the requirement that the solutions smoothly pass through a critical point, and is of the order one. These shocks are therefore only mildly supersonic, and they result in both outward motion immediately behind the shock and accretion onto the point mass at the origin. We refer the reader to Paper I for a much more detailed discussion of the properties of the self-similar solutions.

Importantly, this analysis demonstrates that deviations from the self-similar velocity prescription induce finite perturbations to the fluid quantities. In other words, if the shock velocity varies as anything \emph{other} than Equation \eqref{ssexact}, then perturbations to the fluid quantities must exist in order to satisfy the fluid equations. We therefore parameterize the shock velocity by

\begin{equation}
\frac{GM}{RV^2} = \frac{1}{V_{\rm c}^2}\left\{1-2\zeta\left(\tau\right)\right\}, \label{zetaeq}
\end{equation}
where $\zeta(\tau)$ is assumed to be a small correction that encodes the deviation of the shock velocity from the self-similar one (the sign convention and the factor of 2 will be explained in the next section). Keeping only first order terms in $\zeta$ in Equations \eqref{contfull} -- \eqref{gasenfull} then gives the following three equations for the perturbations:

\begin{equation}
\frac{\partial g_1}{\partial \tau}-ng_1-\xi\frac{\partial g_1}{\partial \xi}+\frac{1}{\xi^2}\frac{\partial}{\partial\xi}\left[\xi^2g_0f_1+\xi^2f_0g_1\right] = 0,
\end{equation}
\begin{equation}
\frac{\partial f_1}{\partial \tau}-\frac{1}{2}f_1-\xi\frac{\partial f_1}{\partial \xi}+\frac{\partial}{\partial \xi}\left(f_0f_1\right)+\frac{1}{g_0}\frac{\partial}{\partial \xi}\left(h_0s_1+\gamma_2 h_0\frac{g_1}{g_0}\right)-\frac{g_1}{g_0^2}\frac{\partial h_0}{\partial \xi} = -f_0\frac{d\zeta}{d\tau}+\frac{2}{V_{\rm c}^2\xi^2}\zeta,
\end{equation}
\begin{equation}
\frac{\partial s_1}{\partial \tau}+\left(f_0-\xi\right)\frac{\partial s_1}{\partial \xi}+f_1s_0' = -2\frac{d\zeta}{d\tau}.
\end{equation}
We can cast these equations in a slightly simplified form if we further define

\begin{equation}
F_1 \equiv g_0\xi^2 f_1, \quad G_1 \equiv \frac{g_1}{g_0}, 
\end{equation}
which are the perturbation to the mass flux and the relative perturbation to the density. Written in matrix form, our final form for the perturbation equations is

\begin{multline}
\frac{\partial}{\partial \tau}\left(
\begin{array}{c}
s_1 \\
F_1 \\
G_1
\end{array}\right)
+
\left(
\begin{array}{ccc}
f_0-\xi & 0 & 0 \\
\xi^2h_0 & f_0-\xi & \gamma_2 \xi^2 h_0 \\
0 & \xi^{-2}g_0^{-1} & f_0-\xi
\end{array} \right)\frac{\partial}{\partial\xi}
\left(
\begin{array}{c}
s_1 \\
F_1 \\
G_1
\end{array}\right)
+ \left(
\begin{array}{ccc}
0 & \xi^{-2}g_0^{-1}s_0' & 0 \\
\xi^2h_0' & \frac{3}{2}-n+2f_0' & \xi^2(\gamma_2-1)h_0' \\
0 & 0 & 0 
\end{array} \right)
\left(
\begin{array}{c}
s_1 \\
F_1 \\
G_1
\end{array}\right)
\\
=\left(
\begin{array}{c}
-2\dot{\zeta} \\
\frac{2g_0}{V_{\rm c}^2}\zeta-f_0g_0\xi^2\dot{\zeta} \\
0
\end{array}\right), \label{eqsoft}
\end{multline}
where primes denote differentiation with respect to $\xi$ and dots with respect to $\tau$. The boundary conditions for $s_1$, $F_1$, and $G_1$ are determined by the shock jump conditions, specifically Equations \eqref{bc1} -- \eqref{bc3}, by inserting our expression for the shock velocity (Equation \ref{zetaeq}) into the right-hand sides, Taylor expanding about $\zeta = 0$, and keeping only first-order terms. The resulting expressions are lengthy in general, but they are of the form

\begin{equation}
s_1(1,\tau) = a_{\rm s} \zeta (\tau), \quad F_1(1,\tau) = a_{\rm F} \zeta(\tau), \quad G_1(1,\tau) = a_{\rm G}\zeta(\tau), \label{bcs}
\end{equation}
where the $a$'s are constants that depend on $\gamma_1$, $\gamma_2$, and $n$ (and $V_{\rm c}$, though this is also a function of $\gamma_1$, $\gamma_2$, and $n$). For the specific case where $\gamma_1 = \gamma_2 = 1+1/n$, which is a good approximation when the shock is propagating through the hydrogen envelope of a massive star (see the discussion at the end of Section 2 of Paper I), these constants reduce to

\begin{equation}
a_{\rm s} = -\frac{2\left(1+2n\right)\left(-2+3nV_{\rm c}^2+2n^2V_{\rm c}^2\right)}{n\left(2+V_{\rm c}^2\right)\left(-1+2nV_{\rm c}^2+2n^2V_{\rm c}^2\right)}, \quad a_{\rm F} = a_{\rm G} = \frac{4}{2+V_{\rm c}^2}.
\end{equation}

Equation \eqref{eqsoft} appears underconstrained, as the number of equations (three) is fewer than the number of unknowns (four). The resolution to this apparent issue is that the determinant of the matrix multiplying the derivatives in Equation \eqref{eqsoft} is 

\begin{equation}
\det \left(
\begin{array}{ccc}
f_0-\xi & 0 & 0 \\
\xi^2h_0 & f_0-\xi & \gamma_2 \xi^2 h_0 \\
0 & \xi^{-2}g_0^{-1} & f_0-\xi
\end{array} \right) = \left(f_0-\xi\right)\left(\left(f_0-\xi\right)^2-\frac{\gamma_2 h_0}{g_0}\right).
\end{equation}
As shown in Paper I, the second factor in this expression equals zero at a critical point $\xi_{\rm c}$, and characterizes the location in the unperturbed, self-similar solution where the gas goes from sub to supersonic infall. We see that this position also marks a special location for the perturbations, and, if the perturbations are to remain finite at the critical point, there is a fourth boundary condition that must be satisfied by the functions at this location within the flow. This fourth boundary condition serves as an additional constraint that closes the system of equations.

{Before moving on to the solutions to the above equations, we would like to stress the fact that, in all of the formalism that we introduced to analyze radial perturbations to self-similar solutions, we \emph{never} invoked or explicitly made reference to the unperturbed shock position or velocity. In particular, Equation \eqref{zetaeq} -- which parameterizes deviations from pure self similarity -- only introduces the dimensionless parameter $\zeta$, which itself is written as a function of $\tau$ (which is just the log of the true shock position). The consistency of the perturbation approach is only dependent on the smallness of $\zeta$. We will return to the relevance of this point and its advantages (compared to working with the perturbations to the shock position and velocity individually) when discussing the previous literature in Section \ref{sec:previous} and when making comparison to simulations in Section \ref{sec:simulations}. }

\section{Eigenmodes}
\label{sec:eigenmodes}
Equation \eqref{eqsoft} describes the general space and time-dependent evolution of the perturbations imposed on top of the self-similar profile. We seek solutions to this equation of the form

\begin{equation}
\zeta = \zeta_{\sigma}e^{\sigma \tau}, \label{zetasigma}
\end{equation}
where $\zeta_{\sigma}$ is a constant and $\sigma$ is an undetermined parameter. We can satisfy the differential equations if we let

\begin{equation}
F_1 = \zeta_\sigma F_\sigma(\xi)e^{\sigma \tau} ,\quad G_1 = \zeta_\sigma G_\sigma(\xi)e^{\sigma \tau},\quad s_1 = \zeta_\sigma s_\sigma(\xi)e^{\sigma \tau}, \label{funssigma}
\end{equation}
where the functions $F_\sigma$, $G_\sigma$ and $s_\sigma$ satisfy the boundary conditions at the shock (Equation \ref{bcs}). Notice, however, that the amplitude $\zeta_\sigma$ scales out of the problem, because the boundary conditions and the right-hand side of Equation \eqref{eqsoft} are all proportional to this parameter. The value of $\sigma$ is then determined solely by the fourth boundary condition, making $\sigma$ an ``Eigenvalue'' from the standpoint that it is uniquely specified by the continuity of the variables through the sonic point. The Eigenvalue equation reads

\begin{multline}
\sigma\left(
\begin{array}{c}
s_\sigma \\
F_\sigma \\
G_\sigma
\end{array}\right)
+
\left(
\begin{array}{ccc}
f_0-\xi & 0 & 0 \\
\xi^2h_0 & f_0-\xi & \gamma_2 \xi^2 h_0 \\
0 & \xi^{-2}g_0^{-1} & f_0-\xi
\end{array} \right)\frac{\partial}{\partial\xi}
\left(
\begin{array}{c}
s_\sigma \\
F_\sigma \\
G_\sigma
\end{array}\right)
+ \left(
\begin{array}{ccc}
0 & \xi^{-2}g_0^{-1}s_0' & 0 \\
\xi^2h_0' & \frac{3}{2}-n+2f_0' & \xi^2(\gamma_2-1)h_0' \\
0 & 0 & 0 
\end{array} \right)
\left(
\begin{array}{c}
s_\sigma \\
F_\sigma \\
G_\sigma
\end{array}\right)
\\
=\left(
\begin{array}{c}
-2\sigma \\
2\frac{g_0}{V_{\rm c}^2}-f_0g_0\xi^2\sigma \\
0
\end{array}\right). \label{eigens}
\end{multline}

Equation \eqref{zetasigma} by itself cannot characterize the general deviation of the shock position from the self-similar one, because at any given time we can measure the shock velocity, acceleration, and all higher order derivatives, and a single value of $\zeta_{\sigma}$ will not be able to accommodate all of these constraints. However, \emph{if there is an infinite number of Eigenvalues}, then we can write

\begin{equation}
\zeta = \sum_{\sigma}\zeta_{\sigma}e^{\sigma\tau},
\end{equation}
and we can relate the set $\{\zeta_{\sigma}\}$ to the properties of the shock at a given time. The total solution for $F_1$ is then

\begin{equation}
F_1 = \sum_{\sigma} \zeta_{\sigma}F_{\sigma}e^{\sigma\tau}, \label{Feq}
\end{equation}
and similarly for the other variables. It is straightforward to show that, if $\sigma$ is a solution to the Eigenvalue equation, then the complex conjugate $\sigma^*$ is also a solution. Therefore, to ensure that the shock position and post-shock fluid quantities are purely real, we require $\zeta_{\sigma} = \zeta_{\sigma}^*$, or that $\zeta_{\sigma}$ also be purely real.

\subsection{Shock Position and Velocity}
{The defining equation for $\zeta$, given by Equation \eqref{zetaeq}, can be rearranged to yield the differential equation for the position of the shock as a function of time:}

\begin{equation}
    \left(1-2\sum_{\sigma}\zeta_{\sigma}e^{\sigma\tau}\right)^{1/2}\frac{d}{dt}\left[e^{\frac{3\tau}{2}}\right] = \frac{3}{2}\frac{V_{\rm c}\sqrt{GM}}{r_{\rm a}^{3/2}}. \label{Rtot}
\end{equation}
{We reiterate that the $\zeta_{\sigma}$ in this expression are determined by the deviations from the shock velocity, acceleration, and all higher-order derivatives at $\tau = 0$. Therefore, \emph{once these quantities are specified}, this differential equation can simply be integrated numerically to solve for the position of the shock given some initial condition.}

{While the numerical integration of the above equation for $R(t)$ is straightforward, since $\zeta(\tau)$ is assumed to be small, } we can also write $R = R_0+R_1$, $V = V_0+V_1$, with $R_0$ and $V_0$ the solutions to the zeroth-order terms in Equation \eqref{Rtot} and $R_1$ and $V_1 = dR_1/dt$ satisfying the first-order corrections induced by finite $\zeta$. Taylor expanding about $\zeta$ and using our solution for $\zeta$ in terms of the Eigenmodes in Equation \eqref{zetaeq} then gives

\begin{equation}
V_0 = V_{\rm c}\sqrt{\frac{GM}{R_0}},
\end{equation}
\begin{equation}
\frac{V_1}{V_0}+\frac{1}{2}\frac{R_1}{R_0} = \sum_\sigma \zeta_{\sigma}e^{\sigma \tau}.
\end{equation}
Due to the arbitrariness of the temporal origin, we can, without loss of generality, define the perturbation to the position of the shock to be zero at some $\tau_0$, which we can also define to be zero owing to the scale invariance of the problem (recall that $\tau = \ln (R / r_{\rm a})$). 
The unperturbed shock position is therefore

\begin{equation}
    R_0(t) = r_{\rm a}\left(1+\frac{3}{2}\frac{V_{\rm c}\sqrt{GM}}{r_{\rm a}^{3/2}}t\right)^{2/3},
\end{equation}
and the solutions for the perturbations to the shock position and velocity are

\begin{equation}
\frac{R_1}{R_0} = \sum_\sigma\frac{\zeta_\sigma}{\frac{3}{2}+\sigma}\left(e^{\sigma\tau}-e^{
-\frac{3}{2}\tau}\right), \label{R1pert}
\end{equation}
\begin{equation}
\frac{V_1}{V_0} = \sum_\sigma \frac{\zeta_\sigma}{\frac{3}{2}+\sigma}\left(\left(1+\sigma\right)e^{\sigma \tau}+\frac{1}{2}e^{
-\frac{3}{2}\tau}\right). \label{V1pert}
\end{equation}

{Note that, in Equation \eqref{R1pert}, the solution for the perturbation to the shock position appears not as a sum of modes, but as a sum of differences between modes with Eigenvalues $\sigma$ and $-3/2$. However, the latter is not a true Eigenvalue, as it is merely a consequence of the scale invariance of the problem -- we can \emph{define} the perturbation to the shock position to be zero at $\tau = 0$, and the second factor in parentheses in Equation \eqref{R1pert} simply enforces this initial condition. The insignificance of the $-3/2$ ``mode'' is even more apparent if we simply numerically integrate the differential equation for $R(t)$ \eqref{Rtot}, as in this case there is no direct appearance of this temporal dependence.} 

\subsection{Post-shock fluid quantities in unperturbed variables}
The total solution for the fluid velocity, density, and pressure behind the shock generated by perturbations to the shock position are

\begin{equation}
v = V(t)\left\{f_0(\xi)+f_1(\xi,\tau)\right\} = V(t)\left(f_0(\xi)+\frac{1}{g_0\xi^2}\sum_{\sigma}\zeta_\sigma e^{\sigma\tau}F_\sigma e^{\sigma\tau}\right), \label{vrtot}
\end{equation}
\begin{equation}
    \rho = R(t)^{-n}\left\{g_0(\xi)+g_1(\xi,\tau)\right\}= R(t)^{-n}\left(g_0(\xi)+g_0(\xi)\sum \zeta_{\sigma}e^{\sigma\tau}G_{\sigma}\right)
\end{equation}
\begin{equation}
    p = R(t)^{-n}V(t)^2\left\{h_0(\xi)+h_1(\xi,\tau)\right\} = R(t)^{-n}V(t)^2\left\{h_0(\xi)+h_0(\xi)\sum_{\sigma}\zeta_{\sigma}e^{\sigma\tau}\left(s_{\sigma}+\gamma G_{\sigma}\right)\right\}, \label{prtot}
\end{equation}
where $V(t)$ is the total shock velocity, $\xi = r/R$ is in terms of the total shock position, and $\tau = \ln (R/r_{\rm a})$ is the log of the true shock position. Again, once $\zeta_{\sigma}$ is specified, we can use our derived expressions for the total shock position and velocity to completely constrain the deviations from self-similarity using the Eigenmodes $F_{\sigma}$, and write the result in terms of the physical coordinates $r$ and $t$.

As we saw in the previous subsection, we can also decompose the shock velocity and shock position into perturbed and unperturbed parts, $V = V_0+V_1$ and $R = R_0+R_1$. We can therefore use these first-order corrections to derive expressions for the velocity, density, and pressure profiles in terms of the unperturbed shock velocity, unperturbed shock position, and unperturbed self-similar variable $\xi_0 = r/R_0(t)$, which are equivalent to using the true shock velocity, true shock position, and total self-similar variable to first order in $\zeta$.  The result is 

\begin{equation}
v(r,t) = V_0\left\{f_0(\xi_0)+\sum_\sigma \zeta_{\sigma}f_{\sigma}(\xi_0,\tau)\right\}, \label{vrt}
\end{equation}
where

\begin{equation}
f_\sigma(\xi_0,\tau) = \left(\frac{\left(1+\sigma\right)f_0-\xi_0f_0'}{\frac{3}{2}+\sigma}+\frac{F_\sigma}{g_0\xi_0^2}\right)e^{\sigma\tau}+\frac{1}{\frac{3}{2}+\sigma}\left(\frac{1}{2}f_0+f_0'\xi_0\right)e^{
-\frac{3}{2}\tau}. \label{fsig}
\end{equation}
Similarly, the density and pressure are

\begin{equation}
\rho(r,t) = \rho_{\rm a}\left(\frac{R_0}{r_{\rm a}}\right)^{-n}\left\{g_0(\xi_0)+\sum_\sigma \zeta_{\sigma}g_{\sigma}\right\},
\end{equation}
\begin{equation}
g_{\sigma}(\xi_0,\tau) = \left(-\frac{ng_0+\xi_0g_0'}{\frac{3}{2}+\sigma}+g_0G_\sigma\right)e^{\sigma\tau}+\frac{ng_0+\xi_0g_0'}{\frac{3}{2}+\sigma}e^{
-\frac{3}{2}\tau}, \label{gsig}
\end{equation}
\begin{equation}
p(r,t) = \rho_0\left(\frac{R_0}{r_{\rm a}}\right)^{-n}V_0^2\left\{h_0(\xi_0)+\sum_\sigma \zeta_\sigma h_{\sigma}(\xi_0,\tau)\right\}, 
\end{equation}
\begin{equation}
h_\sigma(\xi_0,\tau) = \left(-\frac{nh_0+\xi_0h_0'-2\left(1+\sigma\right)h_0}{\frac{3}{2}+\sigma}+\gamma h_0G_{\sigma}+h_0s_{\sigma}\right)e^{\sigma\tau}+\frac{\left(n+1\right)h_0+\xi h_0'}{\frac{3}{2}+\sigma}e^{
-\frac{3}{2}\tau}. \label{hrt}
\end{equation}
Equations \eqref{vrt} -- \eqref{hrt} are the self-consistent, first-order expressions for the post-shock velocity, density, and pressure in terms of the physical radial coordinate $r$. To lowest order in $\zeta$, these expressions are identical to Equations \eqref{vrtot} -- \eqref{prtot}, which use the true shock position, velocity, and self-similar variable.

\subsection{Total energy and energy flux}
The energy equation takes the form

\begin{equation}
\frac{\partial \mathscr{E}}{\partial t}+\frac{\partial \mathscr{F}_{}}{\partial r} = 0,
\end{equation}
where

\begin{equation}
\mathscr{E} = \left(\frac{1}{2}v^2+\frac{1}{\gamma-1}\frac{p}{\rho}-\frac{GM}{r}\right)r^2\rho
\end{equation}
and

\begin{equation}
\mathscr{F}_{} = \left(\frac{1}{2}v^2+\frac{\gamma}{\gamma-1}\frac{p}{\rho}-\frac{GM}{r}\right)r^2\rho v \label{eflux}
\end{equation}
are the energy density and energy flux. Integrating from $r = 0$ to $R+\epsilon$ and taking the limit as $\epsilon \rightarrow 0$ gives 

\begin{equation}
\frac{\partial E_{\rm sh}}{\partial t} = 4\pi V\mathscr{E}_{\rm a}(R)+4\pi \mathscr{F}_{}(0),
\end{equation}
where

\begin{equation}
E_{\rm sh} = 4\pi \int_0^{R}\mathscr{E}dr
\end{equation}
is the total energy behind the shock and 

\begin{equation}
\mathscr{E}_{\rm a}(R) =  GMr_{\rm a}\rho_{\rm a}\frac{1-\gamma}{\gamma}\left(\frac{R}{r_{\rm a}}\right)^{1-n}
\end{equation}
is the energy density of the ambient medium at the location of the shock. After expanding out to first order this expression becomes

\begin{equation}
\frac{\partial E_{\rm sh}}{\partial t} = \frac{1-\gamma}{\gamma}4\pi GMr_{\rm a}\rho_{\rm a}V_0\left(\frac{R_0}{r_{\rm a}}\right)^{1-n}\left(1+\frac{V_1}{V_0}+\left(1-n\right)\frac{R_1}{R_0}\right)+4\pi \mathscr{F}_{0}(0)+4\pi \mathscr{F}_{1}(0), \label{Edot}
\end{equation}
where

\begin{equation}
\mathscr{F}_{ 0} = GMr_{\rm a}\rho_{\rm a}V_0\left(\frac{R_0}{r_{\rm a}}\right)^{1-n}f_0g_0\xi^2\left(\frac{1}{2}V_{\rm c}^2f_0^2+\frac{\gamma}{\gamma-1}V_{\rm c}^2\frac{h_0}{g_0}-\frac{1}{\xi}\right)
\end{equation}
and 

\begin{multline}
\mathscr{F}_{1} = GMr_{\rm a}\left(\frac{R_0}{r_{\rm a}}\right)^{1-n}V_0\xi^2g_0f_0\bigg\{\left(\frac{1}{2}V_{\rm c}^2f_0^2+\frac{\gamma}{\gamma-1}V_{\rm c}^2\frac{h_0}{g_0}-\frac{1}{\xi}\right)\left(\frac{g_1}{g_0}+\frac{f_1}{f_0}+\left(2-n\right)\frac{R_1}{R_0}+\frac{V_1}{V_0}\right) \\ 
+V_{\rm c}^2f_0f_1+V_{\rm c}^2f_0^2\frac{V_1}{V_0}+\frac{\gamma}{\gamma-1}V_{\rm c}^2\frac{h_0}{g_0}\left(2\frac{V_1}{V_0}+\frac{h_1}{h_0}-\frac{g_1}{g_0}\right)+\frac{1}{\xi}\frac{R_1}{R_0}\bigg\}. \label{Flux1}
\end{multline}
The total correction to the energy therefore has two terms, one arising from the additional sweeping up of binding energy from the ambient medium, and another from the change in the flux of energy onto the black hole. 

\subsection{Total mass and accretion rate}
Multiplying the continuity equation \eqref{cont} by $r^2$ and integrating from zero to $R+\epsilon$ gives, after taking the limit as $\epsilon\rightarrow 0$, 

\begin{equation}
\frac{\partial M_{\rm sh}}{\partial t} = 4\pi V\rho_{\rm a}(R)R^2+\dot{M}(0),
\end{equation}
where

\begin{equation}
M_{\rm sh} = \int_0^{R}4\pi r^2\rho dr
\end{equation}
is the total mass contained behind the shock and

\begin{equation}
\dot{M} = 4\pi r^2\rho v
\end{equation}
is the mass flux. Expanding to first order, we find

\begin{equation}
\frac{\partial M_{\rm sh}}{\partial t} = \rho_{\rm a} r_0^2 V_0\left(\frac{R_0}{r_0}\right)^{2-n}\left(1+\frac{V_1}{V_0}+\left(2-n\right)\frac{R_1}{R_0}\right) + \dot{M}_0(0)+\dot{M}_1(0),
\end{equation}
where

\begin{equation}
\dot{M}_0 = 4\pi\rho_{\rm a}r_{\rm a}^2V_0\left(\frac{R_0}{r_{\rm a}}\right)^{2-n}f_0g_0\xi^2
\end{equation}
and

\begin{equation}
\dot{M}_1 = \dot{M}_0\left(\frac{f_1}{f_0}+\frac{g_1}{g_0}+\frac{V_1}{V_0}+\left(2-n\right)\frac{R_1}{R_0}\right). \label{mdot1}
\end{equation}
As was true for the energy, the total mass behind the shock is modified both by the change in the rate at which material is swept up, and also the change in the removal of material at the origin. This expression also shows that the perturbation to the accretion rate onto the black hole is 

\begin{equation}
\dot{M}_{\bullet, 1} = -\dot{M}_0\lim_{\xi \rightarrow 0}\left(\frac{f_1}{f_0}+\frac{g_1}{g_0}+\frac{V_1}{V_0}+\left(2-n\right)\frac{R_1}{R_0}\right).
\end{equation}
The minus sign appears here because the rate at which matter leaves the post-shock fluid equals the rate at which matter is accreted by the black hole.

\subsection{Comparison to previous approaches}
\label{sec:previous}
\citet{vishniac83} analyzed the stability of the Sedov-Taylor blastwave in the limit that $\gamma \rightarrow 1$, in which case the solution is an infinitely-thin shell, and \citet{ryu87} extended this analysis to blastwaves of finite thickness (i.e., the Sedov-Taylor blastwave in a constant-density ambient medium and post-shock adiabatic index $\gamma > 1$). \citet{goodman90} used a Lagrangian approach to analyze the stability of hollow Sedov-Taylor blastwaves, \citet{chevalier90} investigated the stability of accelerating shock waves in an isothermal atmosphere, and \citet{sari00} assessed the stability of the accelerating, type-II solutions obtained by \citet{waxman93}, which are valid when the density profile of the ambient medium falls off more steeply than $\rho \propto r^{-3}$. 

The formalism in all of these past analyses roughly paralleled ours, being to seek solutions to the perturbation equations that are separable in the self-similar variable and $\ln t$. The requirement that the fluid quantities obey a fourth boundary condition then results in the existence of an Eigenvalue that characterizes the growth or decay of the perturbations. The one, major difference is that all of these past investigations chose to initially decompose the perturbations in terms of the perturbed and unperturbed shock position and velocity, and to work in terms of the unperturbed self-similar variable $\xi_0 = r/R_0(t)$. In contrast, our fluid variables were defined in terms of the total shock velocity and position, and we worked in the total self-similar variable $\xi = r/R(t)$; we also defined the perturbation to the shock velocity and position in an indirect way, specifically Equation \eqref{zetaeq}. These choices were motivated from the physical standpoint that, for a shock with arbitrary velocity $V(t)$ and position $R(t)$, one does not \emph{a priori} know if a self-similar solution exists. Nonetheless, we can always describe the fluid variables in terms of these general shock variables (i.e., we are free to make the change of variables $\{t,r\} \rightarrow \{t,r/R\}$ without any knowledge of how to decompose $R$ into unperturbed and perturbed parts); self-similar solutions are merely a subset of solutions for which $R(t)$ and $V(t)$ satisfy very specific relations.

{Moreover, as we demonstrated in Section \ref{sec:equations}, the condition for preserving exact self-similarity is a differential equation between the shock velocity and the position (specifically Equation \ref{ssexact}). Therefore, the most general means of characterizing deviations from self-similarity is by introducing perturbations to this differential equation, which naturally leads to our Equation \eqref{zetaeq}. By pursuing this route, we maintain the scale invariance of the problem (i.e., we are always free to rescale the radial length scale of the ambient medium $r_{\rm a}$ in Equations \ref{rhoambient} and \ref{pambient} and origin of time). By defining the perturbations in this way (as opposed to perturbing the shock position and velocity about the self-similar solutions), we also recover the exact Eigenvalue $\sigma = -3/2$, which is merely a manifestation of the fact that we can always rescale the perturbation to the shock position to be zero at $t = 0$.}

While this choice of coordinates and parameterization of the deviation from self-similarity do not change the fundamental nature of the solutions, they greatly simplify the appearance of the equations and the boundary conditions for the perturbed variables. The latter feature is especially pronounced for the Sedov-Taylor blastwave, where the perturbations satisfy homogeneous boundary conditions at the shock front if we choose to adopt the approach used in this paper. In Appendix \ref{sec:appendixA}, we use this result to show that the Sedov-Taylor blastwave is trivially stable to radial perturbations for all $n$ and $\gamma$, provided that the Sedov-Taylor solution extends to the origin, and we derive analytically the solutions for the perturbations to the fluid quantities. 

Defining the perturbations via Equation \eqref{zetaeq} also gives insight into why, in every past investigation of the stability of blastwaves, there were always \emph{two} Eigenvalues that characterize the behavior of radial perturbations to the shock quantities (see, e.g., Figure 2 of \citealt{sari00}): if we return to Equation \eqref{R1pert}, we see that every Eigenmode that describes the evolution of the shock has two temporal frequencies, being $\sigma$ itself and $-3/2$, and these two frequencies also appear in the definitions of the fluid variables. In our formalism, the decaying mode with Eigenfrequency $-3/2$ is not, in fact, a distinct mode, and is only an artifact of the initial conditions (being that we can always adjust the temporal origin such that the perturbed and unperturbed shock positions coalign), and our freedom to enforce these initial conditions is apparent from the fact that our definition of $\zeta$ is a differential equation that relates $V_1$ and $R_1$. However, we could have also chosen to parameterize the perturbations by $R_1 \propto e^{\sigma\tau}$; in this case the existence of this other mode would not have been obvious, and there would have appeared to have been a solution to the perturbation equations with an Eigenfrequency of exactly $-3/2$. We show in Appendix \ref{sec:appendixA} that the arbitrariness of both the initial shock position \emph{and the initial shock velocity} for the Sedov-Taylor blastwave implies the existence of two, exact solutions for the evolution of radial perturbations to the Sedov-Taylor blastwave, being $\sigma = 0$ and $\sigma = -(5-n)/2$, where $n$ describes the power-law profile of the density of the ambient medium (since the unperturbed shock position scales as $R_0 \propto t^{(5-n)/2}$, the latter mode falls off as $t^{-1}$, identically to that of the CQR solution). 

\section{Solutions}
\label{sec:solutions}
The previous section set out a general formalism for analyzing the temporal and spatial evolution of perturbations to the self-similar shock position and the fluid behind the shock. The general solutions were written in terms of Eigenmodes, with the Eigenfrequencies determined by the continuity of the fluid variables through a sonic point. In this section we use this formalism to derive the Eigenfrequencies and the Eigenfunctions for different values of $n$, $\gamma_1$, and $\gamma_2$ to determine the stability of the self-similar solutions.

\subsection{A specific example: $n = 2.5$, $\gamma_1 = \gamma_2 = 1.4$}
\label{sec:specific}
\begin{figure}[htbp] 
   \centering
   \includegraphics[width=0.495\textwidth]{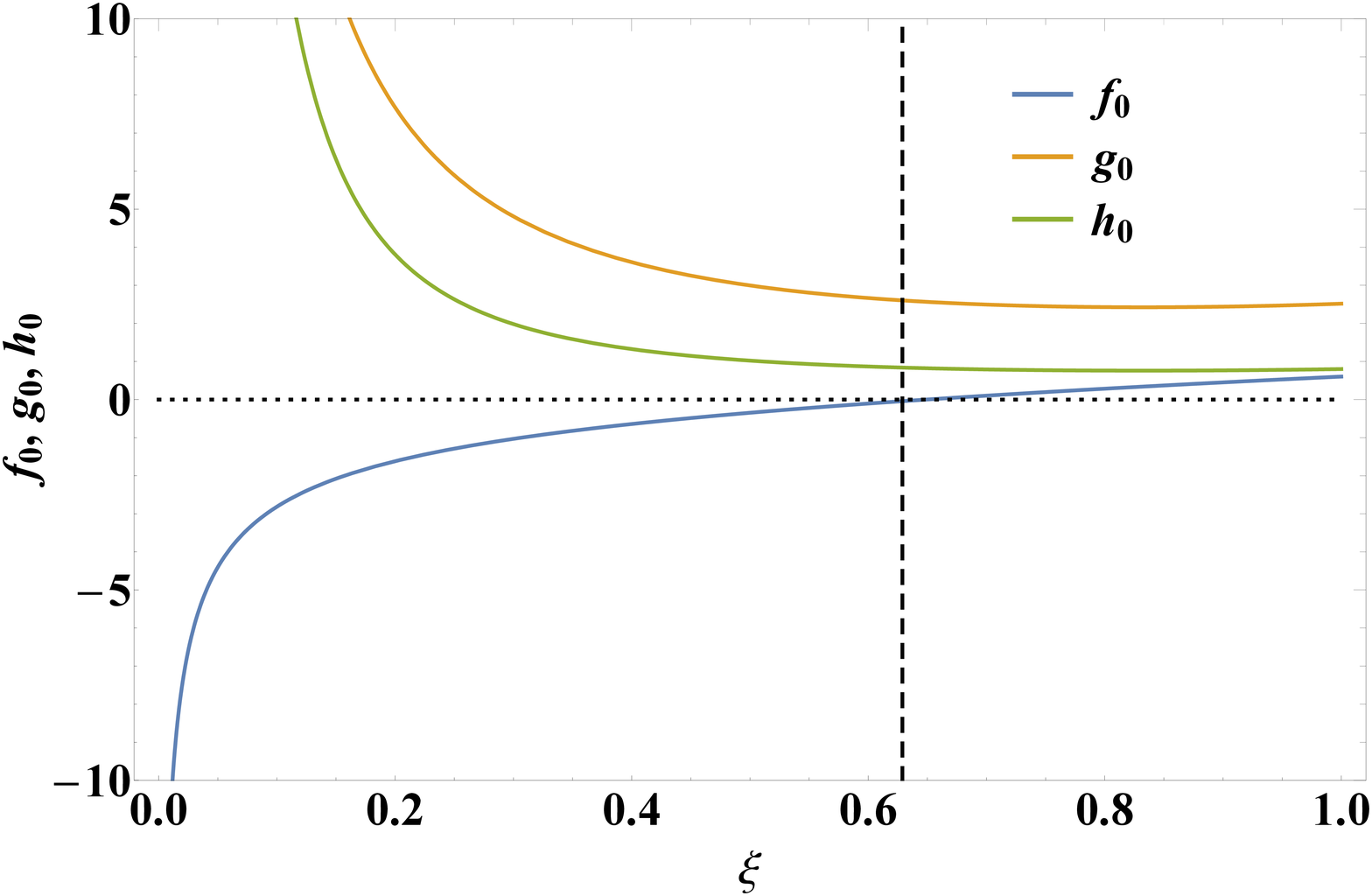} 
    \includegraphics[width=0.495\textwidth]{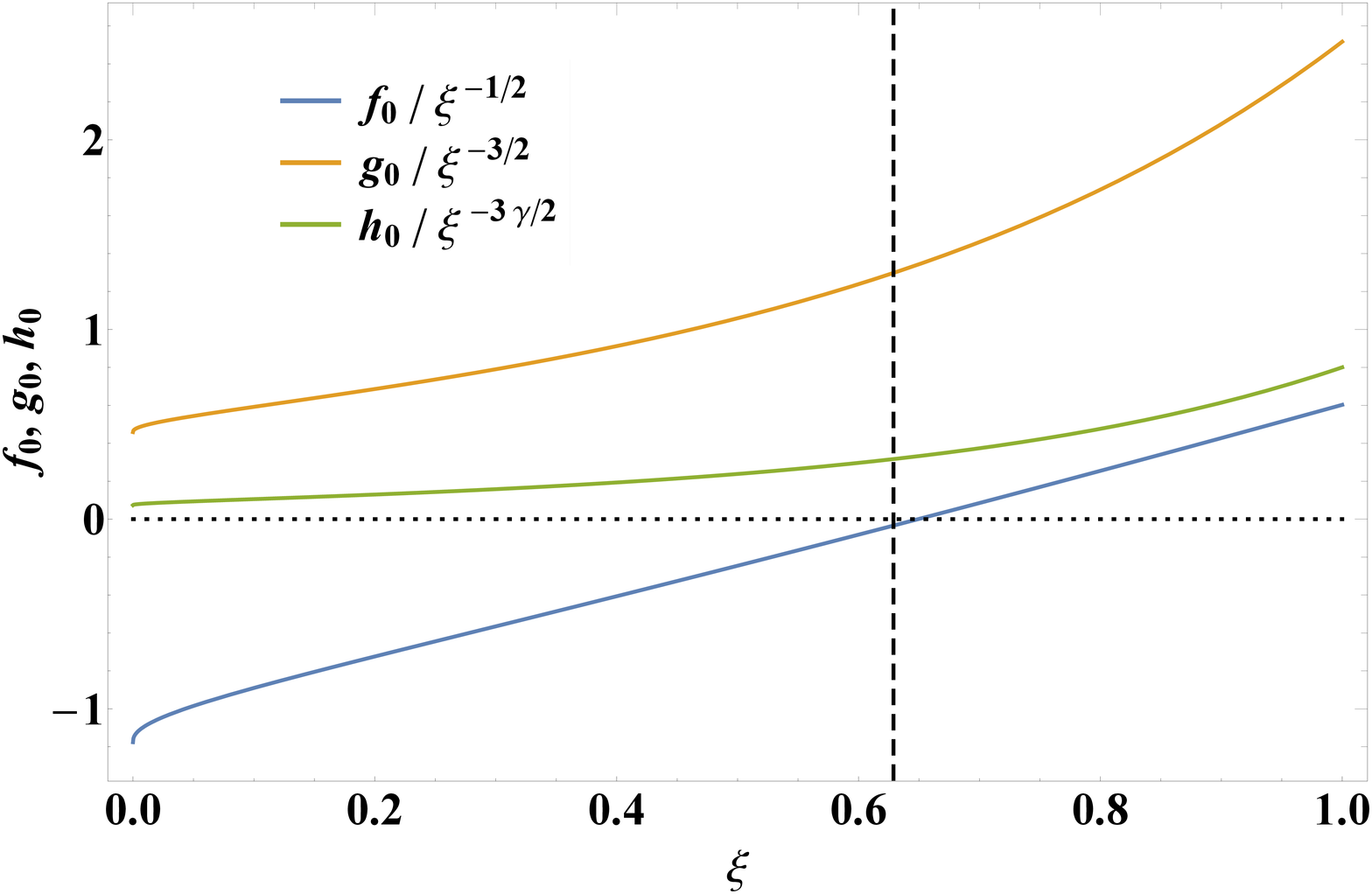} 
     \includegraphics[width=0.495\textwidth]{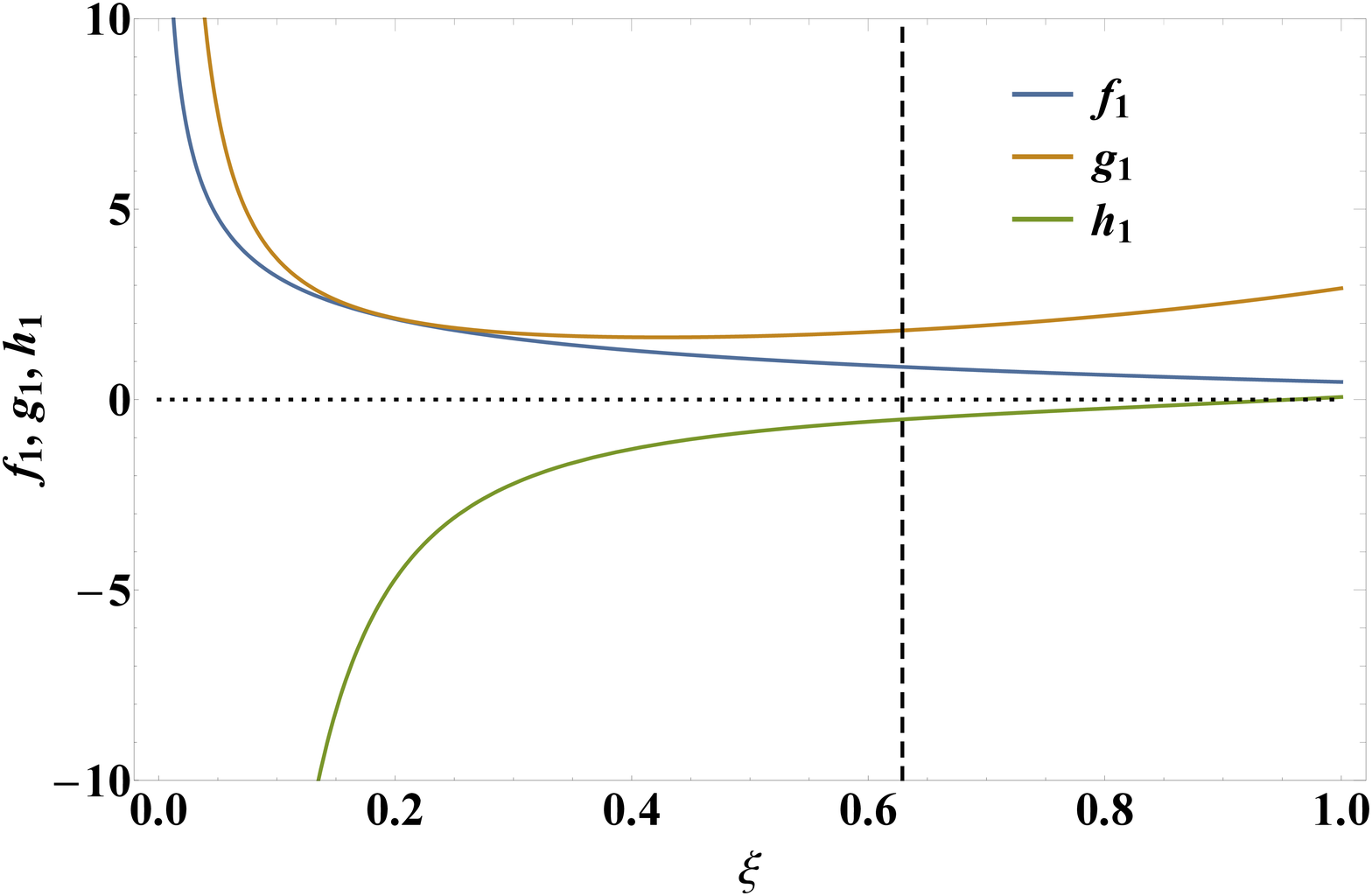} 
    \includegraphics[width=0.495\textwidth]{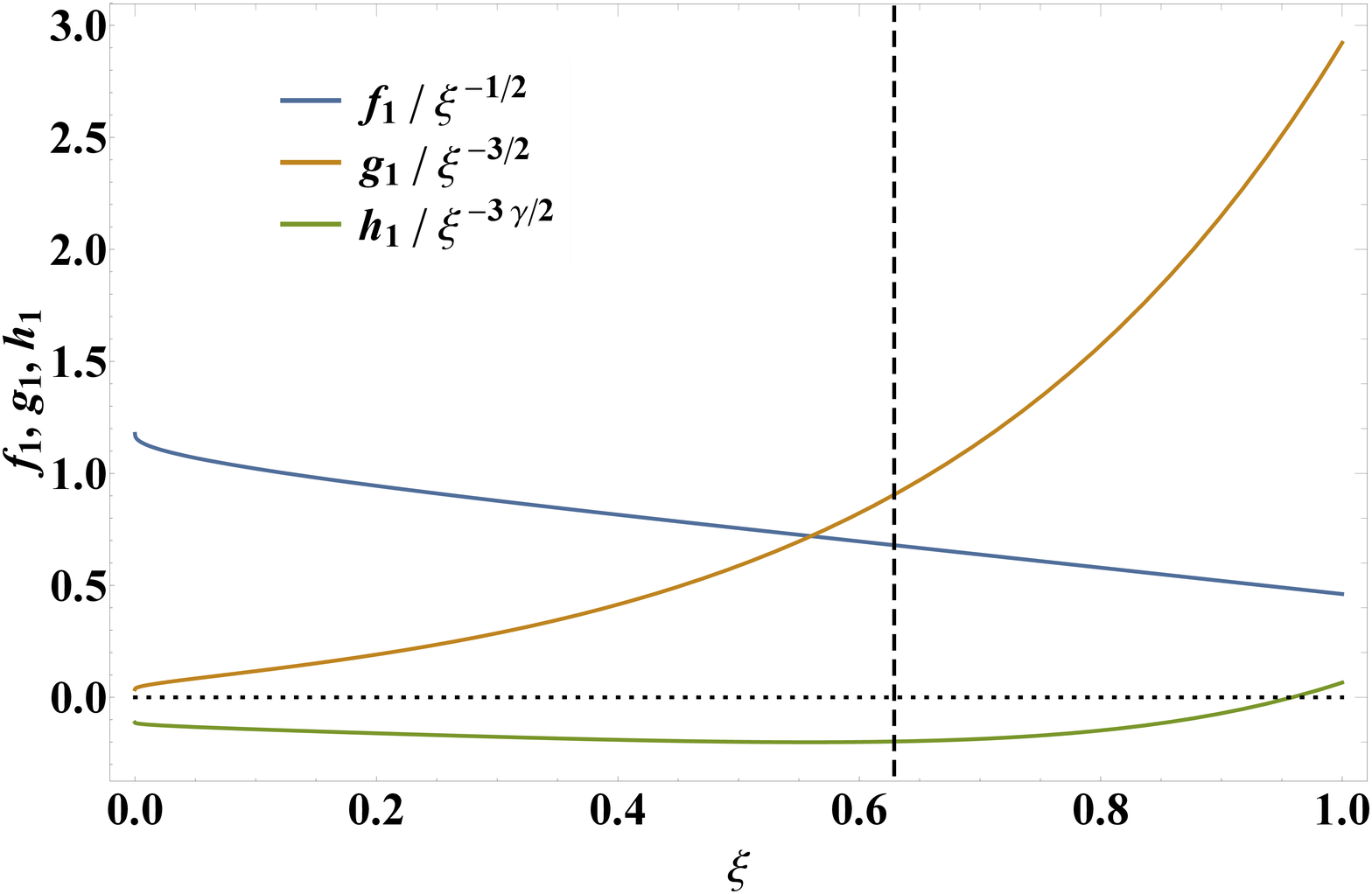} 
   \caption{Top left: The unperturbed, self-similar velocity, $f_0$, density, $g_0$, and pressure $h_0$ of Paper I when the density of the ambient medium falls off as $\rho \propto r^{-2.5}$ and the adiabatic indices of both the pre and post-shock gas are $\gamma_1 = \gamma_2 = 1.4$. The vertical, dotted line shows the location of the sonic point at $\xi_{\rm c} \simeq 0.629$. Top right: The same functions as in the top-left panel, but normalized by their adiabatic, freefall scalings near the origin. Bottom left: The Eigenfunctions $f_{\sigma}$, $g_{\sigma}$, and $h_{\sigma}$ for the growing mode when the density falls off as $\rho \propto r^{-2.5}$ and $\gamma_1 = \gamma_2 = 1.4$. Bottom right: The same functions as in the bottom-left panel, but normalized by their scalings near the origin.}
   \label{fig:fgh}
\end{figure}

Here we focus on the case where the ambient density falls off as $\rho \propto r^{-2.5}$, and the pre- and post-shock adiabatic indices are equal to their polytropic indices of $\gamma = 1+1/n = 1.4$. These parameters were used to compare to weak shock propagation in the hydrogen envelope of a yellow supergiant in Paper I, and therefore have physical relevance. The critical shock velocity is $V_{\rm c} \simeq 1.202$, and the top-left panel of Figure \ref{fig:fgh} shows the solutions for the self-similar velocity, density, and pressure that pass through the critical point $\xi_{\rm c} \simeq 0.629$ and result in accretion at the origin. The asymptotic behavior of these functions near the origin is $f_0(\xi) \propto \xi^{-1/2}$, so the velocity approaches freefall, $g_0(\xi) \propto \xi^{-3/2}$, so the density approaches its time-independent, freefall value, and $h_0(\xi) \propto g_0(\xi)^{\gamma}$, so the flow is isentropic (the latter condition actually holds over the entirety of the post-shock gas, and the flow is exactly isentropic everywhere; this is not true if $\gamma_2 \neq 1+1/n$). These scalings are depicted in the top-right panel of Figure \ref{fig:fgh}. 

\subsubsection{Growing Eigenmode}
To determine the Eigenfrequencies, we integrate Equation \eqref{eigens} numerically from the shock front ($\xi = 1$) inwards. An arbitrarily chosen $\sigma$ will result in the divergence of the solutions near the sonic point, so we iteratively perturb the value of the Eigenfrequency until the fluid quantities smoothly pass through the critical point. We find that there is one purely real and \emph{positive} Eigenfrequency that both satisfies the boundary conditions at the shock and results in the continuity of the functions through the critical point when $n = 2.5$ and $\gamma_1 = \gamma_2 = 1.4$, and its value is

\begin{equation}
\sigma_{} \simeq 0.175311. 
\end{equation}
We do not find any other purely real or growing modes (though there are complex and decaying modes; see below). The bottom-left panel of Figure \ref{fig:fgh} shows the Eigenmodes for the velocity, $f_{\sigma}$, the density, $g_0 G_\sigma$, and the pressure, $h_0(s_{\sigma}+\gamma G_\sigma)$, for this growing mode, which satisfy Equation \eqref{eigens} and smoothly pass through the critical point at $\xi_{\rm c} \simeq 0.629$. The fact that these functions approach finite numbers near the origin imply that the perturbations to the velocity, density, and pressure scale identically to their unperturbed quantities near the black hole. 

Because this Eigenvalue is positive and real, the amplitudes of the perturbations grow with time, and at late times this mode dominates over all of the others. Returning to Equations \eqref{R1pert} and \eqref{V1pert}, good approximations for the late-time evolution of the shock position and velocity are therefore

\begin{equation}
R \simeq R_0\left(1+\zeta\frac{1}{\frac{3}{2}+\sigma}e^{\sigma\tau}\right), \quad V_1 \simeq V_0\left(1+\zeta\frac{1+\sigma}{\frac{3}{2}+\sigma} e^{\sigma\tau}\right). \label{RVtot}
\end{equation}
If we further recall that $\tau = \ln (R_0 / r_{\rm a})$ and that $R_0 \simeq t^{2/3}$, then using $\sigma \simeq 0.175$ allows us to rewrite the above as

\begin{equation}
\frac{\Delta R}{R_0} \propto \frac{\Delta V}{V_0} \propto t^{0.117}, \label{Deltaeqs}
\end{equation}
where $\Delta R$ and $\Delta V$ are the differences between the full and unperturbed shock position and velocity, respectively. These expressions demonstrate that, consistent with the unstable nature of the mode, perturbations to the shock on top of the self-similar solutions grow at late times and cause the solutions to asymptotically deviate from their self-similar prescriptions. However, the rate at which these deviations grow is \emph{extremely slow}, both because the growth is in the form of a power-law, and not an exponential, and the power-law index itself is very close to zero; it would take, for example, roughly 10 orders of magnitude in time for a perturbation to grow by an additional factor of 10. Therefore, for perturbations that are not so large that nonlinear effects start to become important, the shock properties should be very well reproduced by Equation \eqref{RVtot} even for very late times; we validate this notion in Section \ref{sec:simulations} where we compare these predictions to numerical simulations.

\subsubsection{Higher order modes}
As we noted above, the Eigenmode approach is only consistent if there is an infinite number of modes, as this allows us to completely describe the shock position and all higher order derivatives at a given time. Finding the higher order modes numerically is challenging, as the divergence of the functions near the singular point becomes very weak over certain ranges of the Eigenvalue. To minimize the numerical uncertainty related to the divergence of the functions near the critical point, we first reduce the set of Eigenvalue equations \eqref{eigens} analytically to a single, second-order equation for $F_{\sigma}''$ (the equation is only second order when $\gamma = 1+1/n$ and the unperturbed entropy is exactly constant). We then use L'H$\hat{o}$pitals rule to solve for the relationship between $F_{\sigma}$ and $F_{\sigma}'$ at the critical point, and we integrate inward from the shock front and seek solutions that satisfy this constraint. However, near the critical point we replace the full form of the right-hand side of the equation, which diverges for solutions that do not satisfy the critical point condition, with the reduced right-hand side that we obtain from L'H$\hat{o}$pital's rule under the assumption that the function is continuous through the critical point. Therefore, the equation always integrates continuously through the critical point, but its value at the critical point will only be correct if it satisfies the correct relation deduced from L'H$\hat{o}$pitals rule. In this way we assign a meaningful, finite value of the function at the critical point, and minimize the uncertainty related to spurious numerical divergences that arise from maintaining the full right-hand side. 

\begin{figure}[htbp] 
   \centering
   \includegraphics[width=0.495\textwidth]{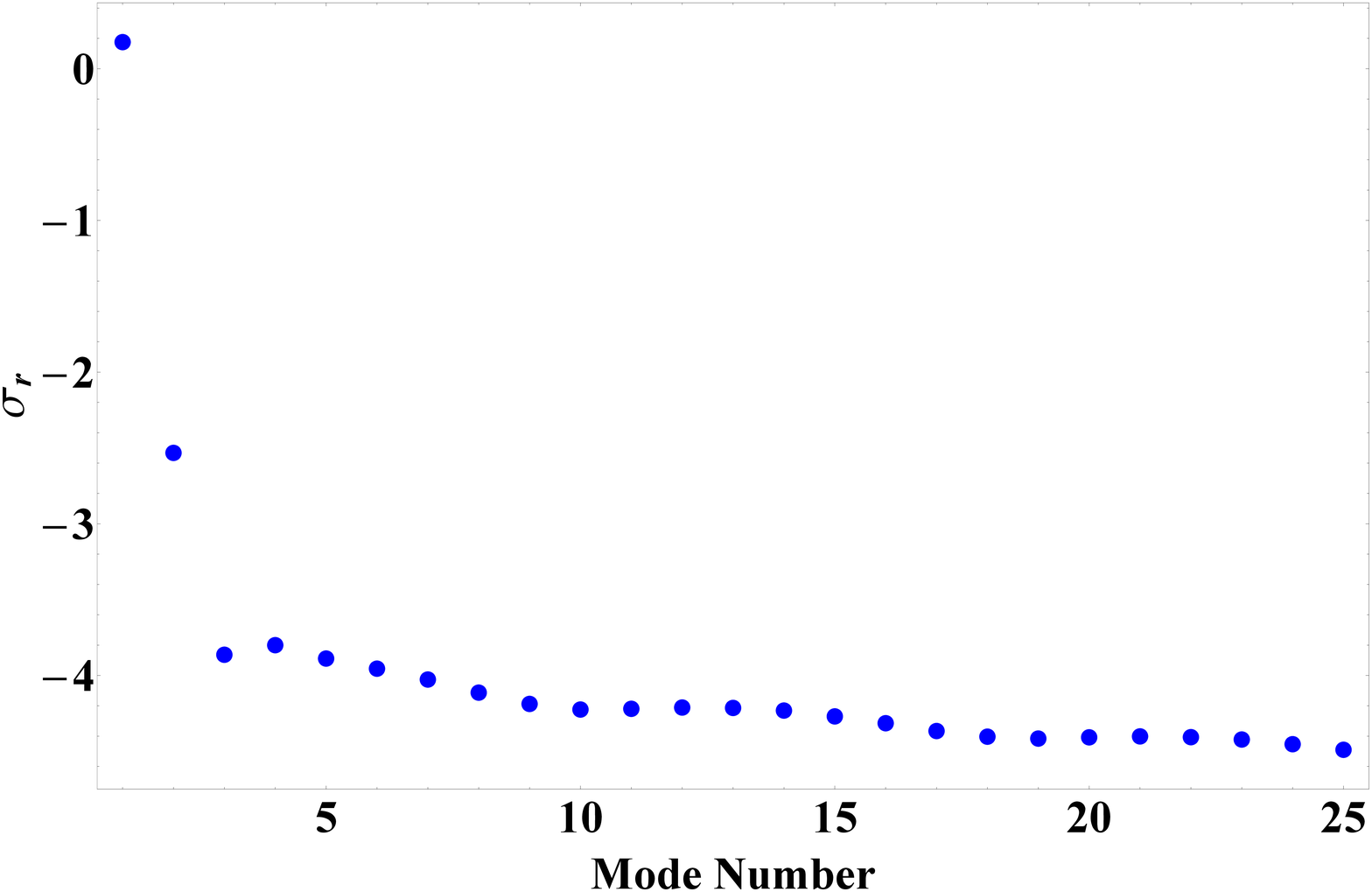} 
   \includegraphics[width=0.495\textwidth]{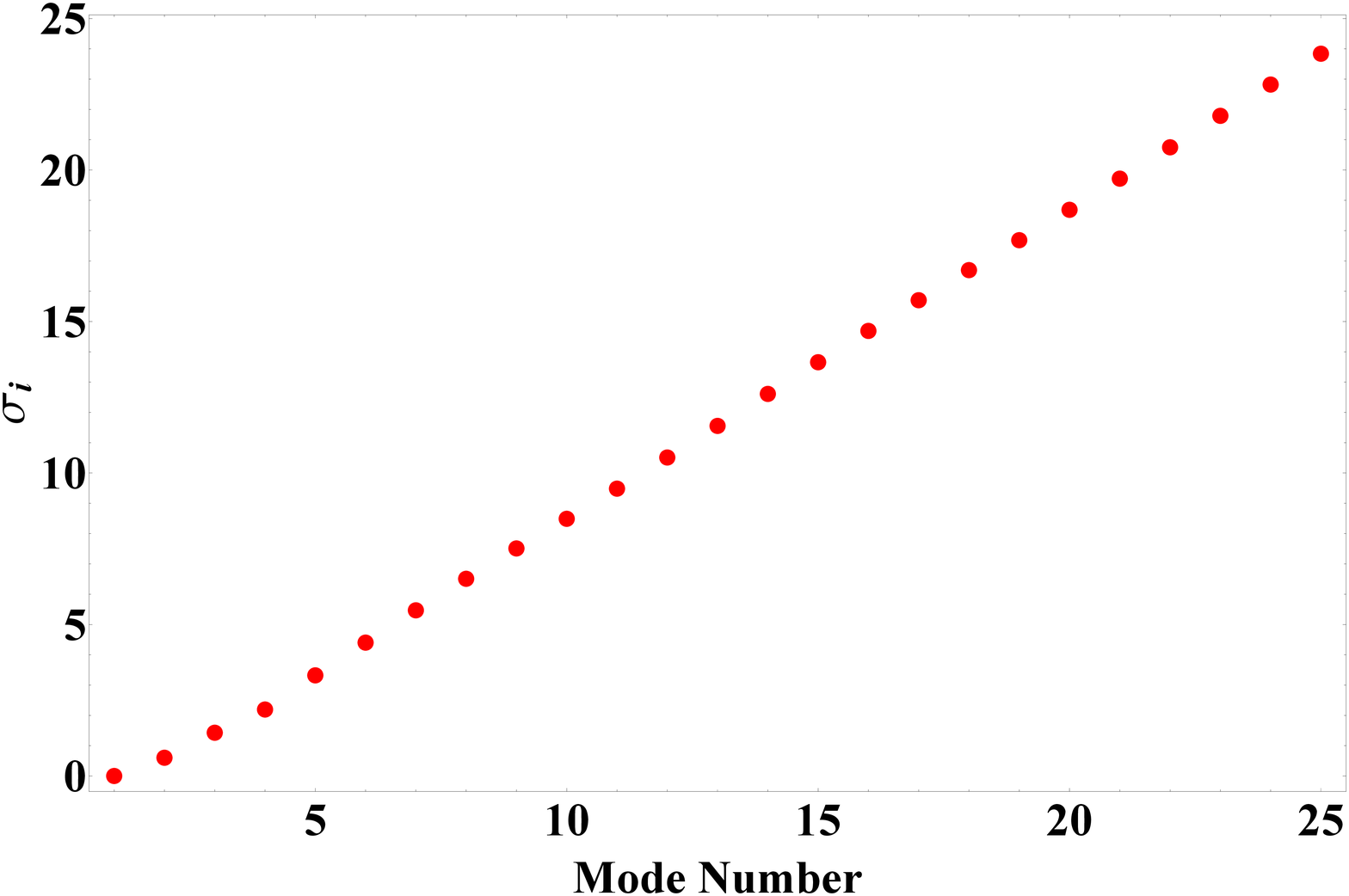} 
   \caption{Left: The real part of the Eigenfrequency for the first 25 modes; we see that the first is slightly positive (and unstable), the next is stable and decays as $\sigma_{\rm r} \simeq -2.5$, and all higher order modes have a frequency that decreases slowly from $\simeq -3.8$. Right: The imaiginary component of the Eigenfrequency, which increases roughly linearly with mode number. }
   \label{fig:sigma}
\end{figure}

Following this approach, we find a series of Eigenmodes that can be ordered by the magnitude of the increasing value of the (absolute value of the) imaginary part of the Eigenvalue; doing so, the next Eigenvalue that we find is $\sigma \simeq -2.53\pm 0.602 i$, which decays and oscillates. The next highest mode we find is $\sigma \simeq -3.86 \pm 1.43 i$; beyond this, the real part of each mode becomes slightly more negative with increasing order, and the imaginary part increases by roughly a factor of one for each higher mode. Defining the Eigenfrequency as $\sigma = \sigma_{\rm r}+i\sigma_{\rm i}$, the left-hand panel of Figure \ref{fig:sigma} shows the real part of the frequency as a function of mode number for the first 25 modes, while the right panel show the imaginary part (there are also solutions with $-\sigma_{\rm i}$). 

\begin{figure}[htbp] 
   \centering
   \includegraphics[width=0.495\textwidth]{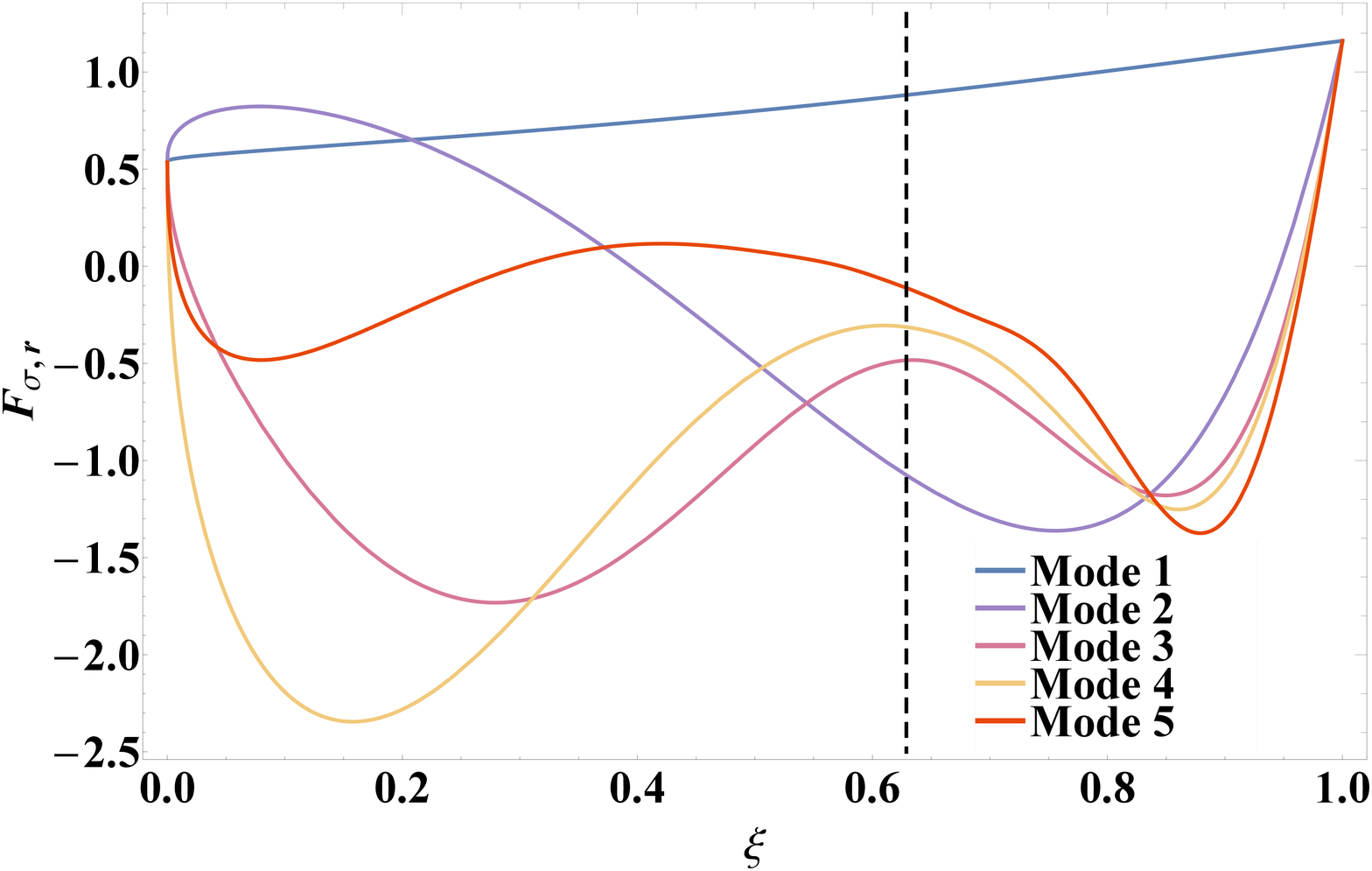} 
   \includegraphics[width=0.495\textwidth]{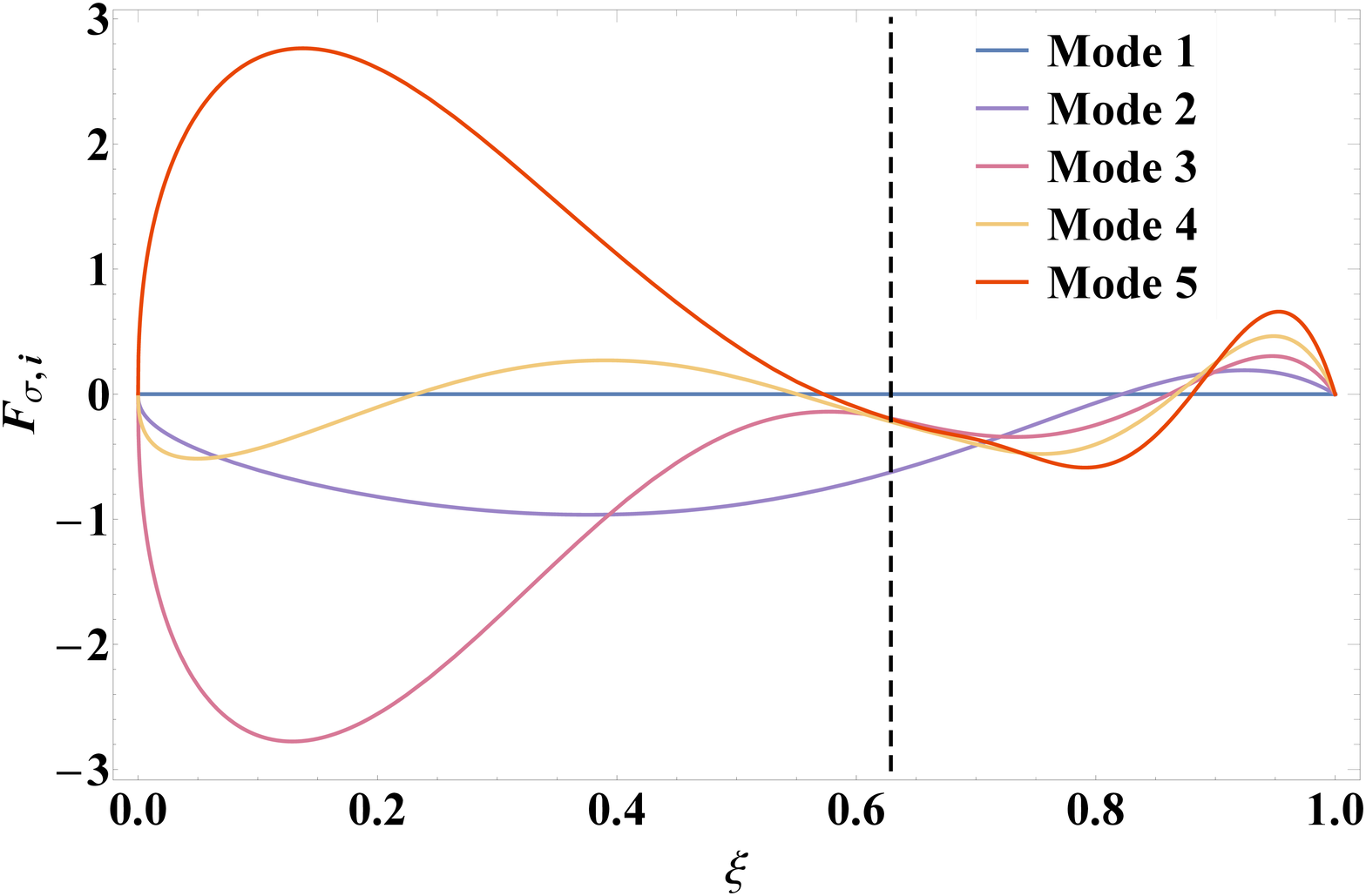} 
   \caption{Left: The real part of the mass flux for the first five Eigenmodes. Right: The imaginary part of the mass flux for the first five Eigenmodes. The vertical, dashed line in each panel shows the location of the sonic point at $\xi_{\rm c} \simeq 0.629$.}
   \label{fig:Fr}
\end{figure}

The Eigenfunctions will also have real and imaginary parts; defining $F_{\sigma} \equiv F_{\sigma, \rm{r}}+i F_{\sigma, \rm{i}}$, the left (right) panel of Figure \ref{fig:Fr} shows the real (imaginary) part of the mass flux $F_{\sigma}$ for the first five Eigenmodes. The fact that these functions all equal a constant near the black hole demonstrates that the perturbations all approach freefall near the origin. Although we have not shown them, the higher order modes for the relative density $G_{\sigma}$ and the entropy $s_{\sigma}$ all show similar behavior, each exhibiting somewhat oscillatory behavior and approaching a constant near the origin. 

Unlike standard Eigenmodes, each higher order mode for the mass flux does not possess an additional zero crossing. The reason for this is that these modes do not form an orthogonal basis; equivalently, the operator equation for $F''_{\sigma}$, while second order, is not in Sturm-Liouville form. 

Because these higher order modes are complex, one expects the shock position to exhibit oscillations in time. Returning to our general expression for the shock position \eqref{R1pert} and expanding $\sigma$ in terms of its real and imaginary parts, we find

\begin{equation}
    R =R_0\left\{1+ \sum_{\sigma}\frac{\zeta_{\sigma}}{\left(\frac{3}{2}+\sigma_{\rm r}\right)^2+\sigma_{\rm i}^2}\left(e^{\sigma_{\rm r}\tau}\left(\left(\frac{3}{2}+\sigma_{\rm r}\right)\cos\left(\sigma_{\rm i}\tau\right)+\sigma_{\rm i}\sin\left(\sigma_{\rm i}\tau\right)\right)-\left(\frac{3}{2}+\sigma_{\rm r}\right)e^{-\frac{3}{2}\tau}\right)\right\},
\end{equation}
where we let $\zeta_\sigma \rightarrow \zeta_\sigma/2$ to maintain consistency with the case where $\sigma_{\rm i} = 0$. In line with our expectations, a non-zero imaginary part of the Eigenfrequency introduces temporal oscillations into the position of the shock. However, unlike the case where the background flow is static, these oscillations proceed \emph{logarithmically} in time (i.e., $\tau = \ln R_0$, so $\cos\left(\sigma_{\rm i}\tau\right) \simeq \cos\left(\sigma_{\rm i}\ln t\right)$, and their periods grow exponentially). 

\begin{figure}[htbp] 
   \centering
   \includegraphics[width=0.495\textwidth]{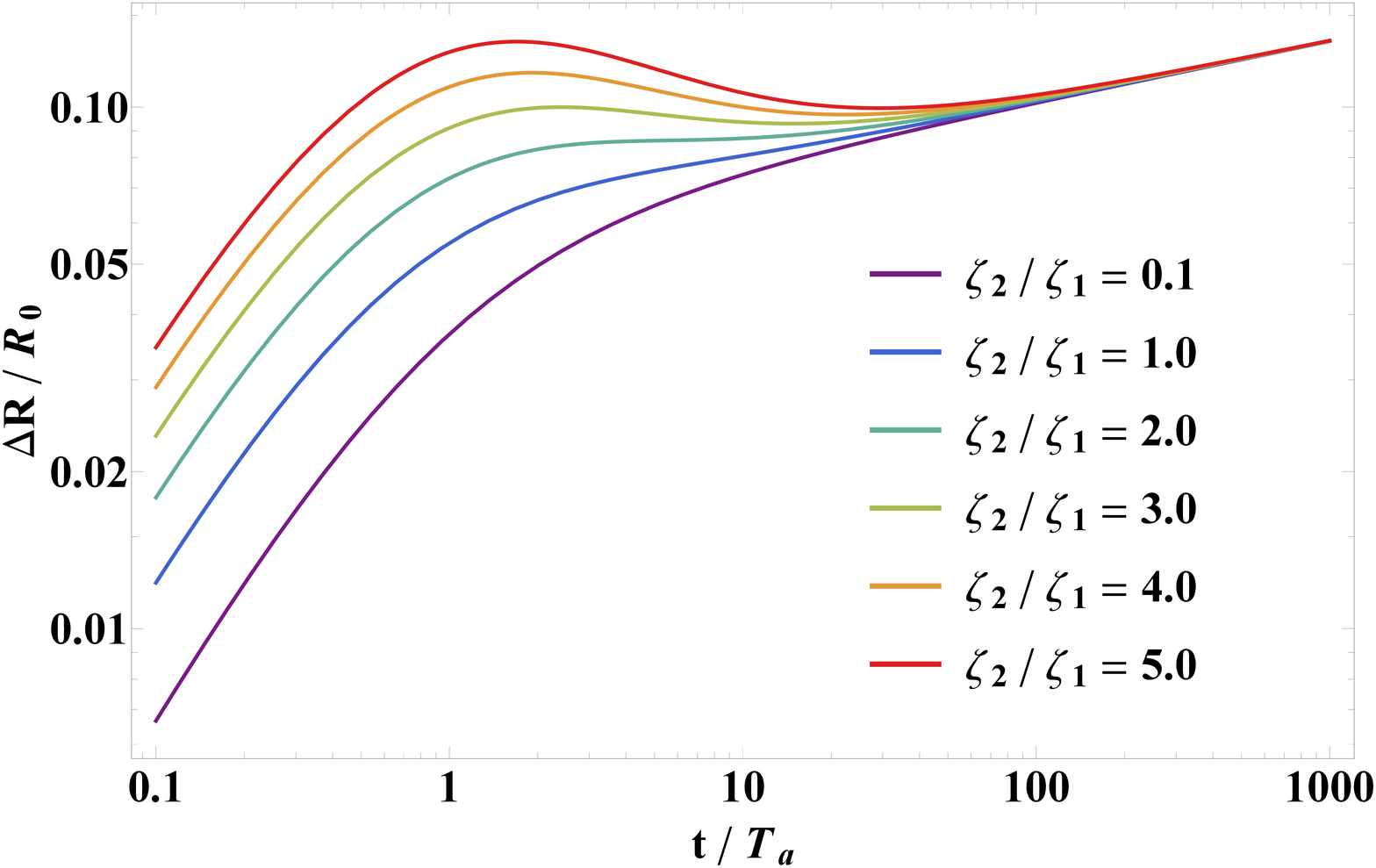} 
   \includegraphics[width=0.495\textwidth]{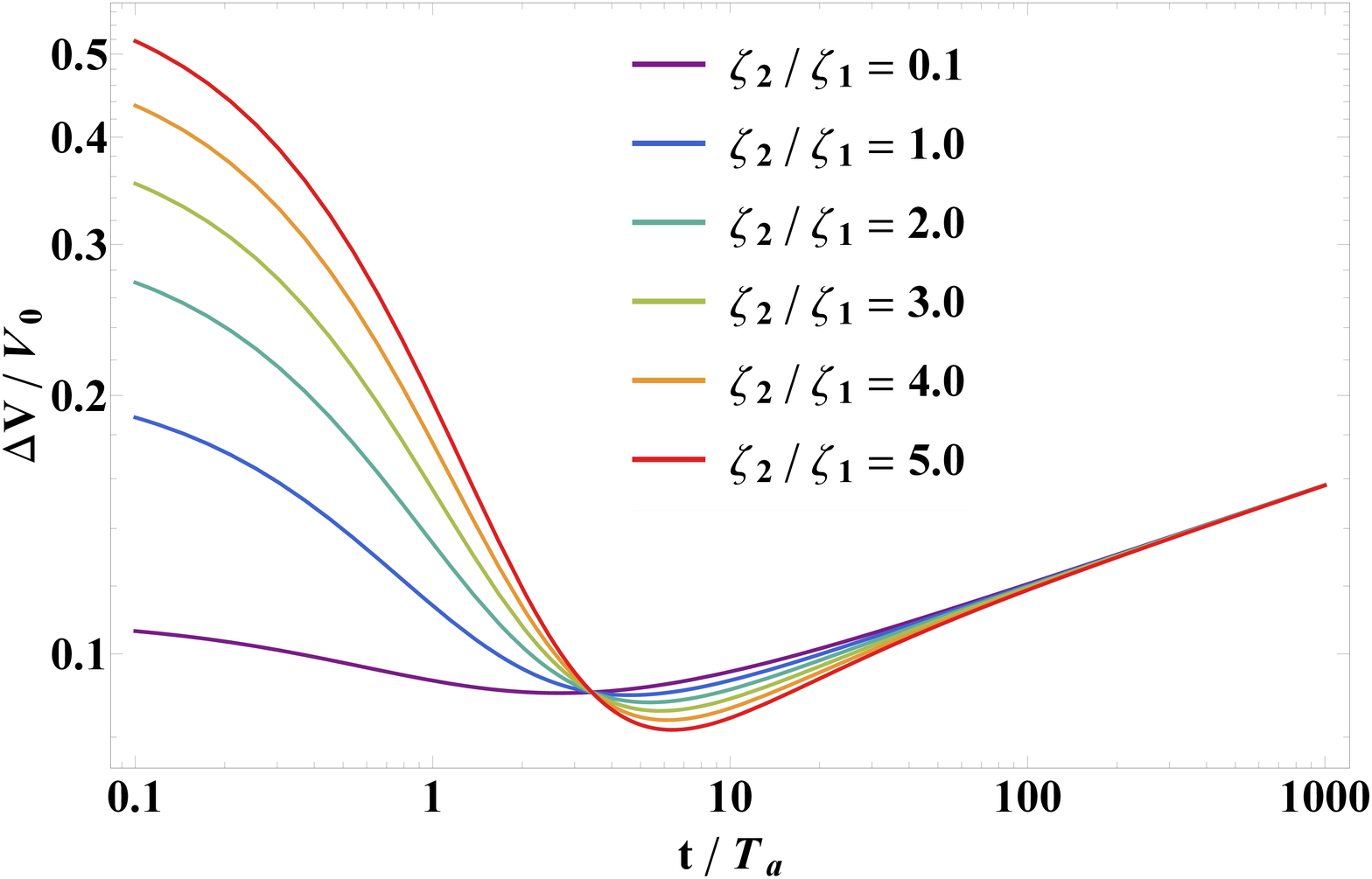} 
   \caption{The normalized difference in the shock position (left panel) and velocity (right panel) as functions of time in units of the dynamical time $T_{\rm a} = 2r_{\rm a}^{3/2}/(3V_{\rm c}\sqrt{GM})$. For each curve we set the perturbation of the growing mode to $\zeta_{1} = 0.1$, and the legend shows the relative contribution of the second order mode (all higher order $\zeta$ were set to zero). We see that, for large ratios of $\zeta_{2}/\zeta_1$, one starts to recover oscillations in both the shock position and velocity.}
   \label{fig:R1_oscillations}
\end{figure}

Figure \ref{fig:R1_oscillations} shows the relative change in the shock position (left) and the shock velocity (right) as we increase the relative importance of the second order mode. As we increase the ratio $\zeta_2 / \zeta_1$ (we set $\zeta_1 = 0.1$ for definiteness, but this only changes the scale of the y axis in the perturbative limit), we see that oscillations start to become more pronounced for both the shock position and the velocity. Because the real component of the second order mode is large, the oscillations are damped fairly quickly and only the growing mode dominates at times $t / T_{\rm a} \gtrsim 10$. 

\subsection{Growing modes for other $n$, $\gamma$}
The previous section demonstrated that, when $n = 2.5$ and $\gamma_1 = \gamma_2 = 1+1/n = 1.4$, the self-similar solutions are weakly unstable, with perturbations growing as $\propto t^{0.117}$ at late times. Here we investigate the nature of the instability for other $n$ and $\gamma_1 = \gamma_2 = \gamma$, and we derive the power-law index for the growing modes as a function of these parameters. {We focus exclusively on the case when $\gamma_1 = \gamma_2$ because, when this condition is violated, the solutions behave qualitatively differently and, as can be seen from the boundary conditions at the shock front \eqref{bc1} -- \eqref{bc3}, the trivial solution when the shock Mach number equals one stops existing (since $\gamma_1 \neq \gamma_2$ is inconsistent with a $\mathscr{M} = 1$ solution in which nothing changes at the ``shock'').}

\begin{figure}[htbp] 
   \centering
   \includegraphics[width=0.995\textwidth]{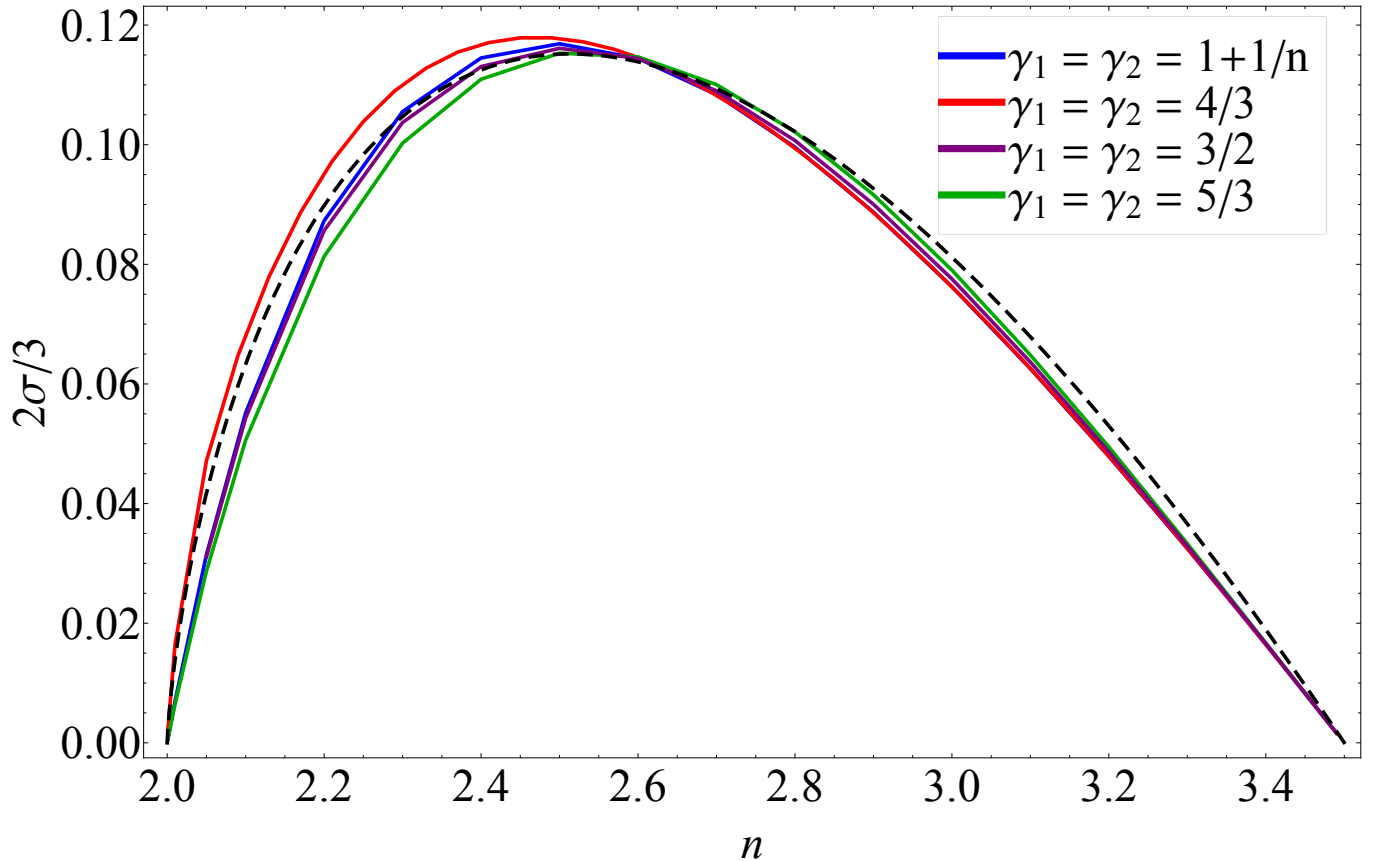}
     \caption{The value of the unstable Eigenmode multiplied by $2/3$, which gives the asymptotic temporal scaling of the perturbations, as a function of the power-law index of the density of the ambient medium, $n$; here the pre and post-shock adiabatic indices are equal, with the specific value appropriate to each curve shown in the legend. The black curve gives a simple fit to the data (see Equation \ref{sigmafit}), that reproduces well the variation in the growth rate as a function of $n$.} 
   \label{fig:sigma_of_gamma}
\end{figure}

Figure \ref{fig:sigma_of_gamma} shows the unstable Eigenvalue $\sigma$, multiplied by 2/3 to show the asymptotic temporal scaling of the perturbations (recall that the perturbations were parameterized as $\propto R^{\sigma} \propto t^{2\sigma/3}$), as a function of $n$ for the $\gamma$ shown in the legend. This plot demonstrates that the growth rate of the perturbations exhibits some dependence on the power-law index of the ambient medium, equaling zero when $n = 2$ and when $n=3.5$; as noted in Paper I, these values of $n$ are special because they are identical to the Sedov-Taylor blastwave ($n=2$) and the shock reduces to a rarefaction wave (i.e., the shock moves at the local sound speed and is an infinitesimal perturbation on the ambient gas; $n = 3.5$). In between these two extremes, the growth rate of the perturbations reaches a relative maximum of $2\sigma/3 \lesssim 0.12$ for $n \simeq 2.5$. 

\begin{table}
\begin{center}
\begin{tabular}{|c|c|c|c|c|}
\hline
 $n$ ($\rho \propto r^{-n}$)& $\gamma_1 = \gamma_2 = 1+1/n$ & $\gamma_1 = \gamma_2 = 4/3$ & $\gamma_1 = \gamma_2 = 3/2$ & $\gamma_1 = \gamma_2 = 5/3$ \\
\hline
$n = 2.01$ & $\{V_{\rm c},\sigma\}$ = \{8.72, 0.0104\} &  \{3.92, 0.0246\} & \{8.78, 0.0103\} & \{11.5, 0.00967\} \\ 
\hline
2.1 & \{2.93, 0.0828\} & \{2.29, 0.103\} & \{3.03, 0.0813\} & \{3.59, 0.0759\}\\
\hline
2.2 & \{2.05, 0.131\} & \{1.76, 0.143\} & \{2.15, 0.128\} & \{2.48, 0.0122\}\\
\hline
2.3 & \{1.64, 0.158\} & \{1.47, 0.165\} & \{1.74, 0.156\} & \{1.97,0.150\}\\
\hline
2.4 & \{1.38, 0.172\} & \{1.28, 0.175\} & \{1.48, 0.170\} & \{1.66,0.166\}\\
\hline
2.5 & \{1.20, 0.175\} & \{1.14, 0.177\} & \{1.30, 0.174\} & \{1.44,0.173\}\\
\hline
2.6 & \{1.07, 0.172\} & \{1.02, 0.172\} & \{1.16, 0.172\} & \{1.28, 0.172\}\\
\hline
2.7 & \{0.960, 0.163\} & \{0.934, 0.162\} & \{1.04, 0.164\} & \{1.14, 0.165\}\\
\hline
2.8 & \{0.874, 0.149\} & \{0.860, 0.149\} & \{0.955, 0.151\} & \{1.04, 0.153\}\\
\hline
2.9 & \{0.801, 0.133\} & \{0.796, 0.133\} & \{0.877, 0.135\} & \{0.953, 0.137\}\\
\hline
3.0 & \{0.740, 0.114\} & \{0.740, 0.114\} & \{0.810, 0.116\} & \{0.876, 0.119\}\\
\hline
3.1 & \{0.688, 0.0938\} & \{0.692, 0.0939\} & \{0.752, 0.0955\} & \{0.809, 0.0973\}\\
\hline
3.2 & \{0.642, 0.0718\} & \{0.649, .0720\} & \{0.701, 0.0730\} & \{0.780, 0.0742\}\\
\hline
3.3 & \{0.602, 0.0487\} & \{0.610, 0.0488\} & \{0.655, 0.0493\} & \{0.698, 0.0500\}\\
\hline
3.4 & \{0.566, 0.0247\} & \{0.576, 0.0247\} & \{0.614, 0.0249\} & \{0.651, 0.0251\}\\
\hline
3.49 & \{0.538, 0.00249\} & \{0.547, 0.00248\} & \{0.581, 0.00248\} & \{0.613, 0.00249\}\\
\hline
\end{tabular}
\end{center}
\caption{The left column gives the power-law index of the ambient medium, and each cell gives the critical shock speed $V_{\rm c}$ of the unperturbed self-similar solution and the growth rate of the unstable mode (the temporal scaling of the perturbations is $t^{2\sigma/3}$). Different columns are for different values of $\gamma_1 = \gamma_2 = \gamma$. Near $n = 2$, $V_{\rm c}$ grows to asymptotically large values, and the growth rate approaches zero. Near $n = 3.5$, $V_{\rm c}$ approaches the sound speed of $\sqrt{\gamma/(n+1)}$, and the growth rate again nears zero.}
\label{tab:1}
\end{table}

Interestingly, we see that the value of the growth rate is almost completely independent of the adiabatic index, with the most noticeable difference being a slight trend with the peak growth rate and the value of $n$ at which that peak is obtained. Specifically, smaller $\gamma$ result in a maximum growth rate at a smaller value of $n$. However, the differences between the curves in this figure are extremely small. This weak dependence on the adiabatic index is not too surprising, as the critical velocity $V_{\rm c}$ is not heavily modified by $\gamma$, and the functions $f_0$, $g_0$, and $h_0$ are also not too affected by this value. The black, dashed curve in this figure is the fit

\begin{equation}
    \frac{2\sigma}{3} = 1.58\left(n-2\right)^{0.87}-1.5\left(n-2\right), \label{sigmafit}
\end{equation}
which, aside from slightly overestimating the growth rate for small and large $n$, does a very good job at reproducing the numerically-obtained curves. Table \ref{tab:1} gives the specific value of the growth rate for a range of $n$ and different values of $\gamma$. 

\section{Comparison to Simulations}
\label{sec:simulations}
The previous subsections outlined a general framework for analyzing the perturbations to the self-similar, weak shock solutions presented in Paper I. We applied this formalism to the specific case when the density of the ambient medium declines as $\rho \propto r^{-2.5}$ and $\gamma_1 = \gamma_2 = 1.4$, and demonstrated that the lowest-order Eigenmode -- which characterizes the long-term evolution of perturbations on top of the self-similar solutions -- grows as a very weak power law in time. A more general analysis showed that this weak instability should be present for other $n$ and $\gamma$ as well. In this section, we investigate the validity of the perturbation theory by comparing to the results of idealized, hydrodynamical simulations.

\subsection{Simulation setup}
We used the hydrodynamics code {\sc flash} (v 4.3; \citealt{fryxell00}) to analyze the propagation of a weak shock into an ambient medium that is in hydrostatic equilibrium with a central mass, is non-self-gravitating, and has a density profile of $\rho \propto r^{-n}$ and adiabatic index $\gamma_1 =1+1/n$. The adiabatic index of the post-shock gas was fixed to $ \gamma_2 = 1+1/n$, and hence the setup is identical to the scenario in which we expect the CQR, self-similar solutions to arise.

We generate an outward-propagating shock by initializing a pressure wave on top of the hydrostatic fluid. 
The pressure is set to  $p(t=0)=(1+\delta p) \times p_1(r_\mathrm{a})$, between the inner boundary, $r_\mathrm{a}$, and wave front $r_{\rm f}=1.5\times r_\mathrm{a}$. The density and velocity are unchanged. This construction enables material below $r_{\rm f}$ to begin falling out of the domain due to the lack of pressure support, while an outward-propagating shock forms at the over-pressurized wave front.

The simulation domain is on a uniform grid that spans between $r_\mathrm{a}=1$ and $r_\mathrm{max}=10^3$; therefore, the initial pressure wave is spatially small. The simulations present here have a minimum grid resolution of $\delta r\simeq 10^{-2}$ except for the case where $\delta p=0.1$ which has $\delta r\simeq 0.25\times10^{-2}$. {Table \ref{tab:tab2} gives the parameters for each simulation analyzed here; the first column gives the initial pressure perturbation $\delta p$, the second column gives the difference between the shock Mach number and one, $\delta \mathscr{M}$, when the shock reaches a distance of 100 $r_{\rm a}$, and the third column is the resolution $\delta r / r_{\rm a}$ (grid spacing) used in the simulation. In the following, we label the simulation by the second column in this table, $\delta \mathscr{M}$. In a forthcoming paper (Paper III) we will present more results of these and other simulations.}

\begin{table}[]
    \centering
    \begin{tabular}{|c|c|c|}
    \hline
        $\delta p$  & $\delta \mathscr{M}$ & $\delta r / r_{\rm a}$\\
        \hline
       0.1 & 0.78 & $6.25\times 10^{-4}$ \\
       0.25 & 0.99 & 0.01\\
       0.275 & 1.0 & 0.01\\
       0.6 & 1.7 & 0.01 \\ 
       1.0 & 2.6 & $2.5\times 10^{-3}$ \\
       \hline
    \end{tabular}
    \caption{The parameters for each simulation done with {\sc flash}: the initial pressure perturbation $\delta p$, 
    the difference between the Mach number and one when the shock reaches a distance of 100 $r_{\rm a}$, $\delta M$, and the grid spacing adopted for each simulation, $\delta r / r_{\rm a}$.  }
    \label{tab:tab2}
\end{table}

\subsection{Results}
\label{sec:simulation_results}
\begin{figure}[htbp] 
   \centering
   \includegraphics[width=0.495\textwidth]{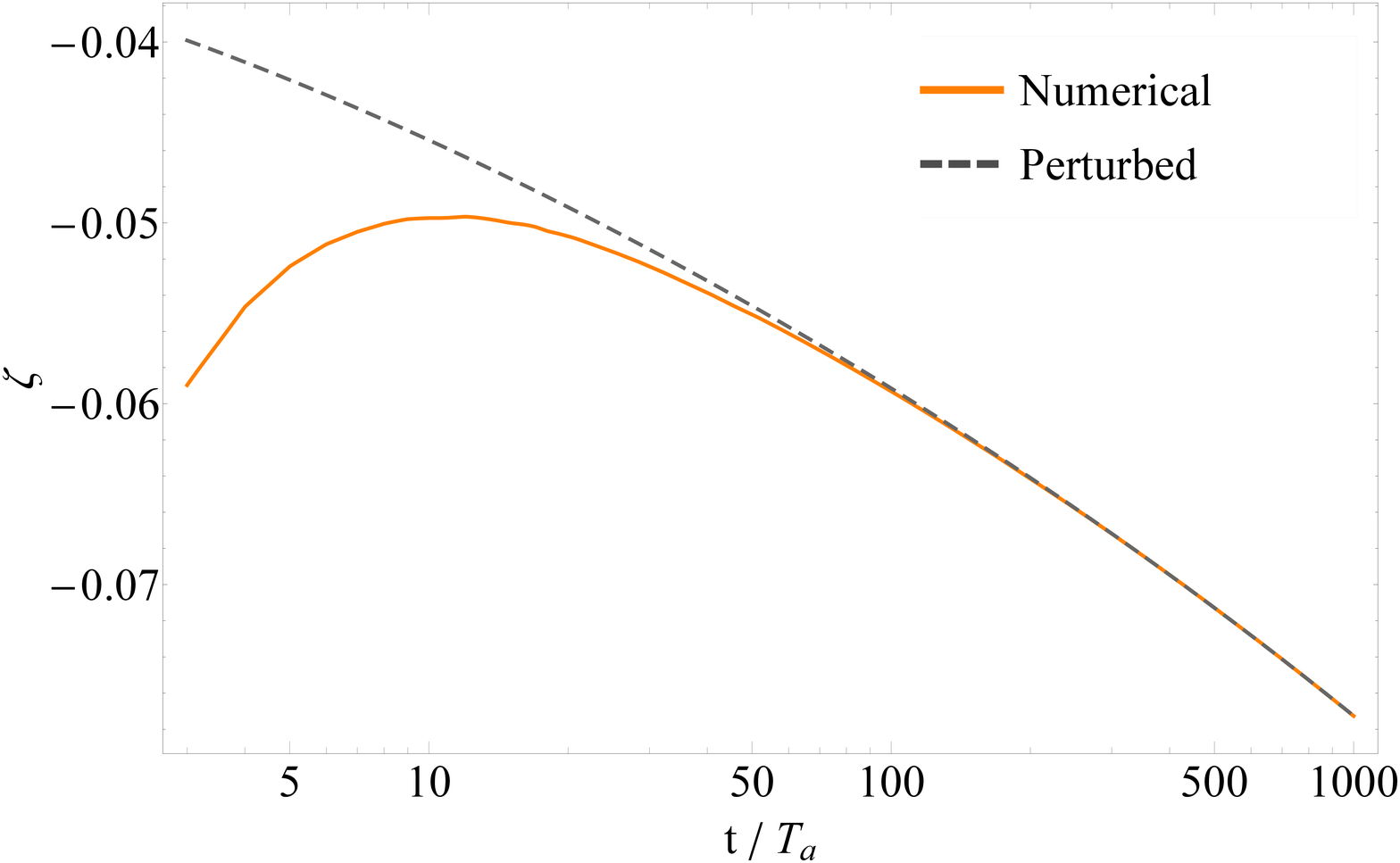}
   \includegraphics[width=0.495\textwidth]{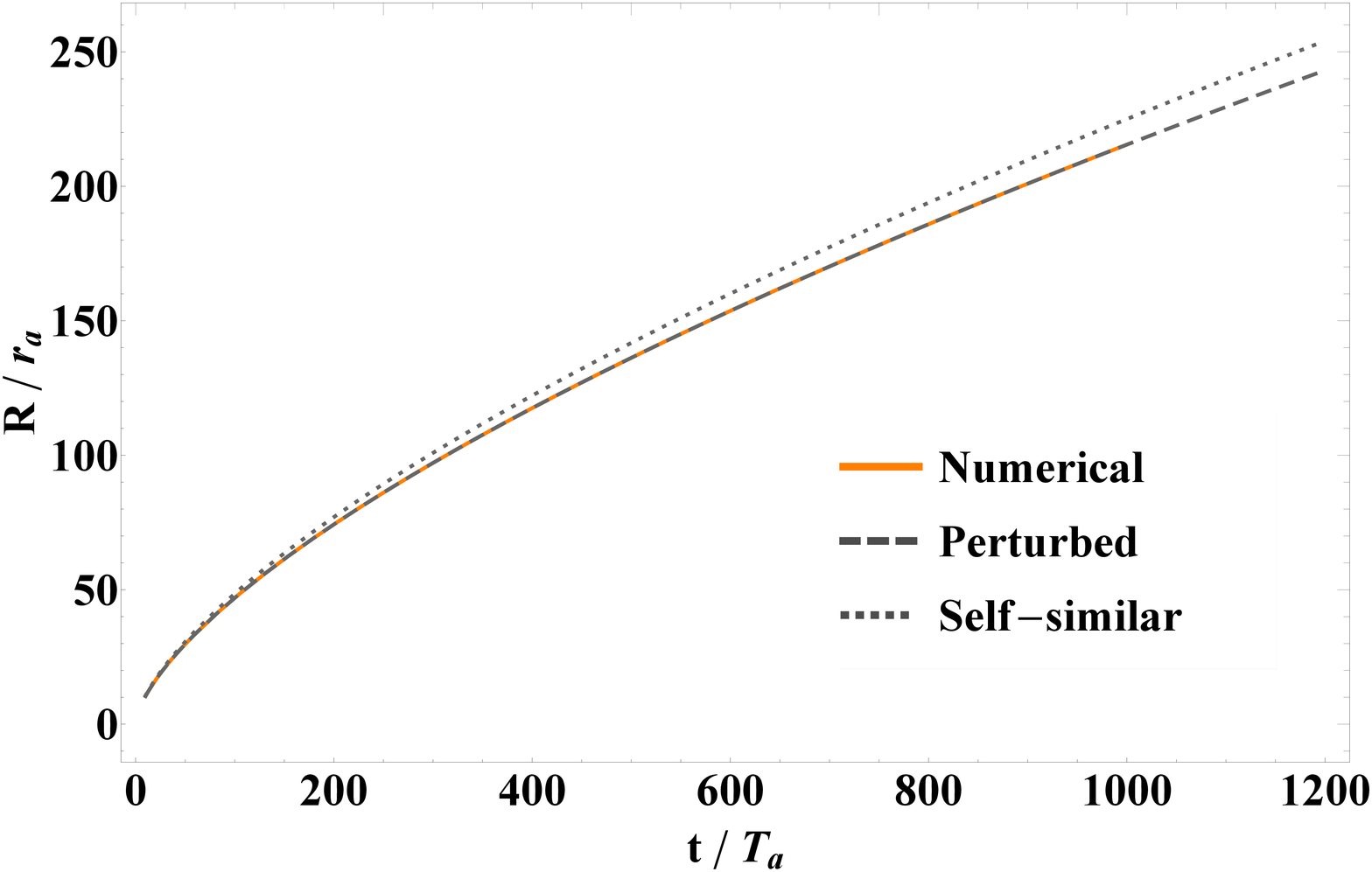}
   \caption{Left: $\zeta$ as a function of the shock position obtained from the time-dependent numerical simulation with {\sc flash} for $\delta M = 0.78$ (orange) and the predicted $\zeta$ from the growing mode (black, dashed). Here time is measured in units of $T_{\rm a} = r_{\rm a}^{3/2}/\sqrt{GM}$. At late times, the two curves approach the power-law predicted by the perturbation analysis, being $\zeta \propto t^{0.117}$. Right: The numerical position of the shock with $\delta M = 0.78$ (orange), the self-similar prediction for the evolution of the shock (black, dotted), and the one-parameter correction that includes the growing mode (black, dashed). }
   \label{fig:R0p275}
\end{figure}

{The left panel of Figure \ref{fig:R0p275} shows $\zeta$ as a function of time in units of $T_{\rm a} = r_{\rm a}^{3/2}/\sqrt{GM}$ measured from the simulation with $\delta \mathscr{M} = 0.78$ (orange, solid curve) and the predicted scaling from the growing mode (black, dashed curve). The orange curve is obtained by using the numerically-obtained shock position and velocity, $R$ and $V$, to construct }

\begin{equation}
    \zeta = \frac{1}{2}\left(1-\frac{GM V_{\rm c}^2}{RV^2}\right).
\end{equation}
{The dashed curve in this figure is given by }

\begin{equation}
    \zeta = \zeta_{\sigma}e^{\sigma \ln R},
\end{equation}
{where $\sigma \simeq 0.175$ is the growing Eigenvalue; $\zeta_{\sigma}$ is chosen to match the asymptotic scaling approached by the numerical solution, and in this case is given by $\zeta_{\sigma} = -0.0301$. The negative sign of $\zeta_{\sigma}$ arises here because the initial perturbation yields a shock with a Mach number that falls below the CQR value. The right panel of this figure gives the numerically-obtained shock position (orange, solid), the self-similar shock position (black, dotted), and the perturbed solution obtained by including the growing mode (black, dashed). For the latter case, which agrees extremely well with the numerical solution, we used $\zeta_{\sigma} = -0.0301$ in Equation \eqref{Rtot}, and integrated the equation numerically to obtain the shock position.}

\begin{figure}[htbp] 
   \centering
   \includegraphics[width=0.495\textwidth]{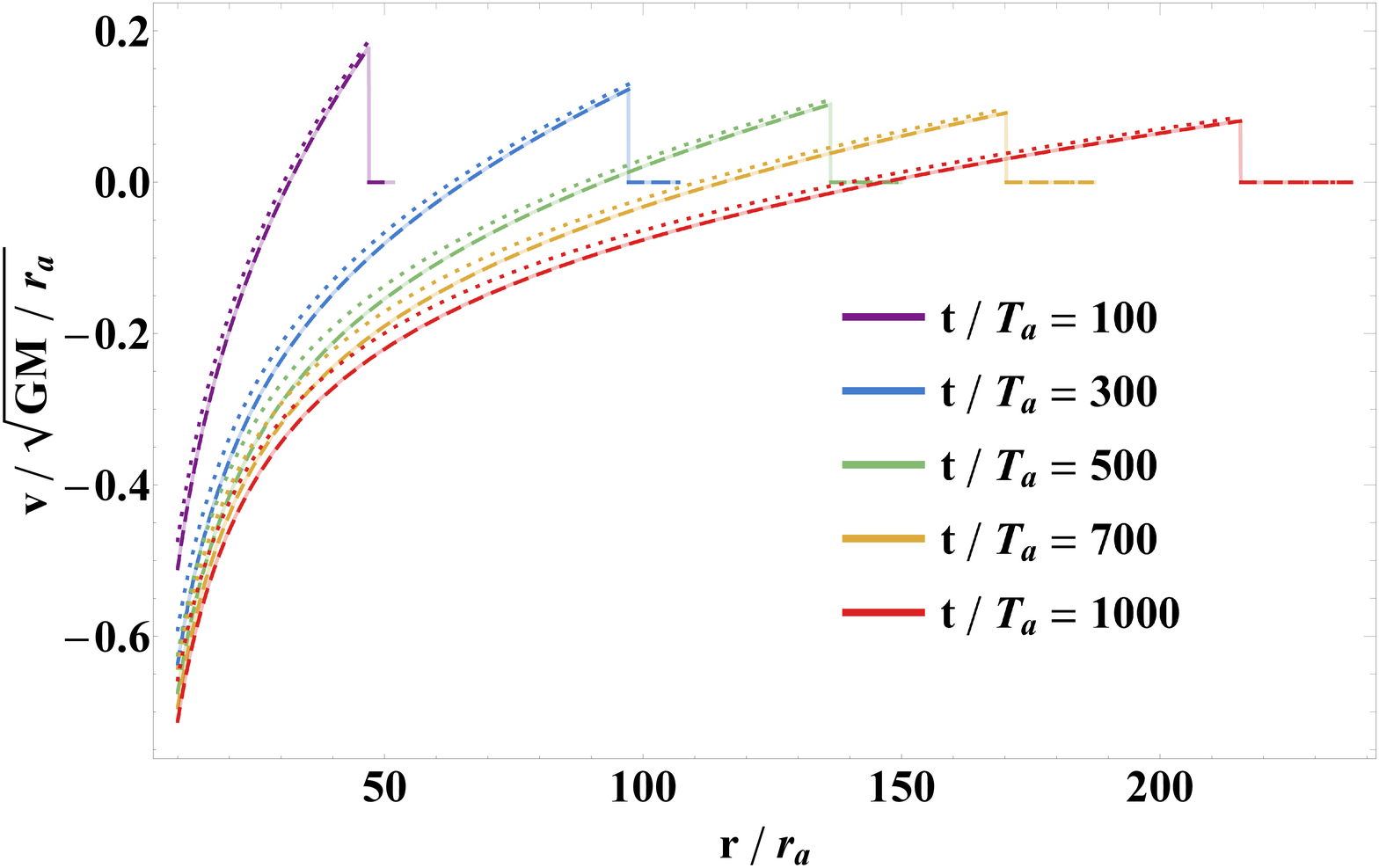}
   \includegraphics[width=0.495\textwidth]{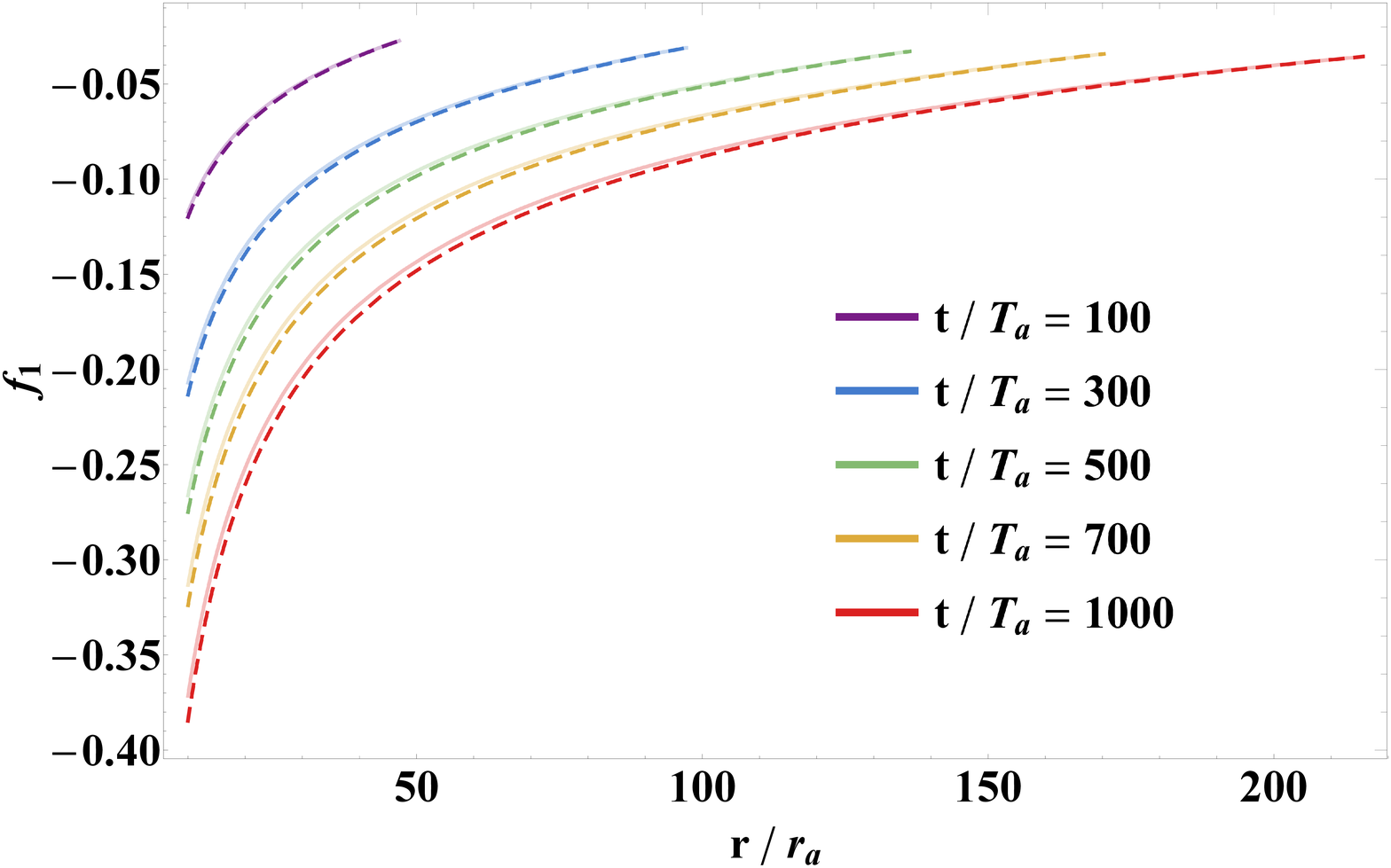}
   \caption{Left: The numerically-obtained velocity profile (light, solid) for $\delta \mathscr{M} = 0.78$, the self-similar solution (dotted), and the prediction from the one-parameter correction that accounts for the unstable mode (dashed) at the times shown in the legend. Right: The normalized difference between the numerical and self-similar velocity profiles (light, solid) and the prediction from the growing mode (dashed). }
   \label{fig:vcomps_0p275}
\end{figure}

In addition to the shock position, the perturbation analysis of Section \ref{sec:eigenmodes} predicts that there should also be corrections to the velocity, density, and pressure of the fluid behind the shock. Moreover, these corrections are determined by the amplitude of the perturbation to the growing mode, $\zeta$, meaning that there is one parameter that should account for the asymptotic deviation of the fluid velocity, density, and pressure from their self-similar forms. The left panel of Figure \ref{fig:vcomps_0p275} gives the numerically-obtained velocity profile as a function of $r / r_{\rm a}$ (solid, lightly-colored curves), the  self-similar solution (dotted curves), and the perturbed solutions with the one-parameter correction (dashed curves) at the times shown in the legend for $\delta \mathscr{M} = 0.78$. The self-similar solution is given by

\begin{equation}
    v_{\rm ss} = V(t)f_0(\xi),
\end{equation}
where $V(t)$ is the numerically-obtained shock velocity and $\xi = r/R(t)$ is the total self-similar variable, while the perturbed velocity profile is

\begin{equation}
    v_{\rm pert} = V(t)\left\{f_0(\xi)+\zeta_{\sigma} e^{\sigma \tau}f_1(\xi)\right\},
\end{equation}
where $\zeta_{\sigma} = -0.0301$, $\sigma = 0.175311$, and $\tau = \ln (R(t)/r_{\rm a})$. From this figure, we see that there are small differences between the numerical and self-similar velocity profiles, and these differences are accounted for well by the contribution from the unstable mode. 

The right panel of Figure \ref{fig:vcomps_0p275} shows the difference between the numerical and self-similar velocity profiles normalized by the shock velocity (solid, lightly-colored curves) and the prediction from unstable mode (dashed curves). From this figure, we see that the discrepancy between the self-similar solution and the numerical one is very accurately reproduced by the growing mode. However, there are small differences that become more pronounced at later times, and these differences likely arise from nonlinear terms that are not accounted for in the linear perturbation analysis.

\begin{figure}[htbp] 
   \centering
   \includegraphics[width=0.495\textwidth]{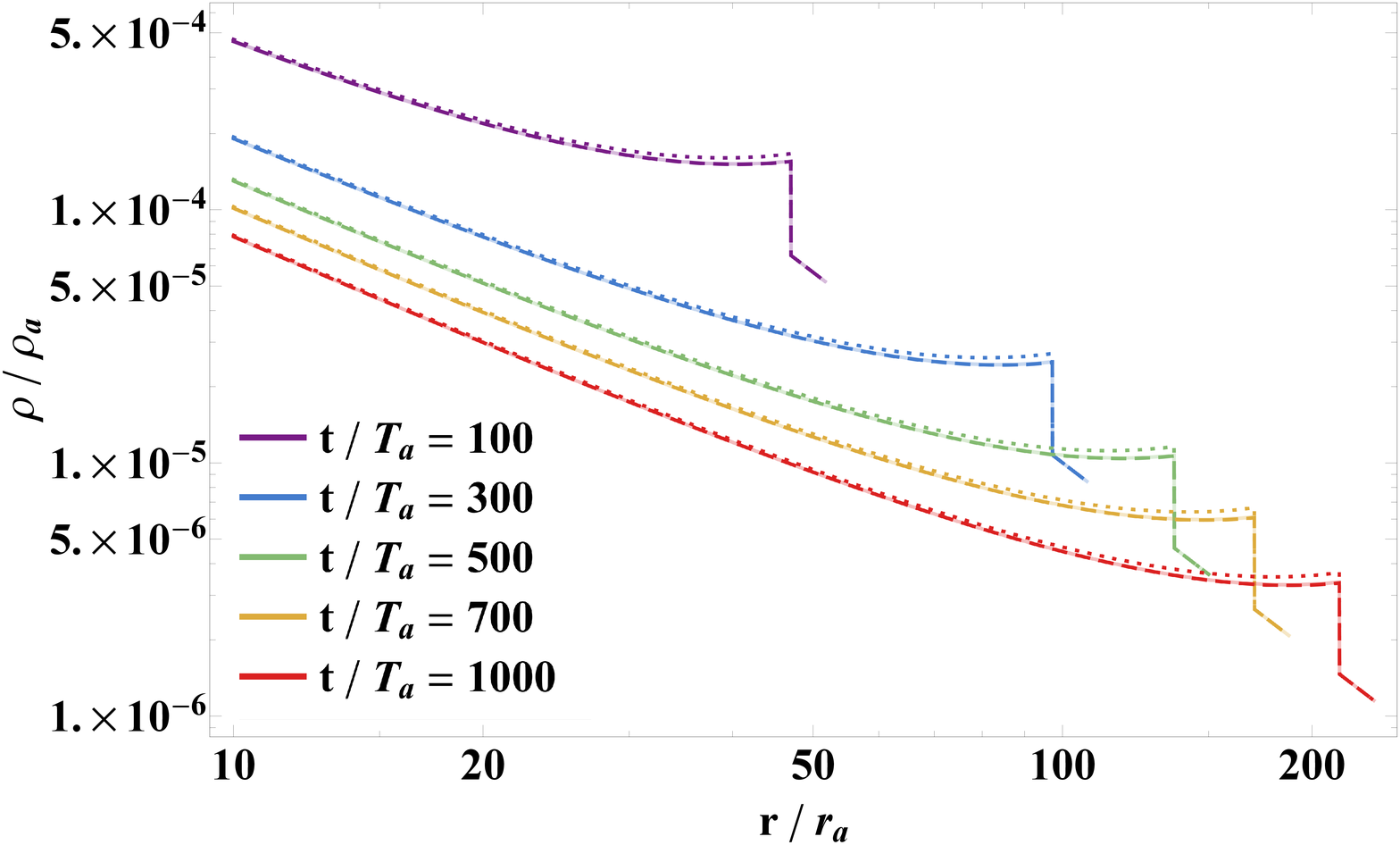}
   \includegraphics[width=0.495\textwidth]{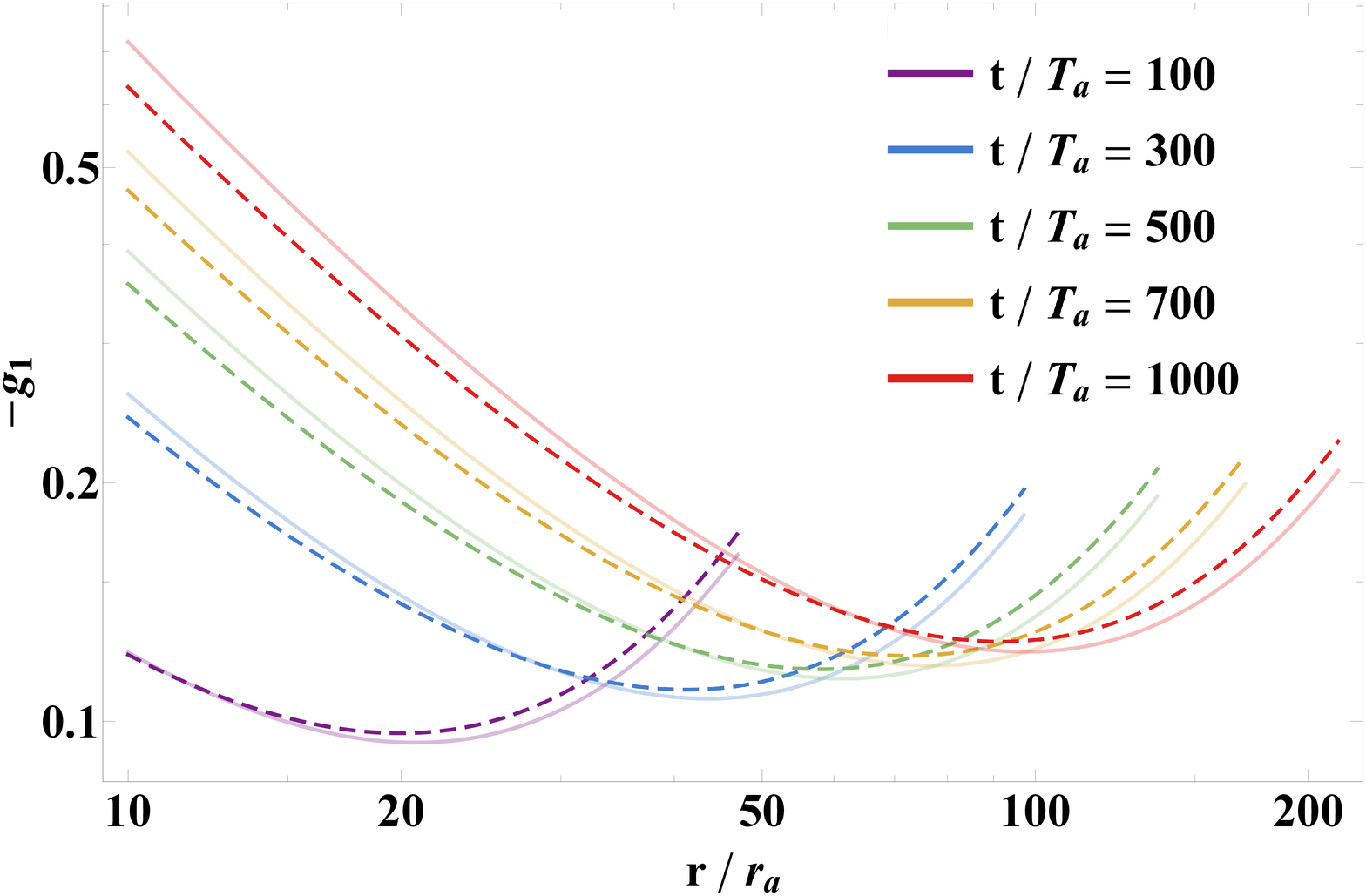}
   \caption{Left: The numerically-obtained density profile (light, solid) for $\delta \mathscr{M} = 0.78$, the self-similar solution (dotted), and the prediction from the one-parameter correction that accounts for the unstable mode (dashed) at the times in the legend. Right: The normalized difference between the numerical and self-similar density profiles (light, solid) and the prediction from the growing mode (dashed). }
   \label{fig:pcomps_0p275}
\end{figure}

\begin{figure}
    \centering
    \includegraphics[width=0.495\textwidth]{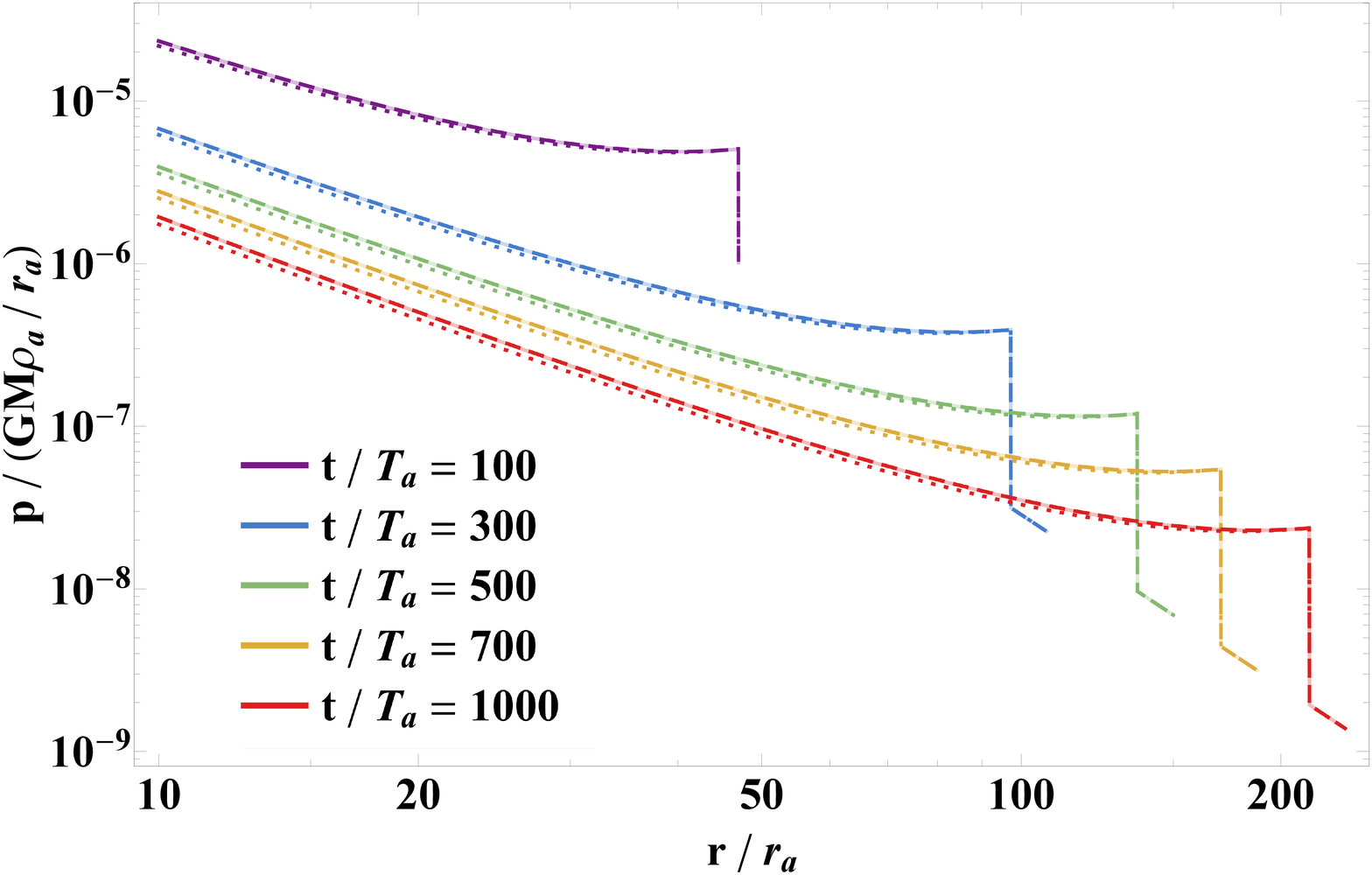}
    \includegraphics[width=0.495\textwidth]{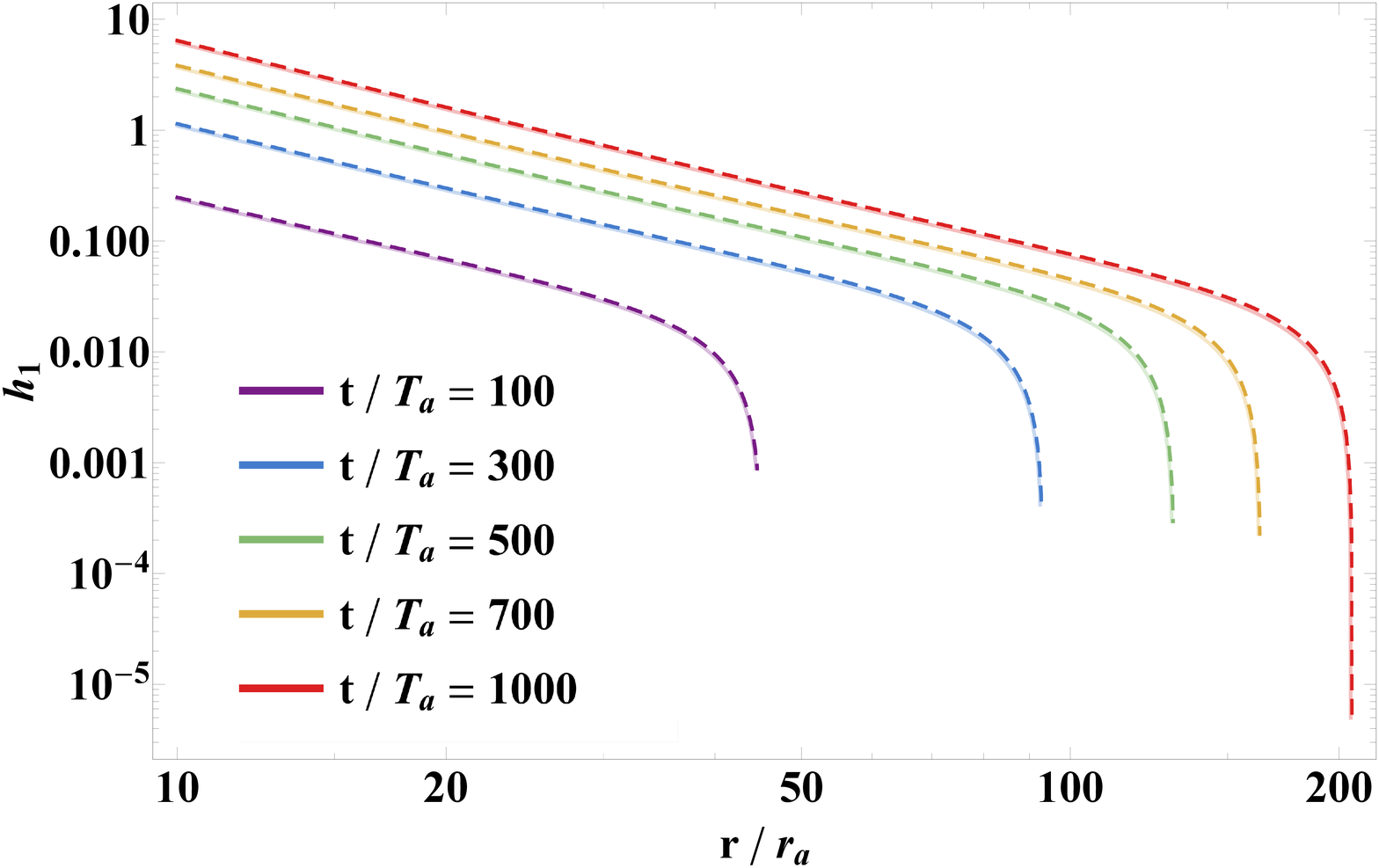}
    \caption{Left: The numerically-obtained pressure profile (light, solid) for $\delta \mathscr{M} = 0.78$, the self-similar solution (dotted), and the prediction from the one-parameter correction that accounts for the growing mode (dashed) at the times indicated in the legend. Right: The normalized difference between the numerical and self-similar pressure profiles (light, solid) and the prediction from the growing mode (dashed). }
    \label{fig:pcomps_0p25}
\end{figure}

The left panel of Figure \ref{fig:pcomps_0p275} shows the numerically-obtained density profile (solid, lightly-colored), the self-similar density profile (dotted), and the one-parameter correction at the times shown in the legend, and the left panel of Figure \ref{fig:pcomps_0p25} gives the analogous curves for the post-shock pressure; the definitions of the self-similar functions are

\begin{equation}
    \rho_{\rm ss} = R^{-n}g_0(\xi), \quad p_{\rm ss} = V^2R^{-n}h_0(\xi),
\end{equation}
while the perturbed functions are

\begin{equation}
    \rho_{\rm pert} = R^{-n}\left\{g_0(\xi)+\zeta_\sigma e^{\sigma\tau}g_1(\xi)\right\}, \quad p_{\rm pert} = V^2 R^{-n}\left\{h_0(\xi)+\zeta_\sigma  e^{\sigma\tau}h_1(\xi)\right\}.
\end{equation}
For these figures $\zeta_\sigma = -0.0301$ is set by the perturbations to the shock position, and $\sigma \simeq 0.175$ is the Eigenvalue of the growing mode. As was true for the velocity profile, we see that there are small, but noticeable differences between the predictions of the self-similar solutions and those obtained from the simulations. The right panels of Figures \ref{fig:pcomps_0p275} and \ref{fig:pcomps_0p25} show the differences between the numerical solutions and the self-similar solutions (lightly-colored, solid) and the prediction from the growing mode (dashed), and demonstrate that the differences are accurately reproduced by the Eigenfunctions of the growing mode.

\begin{figure}[htbp] 
   \centering
   \includegraphics[width=0.495\textwidth]{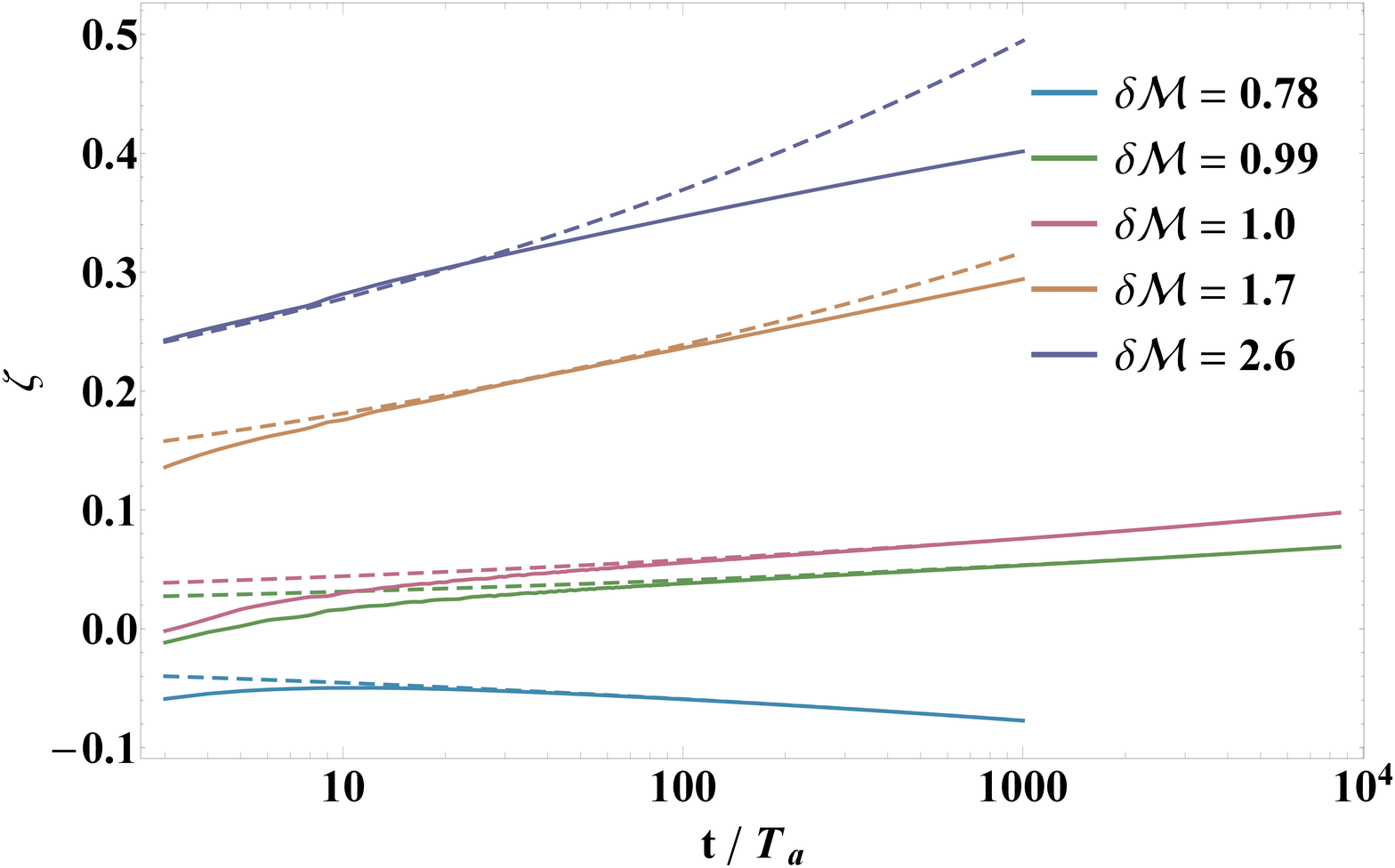}
   \includegraphics[width=0.495\textwidth]{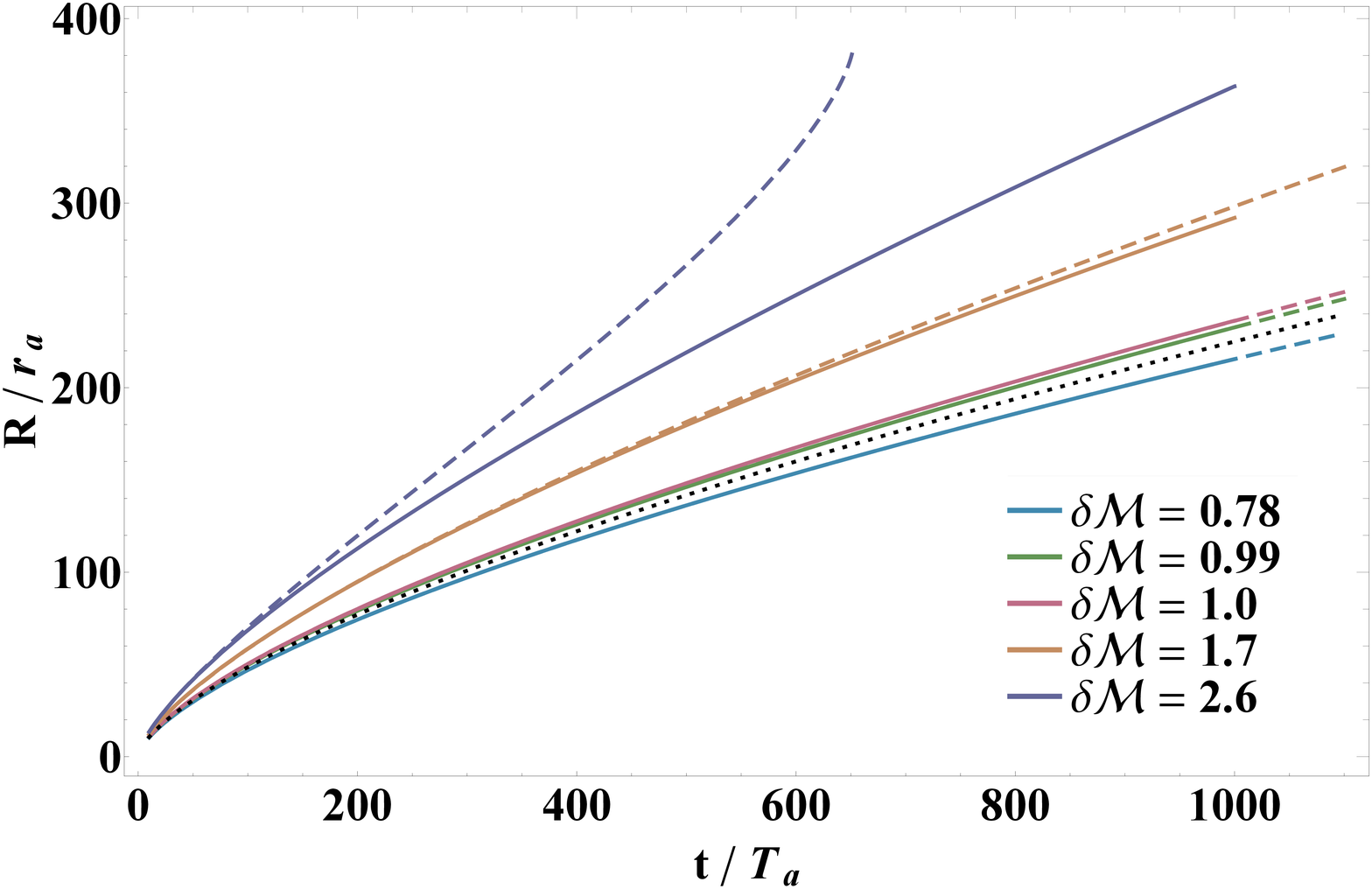}
   \caption{Left: The value of $\zeta$ from the numerical simulations (solid) and the prediction from the growing mode (dashed) as functions of time for the simulations with $\delta \mathscr{M}$ shown in the legend. Right: The numerically-obtained shock position (solid) and the perturbed solution (dashed) as functions of time for the simulations with $\delta \mathscr{M}$ shown in the legend; the dotted, black curve is the self-similar solution. It is clear that, at late times, nonlinear terms are becoming important for simulations with larger $\delta \mathscr{M}$, and the perturbation approach is breaking down.}
   \label{fig:Rcomps_deltap}
\end{figure}

The left panel of Figure \ref{fig:Rcomps_deltap} illustrates $\zeta$ as a function of time obtained numerically (solid curves) and the one-parameter correction (dashed curves) for the range of $\delta \mathscr{M}$ shown in the legend. In this case, the constant $\zeta_{\sigma}$ that determines $\zeta(0)$ increases as $\delta \mathscr{M}$ increases. In particular, denoting the amplitude of the perturbation particular to simulation $\delta \mathscr{M}$ by $\zeta_{\delta \mathscr{M}}$, for this plot we set $\zeta_{0.78} = -0.0301$, $\zeta_{0.99} = 0.0206$, $\zeta_{1.0} = 0.0291$, $\zeta_{1.7} = 0.117$, and $\zeta_{2.6} = 0.176$. These values provide reasonable fits to the numerical position of the shock. At late times, we see that simulations with relatively large $\delta \mathscr{M}$ produce $\zeta$ that differ noticeably from the prediction of the perturbation analysis, and the deviations start to become appreciable at a time where $\zeta \gtrsim \mathrm{few}\times 0.1$. This breakdown is expected, as in these cases the nonlinear terms are no longer ignorable. 

The right panel of Figure \ref{fig:Rcomps_deltap} shows the numerical shock position (solid), the perturbed shock position (dashed), and the self-similar solution (dotted), where the perturbed solution is found by integrating Equation \eqref{Rtot} with the growing mode included. As expected from the left panel of this figure, we see that the numerical solutions match the perturbed solutions well when $\delta\mathscr{M}$ is small, but there are significant deviations from the predictions of the perturbation theory when $\delta \mathscr{M}$ becomes large.

While the amplitude of the growing mode, $\zeta_{\sigma}$, is in principle determined from the Eigenmode expansion of all of the derivatives of the shock position at a given time, we also expect $\zeta_{\sigma}$ to scale monotonically with the difference between the Mach number of the shock and the one enforced by the CQR solution. Taking the Mach number at a given instant in time or an average over some temporal range also should not greatly modify the scaling, as the Mach number only depends very weakly on time (and varies as $t^{2\sigma/3}$) when the solution is close to the CQR one. Performing a quadratic fit to $\zeta(\Delta \mathscr{M})$, where $\Delta \mathscr{M}$ is the difference between the shock Mach number and the CQR value when the shock reaches $R = 100$, we find

\begin{equation}
    \zeta = 0.19\Delta M-0.053\Delta M^2.
\end{equation}
There is only a very small difference (at the 1\% level) if we fit the average $\Delta \mathscr{M}$ over the duration of the simulation.

\section{Rarefaction Wave Solution}
\label{sec:rarefaction}
In the previous section we demonstrated that, when the shock Mach number is larger than the one imposed by the CQR solution, the instability causes the shock position and velocity to asymptotically grow. In this case the shock position and the post-shock quantities should asymptotically approach the energy-conserving, Sedov-Taylor blastwave (or, for sufficiently large $n$, the Waxman-Shvarts solution; \citealt{waxman93}), and we plan to investigate this transition in detail in a followup paper. 

However, we also showed that when the shock Mach number is below the CQR prediction, then the value of $\zeta$ multiplying the growing mode is negative. Therefore, in this regime the difference between the shock Mach number and the CQR value becomes increasingly negative and, if the linear theory remained valid into the nonlinear regime, would predict that the shock becomes subsonic at late times. However, the Mach number cannot become less than unity, resulting in the dissolution of the shock and the formation of a pressure wave of finite thickness, as the conservation of wave luminosity implies that such a wave should steepen back into a shock 
\citep{dewar70, ro17, coughlin18a}. This finding then raises the question as to the asymptotic state of these solutions as they become ever weaker. 

One can show that there is, in fact, a \emph{second, exact solution} to the self-similar Equations (\ref{ss1} -- \ref{ss3}) that have $V_{\rm c} = \sqrt{\gamma/(n+1)}$, or a Mach number of exactly one. In this case, the critical point is at the location of the shock itself, and one of the solutions is the trivial solution $f_0 = 0$, $g_0 = \xi^{-n}$, $h_0 = \xi^{-n-1}/\gamma$ (i.e., the ambient medium retains perfect hydrostatic balance everywhere). However, owing to the nature of the critical point, there is a second solution that satisfies $f_0(1) = 0$, $g_0(1) = 1$, $h_0(1) = 1/\gamma$, meaning that the ``post-shock'' fluid is still in hydrostatic balance precisely at the location of the sound-crossing horizon, but with derivatives that differ from those of the trivial solution.

These solutions adopt the same self-similar variable as the CQR blastwave, and therefore the self-similar equations are identical:

\begin{equation}
-n g_{\rm r}-\xi g_{\rm r}'+\frac{1}{\xi^2}\frac{d}{d\xi}\left[\xi^2g_{\rm r}f_{\rm r}\right] = 0, \label{ss1r}
\end{equation}
\begin{equation}
-\frac{1}{2}f_{\rm r}-\xi f_{\rm r}'+f_{\rm r}f_{\rm r}'+\frac{1}{g_{\rm r}}h_{\rm r}' = -\frac{1}{V_{\rm r}^2\xi^2},
\end{equation}
\begin{equation}
n\gamma-n-1+\left(f_{\rm r}-\xi\right)\frac{d}{dr}\ln\left(\frac{h_{\rm r}}{g_{\rm r}^{\gamma}}\right) = 0. \label{ss3r}
\end{equation}
Here we are letting $\gamma_1 = \gamma_2 = \gamma$, which is necessary for recovering the trivial solution (i.e., the trivial solution mandates that the wave does nothing, and setting $\gamma_1 \neq \gamma_2$ by definition implies that the wave has changed something). The definitions of the self-similar functions $f_{\rm r}$, $g_{\rm r}$, and $h_{\rm r}$ in terms of the velocity, density, and pressure are also the same:

\begin{equation}
    v = V(t)f_{\rm r}(\xi), \quad \rho = \rho_{\rm a}\left(\frac{R(t)}{r_{\rm a}}\right)^{-n}g_{\rm r}(\xi), \quad p = \rho_{\rm a}V(t)^2\left(\frac{R(t)}{r_{\rm a}}\right)^{-n}h_{\rm r}(\xi),
\end{equation}
and the shock velocity and position are likewise related in the same manner:

\begin{equation}
    V^2 = V_{\rm r}^2\frac{GM}{R},
\end{equation}
where here $V_{\rm r} = \sqrt{\gamma/(n+1)}$, i.e., the ``shock'' moves out at the sound speed. The jump conditions at the shock front, Equations \eqref{bc1} -- \eqref{bc3}, also give the boundary conditions obeyed by the functions $f_{\rm r}$, $g_{\rm r}$, and $h_{\rm r}$, and when $V_{\rm r} = \sqrt{\gamma/(n+1)}$ become

\begin{equation}
    f_{\rm r}(1) = 0, \quad g_{\rm r}(1) = 1, \quad h_{\rm r} = 1/\gamma.
\end{equation} 
Using L'Hopital's rule in Equations \eqref{ss1r} -- \eqref{ss3r} shows that the derivatives of the non-trivial solutions are

\begin{equation}
    f'_{\rm r}(1) = \frac{5+2n}{2+2\gamma}, \quad g'_{\rm r}(1) = \frac{5-2n\gamma}{2+2\gamma}, \quad h'_{\rm r}(1) = -\frac{2+2n-3\gamma}{2\gamma+2\gamma^2}. \label{derivs}
\end{equation}
These rarefaction wave (RW) solutions are thus in hydrostatic equilibrium at radii exterior to the sound crossing horizon, and undergo collapse onto the point mass at the origin for radii interior to that radius. These solutions were also found by \citet{kazhdan94}, who focused on the case of $n \le 2$.

\begin{figure}[htbp] 
   \centering
   \includegraphics[width=0.495\textwidth]{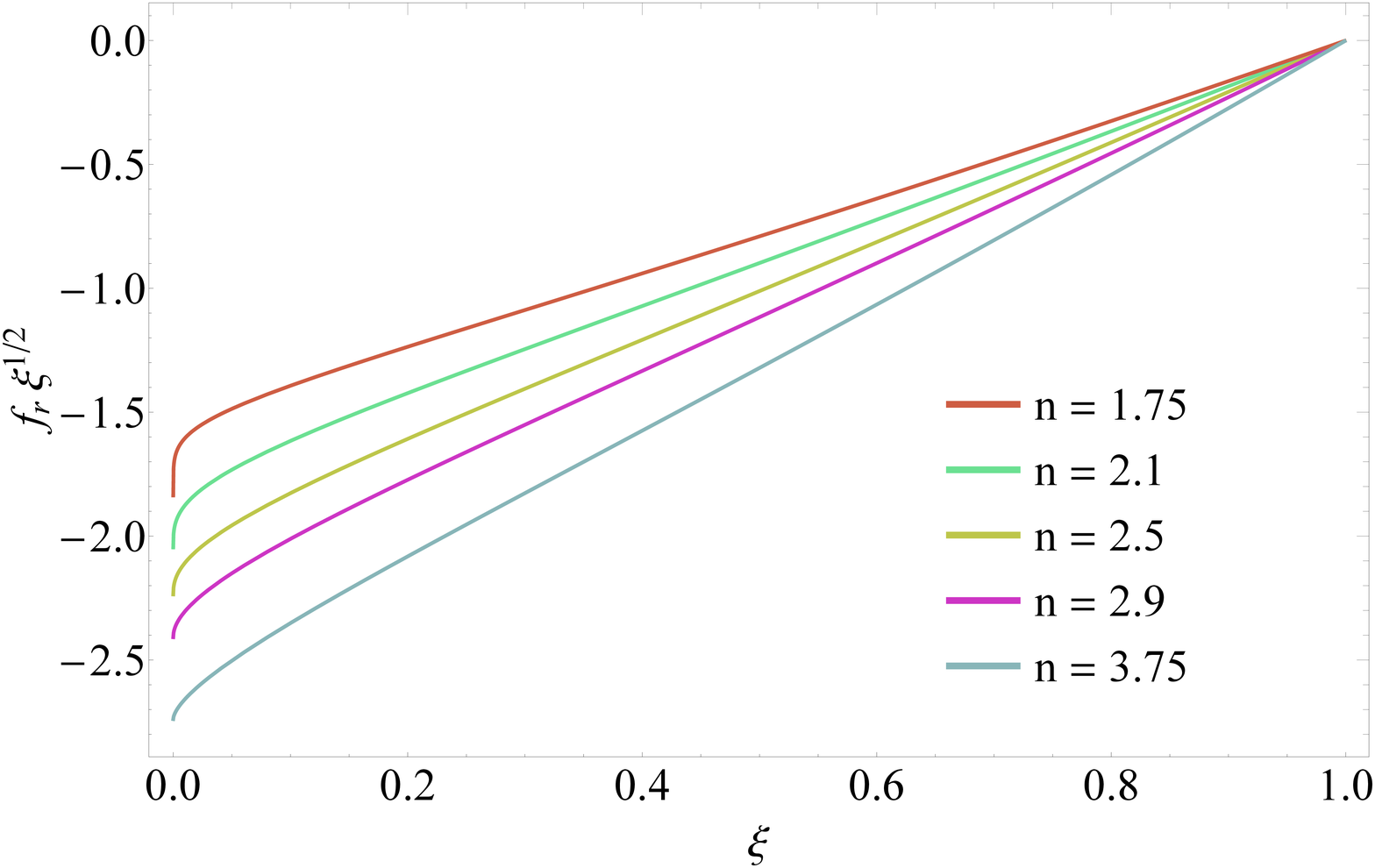}
   \includegraphics[width=0.495\textwidth]{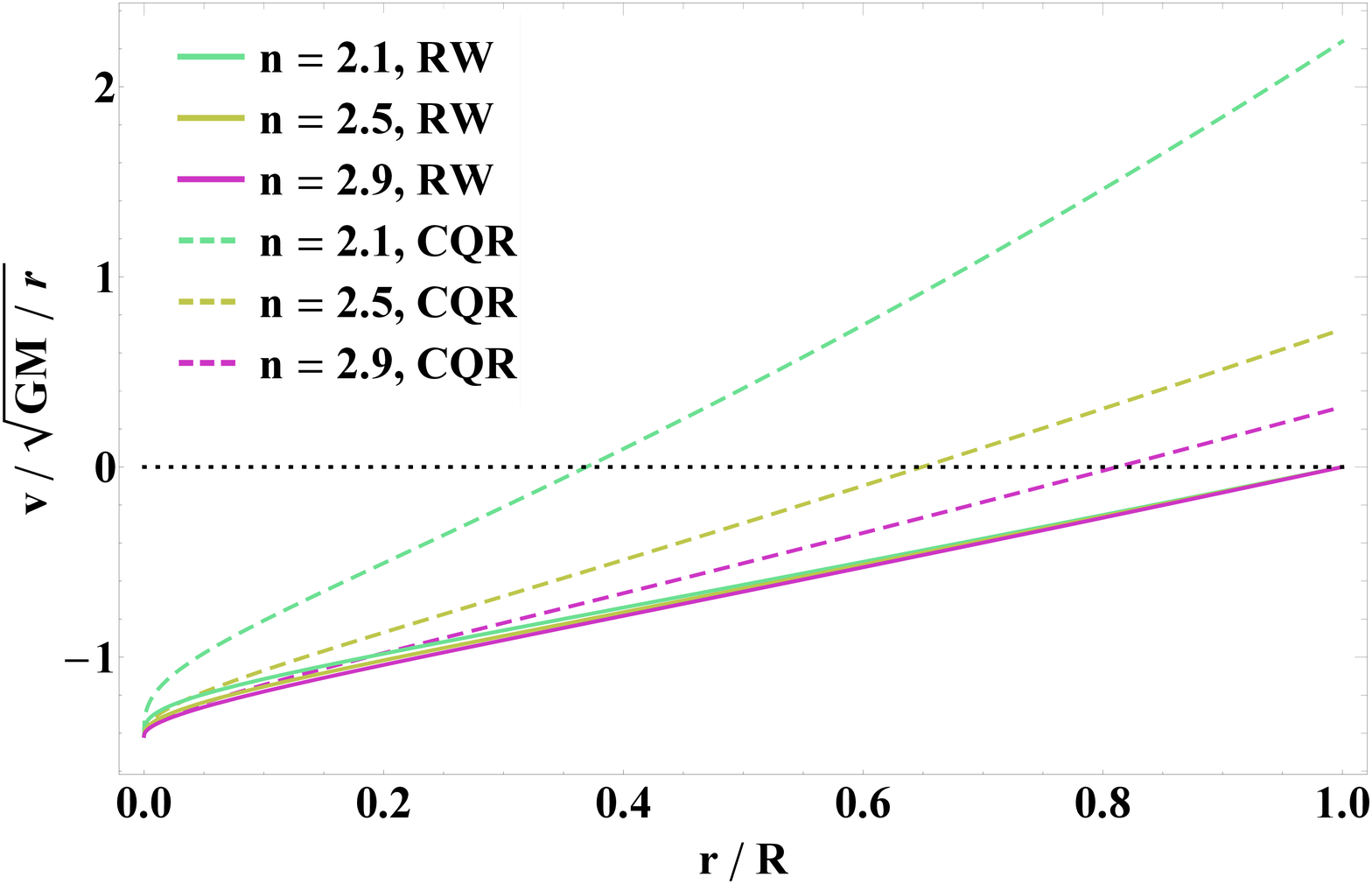}
   \caption{Left: The self-similar velocity profile for the rarefaction wave solutions multiplied by $\xi^{1/2}$ (which gives the asymptotic scaling of the functions near the black hole) for the values of $n$ shown in the legend. These functions all equal zero at $\xi = 1$, signifying that they freefall from rest at a time that corresponds to the sound crossing time from the origin. Right: The velocity profile, normalized by the circular velocity, for the rarefaction wave solution (solid) and the CQR solution (dashed) for the set of $n$ shown in the legend. While the CQR and RW velocity profiles exhibit noticeable differences near the shock front, they all converge to a value of $v / \sqrt{GM / r} = -\sqrt{2}$ near the origin, i.e., in both cases the gas freefalls onto the black hole.}
   \label{fig:vcomps_CQR_RW}
\end{figure}

\begin{figure}[htbp] 
   \centering
   \includegraphics[width=0.495\textwidth]{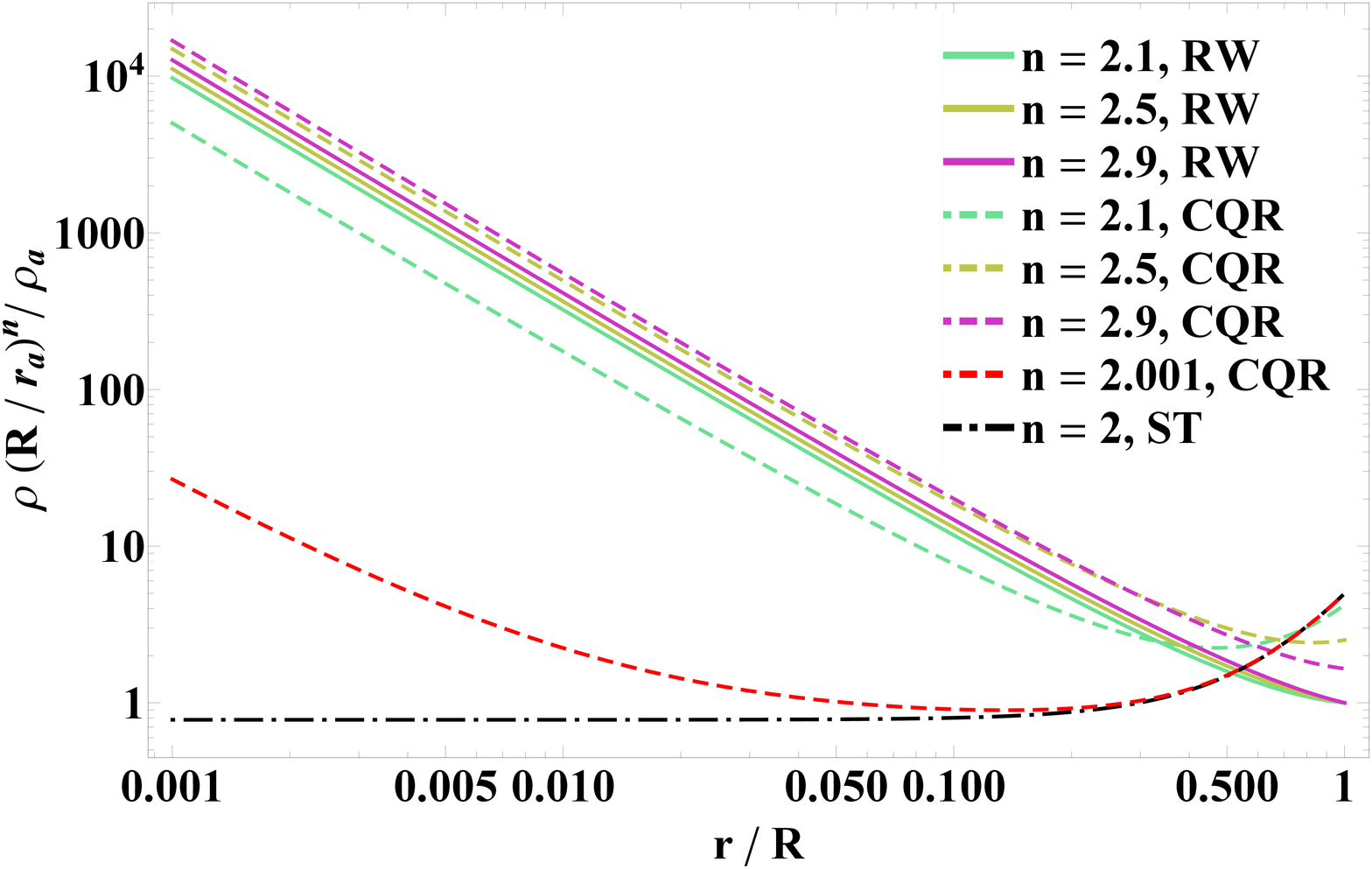}
   \includegraphics[width=0.495\textwidth]{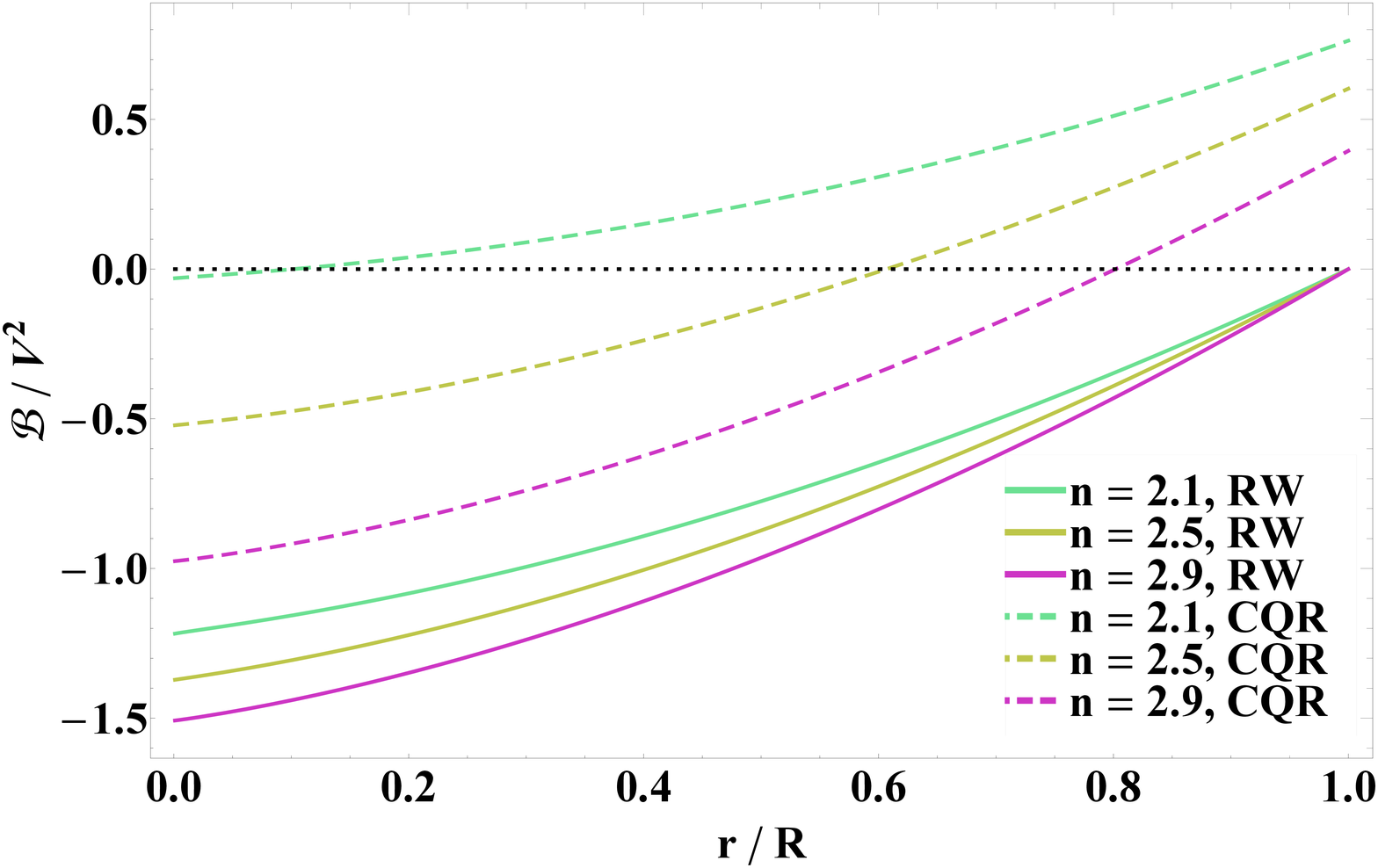}
   \caption{Left: The normalized density profile for the rarefaction wave solutions (solid curves) and the CQR solutions (dashed curves) for the values of $n$ shown in the legend. These functions all scale as $r^{-3/2}$ near the origin, and therefore approach freefall, though the constant of proportionality depends on $n$ and the nature of the solution (i.e., RW or CQR). The dot-dashed, black curve is the Sedov-Taylor solution for $n = 2$ and $\gamma = 3/2$, which is the solution to which the CQR solution tends in the limit of $n \rightarrow 2$; this tendency is shown directly by the dashed, red curve, which is the CQR solution when $n = 2.001$ (and $\gamma = 1+1/2.001$). Right: The Bernoulli parameter of the flow, normalized by the square of the shock velocity, for the rarefaction wave solutions (solid) and the CQR solutions (dashed) for the set of $n$ shown in the legend. The CQR solutions all have a positive Bernoulli parameter near the shock front and a negative one near the point mass, while the Bernoulli parameter of the RW solutions is negative for the entire post-shock flow.}
   \label{fig:bcomps_CQR_RW}
\end{figure}

The left panel of Figure \ref{fig:vcomps_CQR_RW} shows the solution for the self-similar velocity, multiplied by $\xi^{1/2}$, for the range of $n$ shown in the legend, and for this figure we let $\gamma = 1+1/n$. We see that, near the position of the rarefaction wave, the velocity is zero and the gas is in hydrostatic equilibrium. Immediately inside of $\xi = 1$, however, the velocity adopts a non-trivial value, signifying the fact that the sound wave conveys to the overlying envelope that there is accretion at the origin. The fact that these solutions all approach $\sim \xi^{-1/2}$ near the origin confirms the fact that they do, indeed, accrete. These solutions also exist for $n < 2$ and $n > 3.5$, unlike the CQR blastwave.

The right panel of Figure \ref{fig:vcomps_CQR_RW} gives the velocity, normalized by the circular velocity of the point mass, as a function of radius normalized by the shock, or rarefaction wave for this case, radius. The solid curves give the RW solutions for a subset of $n$ in the left panel, while the dashed curves are the CQR solutions for the same $n$. It is apparent that, while the functions $f_{\rm r}$ differ fairly substantially, the \emph{velocity} profile resulting from the different rarefaction waves are all quite similar; this finding implies that the functions $f_{\rm r}$ are almost the same function aside from a scaling factor. We also see that the RW solutions and the CQR solutions predict different behavior near the location of the shock front. However, all of the curves equal $v / \sqrt{GM / r} = -\sqrt{2}$ at the origin, which demonstrates that the velocity always approaches freefall onto the black hole independent of the solution near the shock front.

The left panel of Figure \ref{fig:bcomps_CQR_RW} gives the normalized density as a function of dimensionless radius for the RW (solid) and CQR (solutions) and the set of $n$ shown in the legend. In all cases the density approaches the freefall scaling of $\rho \propto r^{-3/2}$ near the origin, but the proportionality factor changes depending on the type of solution (CQR or RW) and $n$. We also see that, as $n$ nears 2, the density of the CQR solution falls below that of the RW solution. The origin of this behavior is that, as $n \rightarrow 2$, the CQR solution converges to the Sedov-Taylor blastwave, where the kinetic energy is infinite relative to the gravitational binding energy. In this limit, then, the point mass effectively vanishes from the solution, and there is no finite radius at which the solution approaches freefall. The black, dot-dashed and the red, dashed curves in this figure show, respectively, the Sedov-Taylor solution for $n = 2$ and $\gamma = 1.5$ and the CQR solution when $n = 2.001$ (and $\gamma = 1+1/2.001$), which illustrates directly the notion that the freefalling nature of the solutions drops out as $n \rightarrow 2$.

The right panel of Figure \ref{fig:bcomps_CQR_RW} gives the Bernoulli parameter of the fluid, given by

\begin{equation}
    \mathscr{B} = \frac{1}{2}v^2+\frac{\gamma}{\gamma-1}\frac{p}{\rho}-\frac{GM}{r},
\end{equation}
normalized by the square of the shock velocity. We see that, for the CQR solution, the Bernoulli parameter is characterized by positive and negative regions, the former being confined to the fluid near the shock front, the latter being near the point mass. The RW solutions, on the other hand, have a negative Bernoulli parameter everywhere, and only approach exactly zero near the location of the shock front.

Because the rarefaction wave and CQR solutions adopt the same definitions for the self-similar velocity, density, and pressure, and the equations describing the evolution of the fluid quantities are the same, we can employ exactly the same formalism as outlined in Sections \ref{sec:equations} and \ref{sec:eigenmodes} to assess the stability of the rarefaction wave solutions. In particular, the matrix equation for the Eigenmodes describing perturbations to the rarefaction wave is identical to Equation \eqref{eigens} if we define

\begin{multline}
    v(r,t) = V\left\{f_0(\xi)+e^{\sigma\tau}f_\sigma(\xi)\right\}, \quad \rho(r,t) = \rho_{\rm a}\left(\frac{R}{r_{\rm a}}\right)^{-n}\left\{g_0(\xi)+e^{\sigma\tau}g_\sigma(\xi)\right\}, \quad p = V^2\left(\frac{R}{r_{\rm a}}\right)^{-n}\left\{h_0+e^{\sigma\tau}h_\sigma(\xi)\right\}, \\
    \frac{GM}{RV^2} = \frac{1}{V_{\rm c}^2}\left(1-2\zeta_\sigma e^{\sigma\tau}\right),
\end{multline}
with the functions $f_0$, $g_0$, and $h_0$ given by the rarefaction wave solutions, and $V_{\rm c} = \sqrt{\gamma/(n+1)}$. The boundary conditions on the functions $f_{\sigma}$, $g_{\sigma}$, and $h_{\sigma}$ are given by

\begin{equation}
    f_{\sigma}(1) = g_{\sigma}(1) = \frac{4}{\gamma+1}\zeta_{\sigma}, \quad h_{\sigma}(1) = \frac{2(\gamma-1)}{\gamma(\gamma+1)}\zeta_{\sigma},
\end{equation}
which result from the shock jump conditions.

In this case, the fourth boundary condition that determines the Eigenvalue -- being that the solutions be continuous through the sonic point -- also takes place at the location of the rarefaction wave. Therefore, at the sonic point we know the values of the perturbations, the values of the unperturbed functions, and, from Equation \eqref{derivs}, the values of the derivatives of the unperturbed functions, which implies that we can determine the Eigenvalue analytically. Doing so yields

\begin{equation}
    \sigma = \frac{1}{4}\left(n-\frac{7}{2}\right).
\end{equation}
When $n < 3.5$, the Eigenvalue is negative, which implies that the solutions are stable; when $n = 3.5$, the solutions are marginally stable, which agrees with the predictions of the CQR solution for this value of $n$ (see Figure \ref{fig:sigma_of_gamma}); and when $n > 3.5$ the solutions are unstable, meaning that any small perturbation to the amplitude of the rarefaction wave will cause the Mach number to grow. This expression is also completely independent of the adiabatic index of the gas.

\section{Physical origin of the instability}
\label{sec:origin}
The results of Section \ref{sec:solutions} demonstrated that the lowest-order Eigenmode of the CQR blastwave, which characterizes the asymptotic deviation of the shock position and the post-shock velocity, density, and pressure from their self-similar expressions, is weakly unstable, meaning that perturbations imposed on top of the self-similar solution grow as very shallow power laws in time. The numerical simulations performed with {\sc flash} also validated this finding, and differences between the numerically-obtained shock position and the self-similar solution were shown to grow approximately as $\Delta R/R_0 \propto t^{0.117}$ when the density profile of the ambient medium falls off as $\rho \propto r^{-2.5}$ (and the pre and post-shock adiabatic indices are equal to $\gamma_1 = \gamma_2 = 1+1/2.5 = 1.4$). The post-shock fluid quantities were also very well matched by the solutions resulting from the perturbation analysis.

Paper I made the following heuristic argument to suggest that the CQR self-similar solutions are likely stable: the Mach number of the CQR blastwave is uniquely specified by the self-similar solutions, and is on the order of unity when the power-law index of the density profile of the ambient medium is not too close to 2. If the initial Mach number is greater than the one imposed by the self-similar solutions, then the energy of the fluid immediately behind the shock is slightly greater than the self-similar value (by virtue of the jump conditions). However, the \emph{rate} at which binding energy is added to the fluid, being $\dot{E} = E_{\rm a} V$, where $E_{\rm a}$ is the binding energy of the ambient medium, is slightly larger. Therefore, it seemed plausible that the shock would self-regulate by sweeping up binding energy at a greater rate, thereby reducing the energy budget of the post-shock fluid to the required, self-similar value. An analogous argument holds when the Mach number is slightly below the self-similar one. 

With the solutions for the perturbations to the CQR blastwave, we can now directly assess the validity of this argument: from the terms in parentheses on the right side of Equation \eqref{Edot}, we see that there are two contributions to the perturbation to the binding energy swept up by the shock. The first is proportional to $V_1 / V_0$, and corresponds to the term that, in Paper I, was argued to contribute to the self-regulation of the shock -- increases in the velocity correspond to a greater influx of the binding energy of the ambient medium. The second term, proportional to $(1-n)R_1/R_0$, was not considered in the heuristic assessment of Paper I; this perturbation arises from the fact that the binding energy of the ambient medium has a radial dependence, and falls off as $E_{\rm a} \propto GM/r\times\rho r^2 \propto r^{1-n}$. Thus, if we perturb the position of the shock to larger values, then the material at the new location of the shock is \emph{less bound} than it was at the original location, which reduces the rate at which binding energy is entrained by the flow. This term, therefore, has a destabilizing effect on the propagation of the shock. 

The ability of the shock to self-regulate is therefore determined by the ratio of the second term, $(n-1)R_1/R_0$, to the first term, $V_1 / V_0$. In situations where the magnitude of this ratio is greater than (less than) one, we expect that the shock is unstable (stable) to perturbations to the energy flux. For a given mode, this ratio is given by (from Equations \ref{R1pert} and \ref{V1pert})

\begin{equation}
\frac{(n-1)R_1V_0}{V_1R_0} = \frac{n-1}{1+\sigma}.
\end{equation}
We thus see that, for small values of $\sigma$, the magnitude of this ratio can be greater than 1, and for all of the unstable modes we find that this number is, indeed, slightly greater than unity (e.g., for $n = 2.5$, the above ratio is $\simeq 1.28$). Ironically, this situation arises because the instability is so weak, meaning that the perturbation to the velocity is too small to significantly reduce the post-shock energy budget. Indeed, for larger values of $\sigma$, which induce correspondingly larger variations in the velocity, the magnitude of this ratio is less than one (e.g., for the second mode for $n = 2.5$, which has $\sigma \simeq -2.5+0.6i$, the modulus of this ratio is $\simeq 0.92$).

For the unstable solutions, when $\zeta$ is positive, the total energy entrained by the shock as it moves out increases asymptotically compared to the self-similar solution, which can be shown directly by integrating the differential equation for $\dot{E}$ (Equation \ref{Edot}). In this case the shock position and the post-shock quantities should asymptotically approach the energy-conserving, Sedov-Taylor blastwave. In the other limit, when $\zeta < 1$, the relative energy decreases asymptotically to ever-more negative values, and in this regime the shock likely transitions to the rarefaction wave solutions presented in Section \ref{sec:rarefaction}. We plan to investigate these transitions in detail in a followup paper.

\begin{figure}[htbp] 
   \centering
   \includegraphics[width=0.495\textwidth]{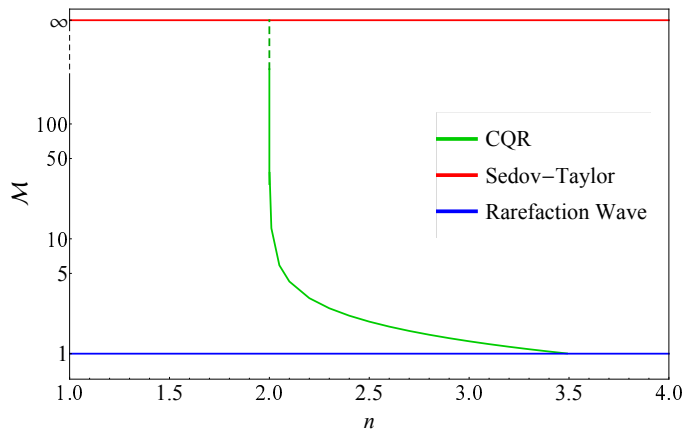}
   \caption{A schematic representation of the shock Mach number that permits self-similar solutions when the density profile of the ambient medium falls off as $\rho \propto r^{-n}$. The red line gives the Sedov-Taylor solution (or the Waxman-Shvarts solution for $n > 3$; \citealt{waxman93}), which has $\mathscr{M} = \infty$; the blue curve is the rarefaction wave, which has $\mathscr{M} = 1$; and the green curve is the CQR value, where here we used the solution for the Mach number as a function of $n$ when $\gamma_1 = \gamma_2 = 1+1/n$. When $2 < n < 3.5$, there are three possible self-similar solutions that describe the propagation of the shock. The CQR solution is weakly unstable, and in infinite time in an infinite medium would approach either the Sedov-Taylor or rarefaction wave solution, depending on the initial Mach number of the shock.}
   \label{fig:machs}
\end{figure}

For the range of $n$ that satisfy $2 < n < 3.5$, there are \emph{three}, exact, self-similar solutions that describe the propagation of shocks in the vicinity of a compact object, and only one -- the CQR solution -- has a Mach number that is both greater than one and less than infinity. Figure \ref{fig:machs} illustrates this situation schematically. The green curve in this figure uses the solution for the CQR Mach number when $\gamma_1 = \gamma_2 = 1+1/n$ (though, as shown by Figure \ref{fig:sigma_of_gamma}, there is not much dependence on the adiabatic index). There is a tempting analogy to draw between the existence of these three solutions and the three types of orbits that are possible in a Keplerian potential, being unbound (positive energy), bound (negative energy), and marginally-bound (zero energy). The last type of orbit is unstable, as infinitesimal perturbations to the energy either generate bound or unbound orbits. Furthermore, the Sedov-Taylor blastwave has positive energy and a positive Bernoulli parameter everywhere; the rarefaction wave has a negative energy and a negative Bernoulli parameter everywhere; and the energy of the CQR solution approaches exactly zero in the limit of large time and has both a positive and negative Bernoulli parameter throughout the post-shock fluid. The CQR solution therefore appears to represent the fluid analog of a marginally-bound orbit in a Keplerian potential.

\section{Summary and Conclusions}
\label{sec:summary}
This paper is the second of a series that investigates the propagation of weak shocks in the gravitational field of a compact object. Here we have analyzed the stability of new, self-similar solutions presented in \citet{coughlin18b} (Paper I). These solutions, which for brevity we refer to as the ``CQR solutions,'' differ from the Sedov-Taylor solutions in a number of ways; for one, the Mach number of the CQR shock is not infinite, but is a number -- of order one -- that is set by the smooth passage of the fluid quantities through a sonic point within the flow. In addition to outward motion immediately behind the shock, the CQR self-similar solutions contain a stagnation point within the flow, interior to which the fluid falls back onto the point mass at the origin. These solutions therefore simultaneously describe the outward motion of fluid behind a shock and the accretion of matter onto a compact object, and the accretion rate is a function of the properties of the ambient gas. We refer the reader to Paper I for more details of the self-similar solutions.

In this paper we first developed a general formalism for describing the radial perturbations to the shock position and post-shock fluid quantities (the velocity, density, and pressure), and showed that the deviations of the shock properties from self-similarity can be characterized in terms of ``Eigenmodes.'' While our approach is similar to that adopted by previous authors, it differs in a number of notable ways (see Section \ref{sec:previous}) and more fully exploits the scale invariance of the problem (which simplifies the stability analysis, not just for the self-similar solution of Paper I, but also for the Sedov-Taylor blastwave; see Appendix \ref{sec:appendixA}). Using this methodology, we showed that the CQR solutions are weakly unstable, with perturbations to the shock position and velocity growing as power-laws in time with power-law index $\lesssim 0.1$ (i.e., perturbations grow with time $t$ as $t^{\alpha}$, with $\alpha \lesssim 0.1$). 

To investigate the validity of our analytical formalism and results, we ran a suite of one-dimensional, high resolution simulations with {\sc flash} \citep{fryxell00}, in which we followed the evolution of a shock propagating through a non-self-gravitating atmosphere with density profile $\rho \propto r^{-2.5}$ in equilibrium with a point mass. These simulations confirmed the general predictions of the perturbation theory: when the shock Mach number was very close to that predicted by the CQR solution, the shock evolution closely tracked the predicted $\propto t^{2/3}$ shock scaling, and the post-shock fluid velocity, density, and pressure were described well by the CQR self-similar solutions. However, small deviations between the numerically-obtained shock position and the CQR prediction grow with time at a rate that agrees very well with the linear perturbation theory prediction. 

When the Mach number of the shock at late times is greater than the CQR value, the perturbation theory predicts that the Mach number should asymptotically grow at a very slow rate, and the solution likely eventually approaches the energy-conserving, Sedov-Taylor blastwave. On the other hand, when the Mach number is less than the CQR value, the linear theory predicts that the Mach number should eventually fall below unity. To understand the long-term behavior of such solutions, we showed that there is a \emph{third solution} to the self-similar equations with a ``shock'' Mach number of exactly one, which corresponds to a rarefaction wave that travels outward at the local sound speed and informs the hydrostatic medium of the infall at the center. It is likely that these rarefaction wave solutions characterize the nonlinear, asymptotic evolution of weak shocks with Mach numbers below the CQR value.  

Paper I argued that the CQR solutions are likely stable, as larger shock velocities (compared to the CQR value) correspond to greater post-shock energy budgets, but the rate at which binding energy is swept up by the ambient medium is also amplified; it therefore seemed reasonable that these shocks would be capable of self-regulating energetically. However, in addition to this stabilizing term, there is an additional, destabilizing term in the energy equation that is related to the fact that the ambient medium is less bound at larger radii (the ambient energy density falls off as $ \propto r^{n-1}$ when the density scales as $\rho \propto r^{-n}$). Thus, if one perturbs the shock position to a larger value, the binding energy at that new radius is slightly reduced, which inhibits the ability of the shock to self-regulate. We find that the magnitude of this second, destabilizing term can slightly exceed that of the stabilizing, velocity-dependent term. It is this slight difference that renders the CQR solutions unstable.

In Paper I, it was shown that the CQR self-similar solution \emph{very accurately} reproduced the simulation results of \citet{fernandez18}, who numerically studied the propagation of a weak shock through the envelope of a yellow supergiant (evolved from the zero-age main sequence with the stellar evolution code {\sc mesa}; \citealt{paxton13,paxton15,paxton18}) following its failure to explode in a traditional supernova. In their simulation, the shock was generated by the sudden decrease of the gravitational field that follows the formation of the proto-neutron star during core-collapse. The failed explosion of a supergiant therefore represents a generic, physical scenario in which one expects such weak shocks to arise. In addition to the fact that the density and pressure profiles of the hydrogen envelope of the yellow supergiant were not \emph{exactly} power-laws, the numerical simulation of \citet{fernandez18} included the effects of a non-adiabatic equation of state and self-gravity, which can all be considered perturbations on top of the conditions necessary to maintain the CQR solution. How, then, did the CQR self-similar solution so accurately describe the propagation of the shock, and the time and space-dependent evolution of all the post-shock fluid variables, over three decades in radius (and four decades in time) when, as we showed here, it is linearly unstable?

The resolution of this apparent contradiction is that the perturbations grow extremely slowly; therefore, as long as the deviations from self-similarity are initially small (e.g., the difference between the shock Mach number and the CQR value are not too large), then the time for the perturbations to grow into the nonlinear regime can be much longer than the time for the shock to propagate through the hydrogen envelope of a supergiant. Thus, the CQR solutions are \emph{effectively stable}, provided that the initial energy involved in the explosion is not too much greater or less than the binding energy of the hydrogen envelope of the star. Indeed, for the YSG analyzed in \citet{fernandez18}, the neutrino-induced mass loss generates a predicted shock energy of $E \simeq 5\times 10^{47}$ erg, while the total binding energy of the hydrogen envelope is $E_{\rm bind} \sim 8\times 10^{47}$ erg (see the left panel of Figure 14 of Paper I).

In this paper, we restricted our perturbation analysis to purely radial modes, to which the CQR solutions -- in contrast to the Sedov-Taylor solutions (see Appendix \ref{sec:appendixA}) -- are unstable. One can generalize the perturbation approach to include non-spherical modes by decomposing the shock front into spherical harmonics, which in turn induce aspherical, post-shock perturbations on the fluid quantities (see, e.g., \citealt{vishniac83}). However, the physical origin of the instability found in this paper for radial modes is related to the fact that, as the shock moves outward, it sweeps up progressively less binding energy from the ambient medium. This effect is maximized geometrically when the perturbations are spherical. For example, an $\ell = 1$ mode generates a ``sloshing'' motion of the shock, meaning that roughly half of the shock receives a positive perturbation in radial velocity, while the other half receives a negative radial velocity perturbation. Therefore, the net reduction in the rate at which binding energy is supplied to the post-shock fluid is smaller in this case, which will likewise result in a smaller growth rate of any potential instability. We therefore expect that, while non-radial modes may be unstable, radial perturbations are likely the most unstable from the energy standpoint considered in this paper (which neglects, e.g., buoyancy, which can lead to distinct, unstable modes). We do, however, plan to analyze the stability of the CQR solution to non-spherical perturbations in a future investigation.

\acknowledgements
ERC acknowledges support from
NASA through the Einstein Fellowship Program, grant PF6-170170. This work was supported in part by a Simons
Investigator award from the Simons Foundation (EQ) and the Gordon and Betty Moore Foundation through Grant
GBMF5076.

\appendix
\section{Stability of the Sedov-Taylor blastwave to radial perturbations}
\label{sec:appendixA}
In this appendix, we use the formalism for analyzing the stability of the CQR solution to assess the stability of the Sedov-Taylor blastwave to radial perturbations, and we compare our results to those of \citet{ryu87}. The Sedov-Taylor blastwave adopts the same functional form for the self-similar variables as the CQR solution, namely that the velocity, $v$, the density, $\rho$, and the pressure, $p$, are written self-similarly as:

\begin{equation}
v = V\left(f_0+f_1\right), \quad \rho = \rho_{\rm a}\left(\frac{R}{r_{\rm a}}\right)^{-n}\left(g_0+g_1\right), \quad p = \rho_{\rm a}\left(\frac{R}{r_{\rm a}}\right)^{-n}V^2\left(h_0+h_1\right), \label{ssst}
\end{equation}
\begin{equation*}
\Rightarrow \quad s = \frac{p}{\rho^{\gamma_2}} = \ln\left(V^2R^{-n+n\gamma_2}\right)+\ln\left(\frac{h_0}{g_0^{\gamma_2}}\right)+\frac{h_1}{h_0}-\frac{\gamma_2g_1}{g_0} \equiv \ln\left(V^2R^{-n+n\gamma_2}\right)+s_0+s_1,
\end{equation*}
where $R$ is the true shock position and $V = dR/dt$ is the shock velocity. Since the change of variables is the same, the governing equations are identical except that we ignore the gravitational term on the right-hand side of the momentum equation:

\begin{equation}
\frac{\partial g_1}{\partial \tau}-n\left(g_0+g_1\right)-\xi\left(g_0'+\frac{\partial g_1}{\partial \xi}\right)+\frac{1}{\xi^2}\frac{\partial}{\partial \xi}\left[\xi^2\left(g_0+g_1\right)\left(f_0+f_1\right)\right] = 0. \label{contfullst}
\end{equation}
\begin{equation}
\frac{\partial f_1}{\partial \tau}+\left(\frac{1}{2}\frac{d}{d\tau}\ln V^2\right)\left(f_0+f_1\right)-\xi\left(f_0'+\frac{\partial f_1}{\partial \xi}\right)+\left(f_0+f_1\right)\frac{\partial}{\partial \xi}\left(f_0+f_1\right)+\frac{1}{g_0}\left(1+\frac{g_1}{g_0}\right)^{-1}\frac{\partial}{\partial \xi}\left(h_0+h_1\right) = 0, \label{rmomfullst}
\end{equation}
\begin{equation}
\frac{\partial s_1}{\partial \tau}+
\frac{d}{d\tau}\ln V^2-n+n\gamma_2-\xi\frac{\partial}{\partial \xi}\left[s_0+s_1\right]+\left(f_0+f_1\right)\frac{\partial}{\partial\xi}\left[s_0+s_1\right] = 0. \label{gasenfullst}
\end{equation}
The boundary conditions are still governed by the jump conditions at the shock front, namely Equations \eqref{bc1} -- \eqref{bc3}, but here we adopt the strong-shock limit, which makes them extremely simple:

\begin{equation}
f_0(\tau,1)+f_1(\tau,1) = \frac{2}{\gamma+1}, \label{bc1st}
\end{equation}
\begin{equation}
g_0(\tau,1)+g_1(\tau,1) = \frac{\gamma+1}{\gamma-1}, \label{bc2st}
\end{equation}
\begin{equation}
h_0(\tau,1)+h_1(\tau,1) = \frac{2}{\gamma+1}. \label{bc3st}
\end{equation}
Notice that these boundary conditions are this simple only because we are defining the self-similar variable in terms of the \emph{total} shock position and velocity; if one instead opts to initially break up the shock velocity and position into its unperturbed and perturbed quantities, these become more complicated (see Equation 5.17 of \citealt{ryu87}).

Investigating Equations \eqref{contfullst} -- \eqref{bc3st}, we see that we can satisfy both the differential equations and the boundary conditions with all subscript-1 quantities identically equal to zero if

\begin{equation}
V^2 \propto R^{\alpha},
\end{equation}
with $\alpha$ any arbitrary constant. However, the energy behind the shock must be conserved for any realistic system (under this set of assumptions, where we ignore gravity and the Mach number is infinite), and we can write this energy as:

\begin{equation}
E = 4\pi\int_{0}^{R(t)}\left(\frac{1}{2}v^2+\frac{1}{\gamma-1}\frac{p}{\rho}\right)\rho r^2 dr = 4\pi \rho_{\rm a} r_{\rm a}^3V^2\left(\frac{R}{r_{\rm a}}\right)^{3-n} \int_{0}^{1}\left(g_0+g_1\right)\left(\left(f_0+f_1\right)^2+\frac{1}{\gamma-1}\frac{h_0+h_1}{g_0+g_1}\right)\xi^2d\xi.
\end{equation}
There are combinations of $n$ and $\gamma$ for which the Sedov-Taylor blastwave only extends to a finite inner radius that serves as a contact discontinuity (see \citealt{goodman90}), and in this case one replaces the lower bound in the integral with some $\xi_0$; for simplicity, however, we focus here on the case when the entire post-shock fluid can be described by the Sedov-Taylor self-similar solutions. We therefore see that if we let

\begin{equation}
4\pi\rho_{\rm a}r_{\rm a}^3 V^2\left(\frac{R}{r_{\rm a}}\right)^{3-n} = E_0,
\end{equation}
with $E_0$ a constant, then the energy is conserved and all of the perturbations can be set to zero. Thus, for the Sedov-Taylor blastwave we write

\begin{equation}
4\pi\rho_{\rm a}r_{\rm a}^3 V_{}^2\left(\frac{R}{r_{\rm a}}\right)^{3-n} = E_0\left(1+2\zeta_{}(\tau)\right) \label{zetadefst}
\end{equation}
\begin{equation*}
\Rightarrow \ln V^2 \simeq \left(n-3\right)\tau+2\zeta_{}(\tau),
\end{equation*}
where $\zeta(\tau)$ is a small quantity that encodes the perturbations to the position and velocity of the shock, and the second line follows in the perturbative limit (i.e., we dropped all nonlinear terms in $\zeta$). The equations for the unperturbed quantities are then

\begin{equation}
-ng_0-\xi g_0'+\frac{1}{\xi^2}\frac{d}{d \xi}\left(\xi^2g_0f_0\right) = 0,
\end{equation}
\begin{equation}
\frac{1}{2}\left(n-3\right)f_0+\left(f_0-\xi\right)f_0'+\frac{h_0'}{g_0} = 0,
\end{equation}
\begin{equation}
-3+n\gamma+(f_0-\xi)\frac{d}{d\xi}\ln\left(\frac{h_0}{g_0^{\gamma}}\right) = 0,
\end{equation}
while those for the perturbations read

\begin{equation}
\frac{\partial g_1}{\partial \tau}-ng_1-\xi\frac{\partial g_1}{\partial \xi}+\frac{1}{\xi^2}\frac{\partial}{\partial \xi}\left[\xi^2g_0f_1+\xi^2f_0g_1\right] = 0, \label{contpertsst}
\end{equation}
\begin{equation}
\frac{\partial f_1}{\partial \tau}+\frac{1}{2}(n-3)f_1-\xi\frac{\partial f_1}{\partial \xi}+\frac{\partial}{\partial \xi}\left(f_0f_1\right)+\frac{1}{g_0}\frac{\partial}{\partial \xi}\left(h_0s_1+\gamma_2h_0\frac{g_1}{g_0}\right)-\frac{g_1}{g_0^2}h_0' = -\dot{\zeta}f_0, \label{rmompertsst}
\end{equation}
\begin{equation}
\frac{\partial s_1}{\partial \tau}+
\left(f_0-\xi\right)\frac{\partial s_1}{\partial \xi}+f_1s_0'= -2\dot{\zeta}. \label{gasenpertsst}
\end{equation}
The boundary conditions for the unperturbed quantities are the usual strong-shock jump conditions:

\begin{equation}
f_0(1) = h_0(1) = \frac{2}{\gamma+1}, \quad g_0(1) = \frac{\gamma+1}{\gamma-1},
\end{equation}
while the perturbations satisfy homogeneous boundary conditions:

\begin{equation}
f_1(1,\tau) = g_1(1,\tau) = h_1(1,\tau) = 0.
\end{equation}
Equations \eqref{contpertsst} -- \eqref{gasenpertsst} can be written in matrix form as

\begin{multline}
\frac{\partial}{\partial \tau}\left(
\begin{array}{c}
s_1 \\
F_1 \\
G_1
\end{array}\right)
+
\left(
\begin{array}{ccc}
f_0-\xi & 0 & 0 \\
\xi^2h_0 & f_0-\xi & \gamma \xi^2 h_0 \\
0 & \xi^{-2}g_0^{-1} & f_0-\xi
\end{array} \right)\frac{\partial}{\partial\xi}
\left(
\begin{array}{c}
s_1 \\
F_1 \\
G_1
\end{array}\right)
+ \left(
\begin{array}{ccc}
0 & \xi^{-2}g_0^{-1}s_0' & 0 \\
\xi^2h_0' & \frac{1}{2}-\frac{n}{2}+2f_0' & \xi^2(\gamma-1)h_0' \\
0 & 0 & 0 
\end{array} \right)
\left(
\begin{array}{c}
s_1 \\
F_1 \\
G_1
\end{array}\right)
\\
=\left(
\begin{array}{c}
-2\dot{\zeta} \\
-f_0g_0\xi^2\dot{\zeta} \\
0
\end{array}\right). \label{eqsoftST}
\end{multline}
Notice that this is exactly the same as Equation \eqref{eqsoft} if we make the substitutions $\zeta \rightarrow -\zeta$, $n\rightarrow -1+2n$, and the gravitational term on the right-hand side is missing.

As was true for the perturbations to the CQR self-similar solutions, Equation \eqref{eqsoftST} appears over-constrained because of the introduction of the variable $\zeta$. There is, however, an additional constraint that these variables must satisfy: taking the energy equation,

\begin{equation}
\frac{\partial \mathscr{E}}{\partial t}+\frac{\partial \mathscr{F}}{\partial r} = 0,
\end{equation}
where

\begin{equation}
\mathscr{E} = \left(\frac{1}{2}v^2+\frac{1}{\gamma-1}\frac{p}{\rho}\right)\rho r^2
\end{equation}
is the energy density and

\begin{equation}
\mathscr{F} = \left(\frac{1}{2}v^2+\frac{\gamma}{\gamma-1}\frac{p}{\rho}\right)\rho r^2 v
\end{equation}
is the energy flux, and integrating from zero to $R(t)+\epsilon$ and taking the limit as $\epsilon \rightarrow 0$ gives

\begin{equation}
\frac{\partial E}{\partial t} = \mathscr{F}(0),
\end{equation}
where

\begin{equation}
E = \int_0^{R}\left(\frac{1}{2}v^2+\frac{1}{\gamma-1}\frac{p}{\rho}\right)\rho r^2dr
\end{equation}
is the total -- including the perturbations -- energy behind the shock. Since we are not introducing any sources or sinks of energy, this quantity must be \emph{exactly} conserved, meaning that the perturbations to this quantity must also be identically zero. Thus, if we expand the flux into its perturbed and unperturbed components, we recover the boundary condition

\begin{equation}
\lim_{\xi \rightarrow 0}\left(\frac{F_{\sigma}}{g_0}+f_0\xi^2s_\sigma+\gamma f_0\xi^2G_\sigma\right) = \lim_{\xi \rightarrow 0} \left(\xi^2f_{\sigma}+f_0\xi^2h_{\sigma}\right) = 0, \label{bcst}
\end{equation}
where in the last line we wrote this condition in terms of the self-similar velocity and pressure (Equation \ref{ssst}). Notice that, if the first term in parentheses and the geometric factors are ignored, this boundary condition is in agreement with that of \citet{ryu87}, being that the correction to the pressure vanishes at the origin. However, their boundary condition is not in fact correct since $f_0\xi^2$ goes to zero at the origin for the Sedov-Taylor blastwave; Equation \eqref{bcst} is more general, and \emph{any}, weakly diverging correction to the pressure (one that does not diverge faster than $\propto 1/\xi^3$, since the unperturbed, Sedov-Taylor velocity satisfies $f_0 \propto \xi$) will satisfy the conservation of energy provided that the velocity does not diverge as $1/\xi^2$ or faster. 

We now search for Eigenmodes of the form

\begin{equation}
\zeta = \zeta_{\sigma}e^{\sigma\tau}, \quad F_1 = \zeta_{\sigma}e^{\sigma \tau} F_{\sigma}, \quad G_1 = \zeta_{\sigma}e^{\sigma\tau}G_{\sigma}, \quad s_1 = \zeta_{\sigma}e^{\sigma\tau} s_{\sigma}.
\end{equation}
As was true for the CQR solution, $\zeta_{\sigma}$ scales out of the problem, and the value of $\sigma$ is determined by the requirement that the functions satisfy the fourth boundary condition at the origin. The Eigenvalue equation is

\begin{equation}
\sigma\left(
\begin{array}{c}
s_\sigma \\
F_\sigma \\
G_\sigma
\end{array}\right)
+
\left(
\begin{array}{ccc}
f_0-\xi & 0 & 0 \\
\xi^2h_0 & f_0-\xi & \gamma_2 \xi^2 h_0 \\
0 & \xi^{-2}g_0^{-1} & f_0-\xi
\end{array} \right)\frac{\partial}{\partial\xi}
\left(
\begin{array}{c}
s_\sigma \\
F_\sigma \\
G_\sigma
\end{array}\right)
+ \left(
\begin{array}{ccc}
0 & \xi^{-2}g_0^{-1}s_0' & 0 \\
\xi^2h_0' & \frac{1}{2}-\frac{n}{2}+2f_0' & \xi^2(\gamma_2-1)h_0' \\
0 & 0 & 0 
\end{array} \right)
\left(
\begin{array}{c}
s_\sigma \\
F_\sigma \\
G_\sigma
\end{array}\right)
=\left(
\begin{array}{c}
-2\sigma \\
-f_0g_0\xi^2\sigma \\
0
\end{array}\right). \label{eigensST}
\end{equation}
Notice that, if $\sigma = 0$, the right-hand side of this matrix equation is zero. Moreover, if we let $F_{\sigma} = G_{\sigma} = s_{\sigma} = 0$, then the differential equations and, since they are homogeneous, the boundary conditions are satisfied. This is therefore an exact, neutrally stable solution for the perturbations to the Sedov-Taylor blastwave. From our definition of $\zeta$ \eqref{zetadefst}, we find that the unperturbed and perturbed shock position satisfy

\begin{equation}
4\pi \rho_{\rm a}r_{\rm a}^3V_0^2\left(\frac{R_0}{r_{\rm a}}\right)^{3-n}\left(1+2\frac{V_1}{V_0}\right)\left(1+\left(3-n\right)\frac{R_1}{R_0}\right) = E_0\left(1+2\zeta\right),
\end{equation}
\begin{equation}
4\pi\rho_{\rm a}r_{\rm a}^3 V_0^2 \left(\frac{R_0}{r_{\rm a}}\right)^{3-n} = E_0, \label{r0steq}
\end{equation}
\begin{equation}
\frac{R_1}{R_0} = \frac{2\zeta}{5-n}\left(1-e^{-\frac{5-n}{2}\tau}\right),
\end{equation}
\begin{equation}
\frac{V_1}{V_0} = \frac{2\zeta}{5-n}\left(1+\frac{3-n}{2}e^{-\frac{5-n}{2}\tau}\right).
\end{equation}
{From these expressions, we see that, similar to the CQR solution, the perturbations to the shock position and velocity are written as the difference between the Eigenmode (which here has an Eigenvalue of $\sigma = 0$) and a second ``mode'' that has an Eigenvalue of $-(5-n)/2$. However, this second mode is just a consequence of the initial conditions, and is not a true Eigenvalue from the standpoint that it does not characterize the temporal dependence of $\zeta$. If we further use the fact that the unperturbed shock position for the Sedov-Taylor blastwave scales as $R_0(t) \propto t^{(5-n)/2}$ (this can be verified directly by integrating Equation \ref{r0steq}), then this second mode decays as $t^{-1}$, which is identical to the temporal scaling of the trivial, decaying mode of the CQR solution.}

While the perturbations to the post-shock fluid quantities may appear to be trivial (i.e., $F_\sigma = G_{\sigma} = s_{\sigma} \equiv 0$), this is in fact not the case because of our definitions of the velocity, density, and pressure in terms of the total shock velocity and position. Specifically, the solutions to first order in the unperturbed self-similar variable, or physical radius $r$, are

\begin{equation}
v(\xi_0,\tau) = V_0\left\{f_0+\frac{2\zeta}{5-n}\left(1+\frac{3-n}{2}e^{-\frac{5-n}{2}\tau}\right)f_0-\frac{2\zeta}{5-n}\left(1-e^{-\frac{5-n}{2}\tau}\right)\xi_0 f_0'\right\},
\end{equation}
\begin{equation}
\rho(\xi_0,\tau) = \rho_{\rm a}\left(\frac{R_0}{r_{\rm a}}\right)^{-n}\left\{g_0-\frac{2\zeta}{5-n}\left(1-e^{-\frac{5-n}{2}\tau}\right)\left(ng_0+\xi_0 g_0'\right)\right\},
\end{equation}
\begin{equation}
p(\xi_0,\tau) = \rho_{\rm a}V_0^2\left(\frac{R_0}{r_{\rm a}}\right)^{-n}\left(h_0+\frac{4\zeta}{5-n}\left(1+\frac{3-n}{2}e^{-\frac{5-n}{2}\tau}\right)h_0-\frac{2\zeta}{5-n}\left(1-e^{-\frac{5-n}{2}\tau}\right)\left(nh_0+\xi_0 h_0'\right)\right).
\end{equation}
These solutions can be shown to satisfy Equations 4.18a -- 4.18d of \citet{ryu87} and the boundary conditions at the shock in terms of the unperturbed self-similar variable (their Equations 5.17a -- 5.17d) when the perturbations are spherically symmetric ($\ell = 0$). 

\begin{figure}
    \centering
    \includegraphics[width=0.495\textwidth]{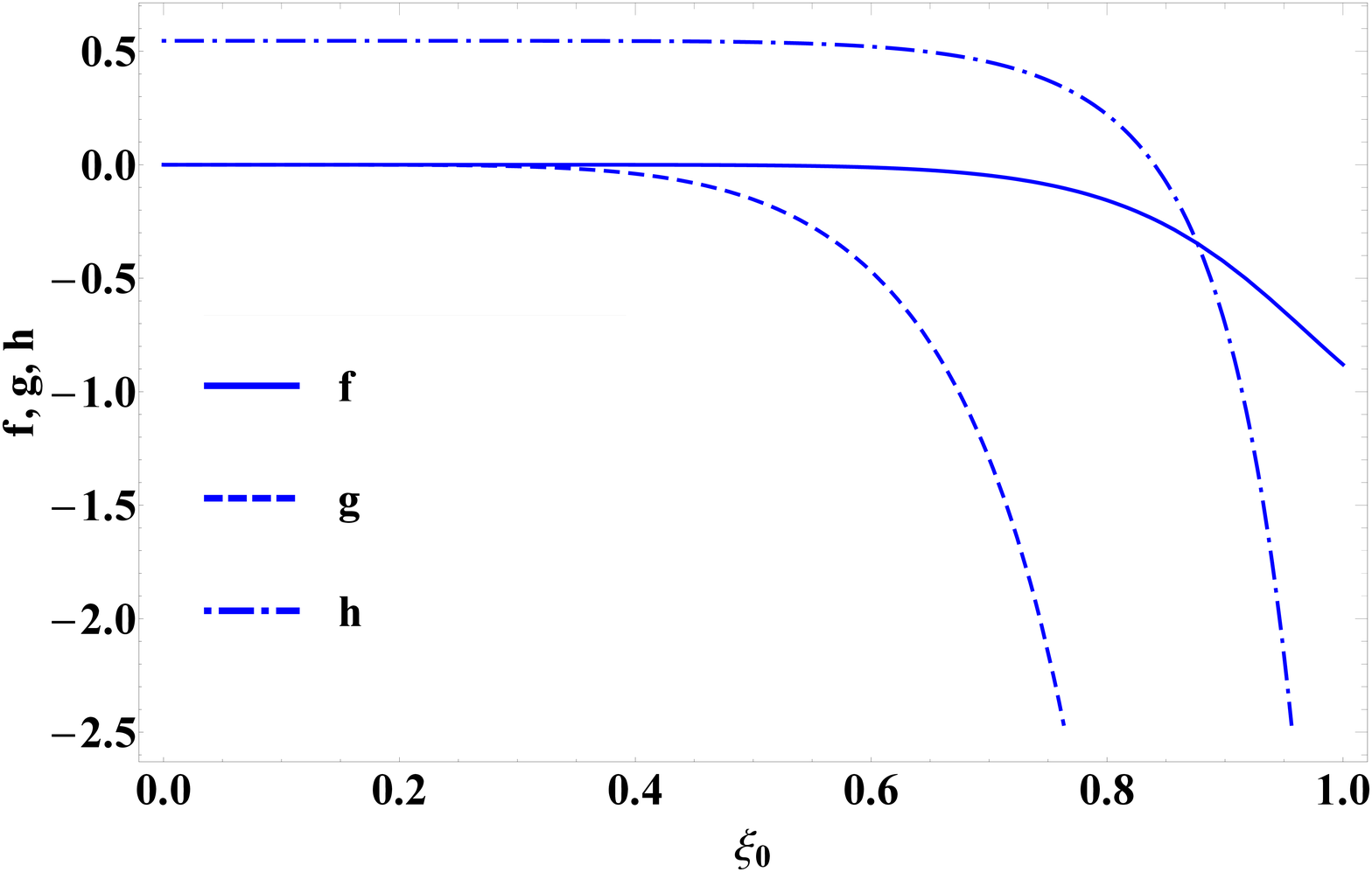}
    \includegraphics[width=0.495\textwidth]{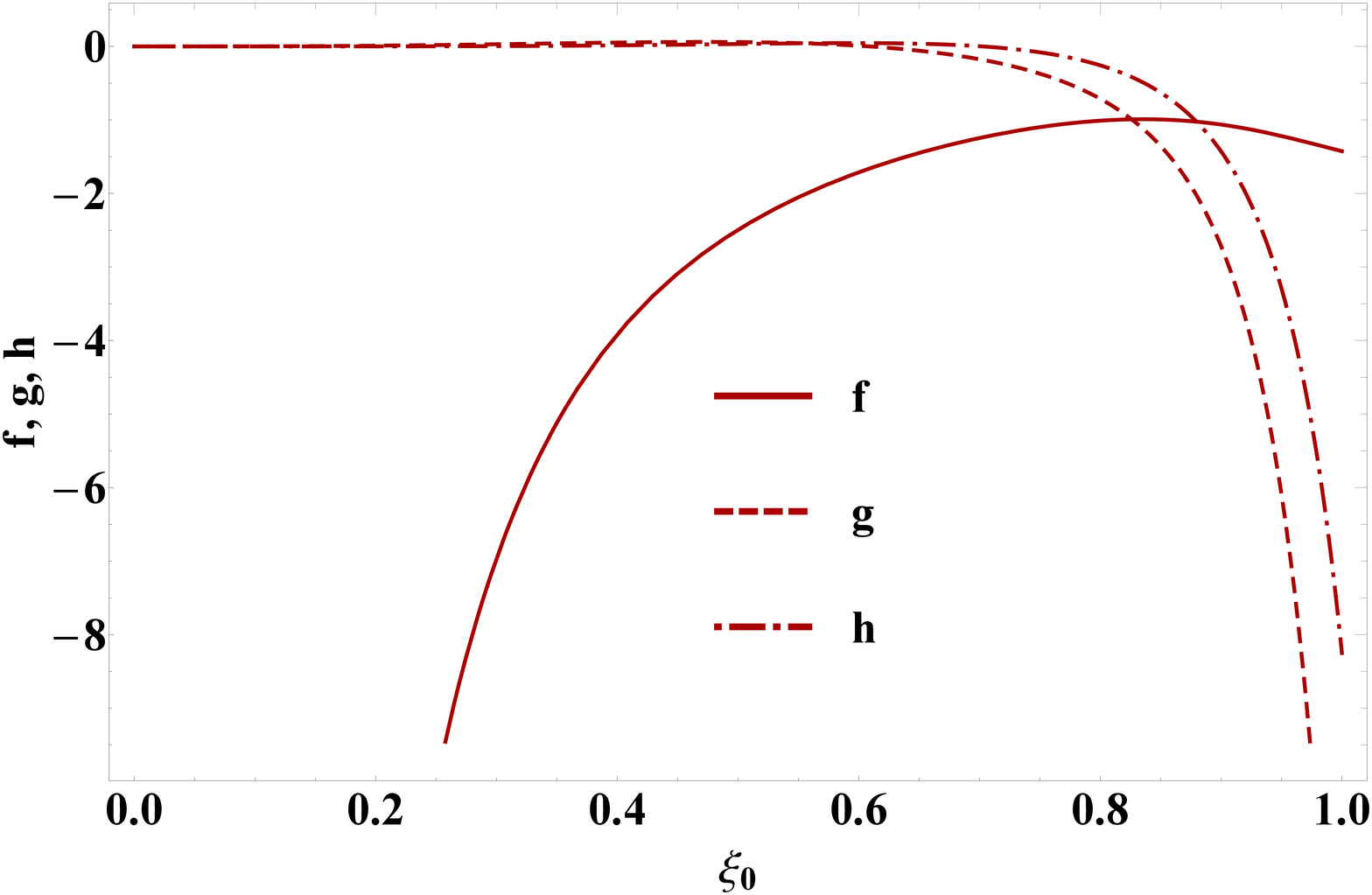}
    \caption{Left: The perturbation to the velocity (solid), density (dashed), and pressure (dot-dashed) of the Sedov-Taylor blastwave as functions of the unperturbed self-similar variable $\xi_0$, when $n = 0$, $\gamma = 1.5$ and $\sigma = 0$. These solutions maintain exact global energy conservation (i.e., the perturbation to the energy does not change in time) and include a finite correction to the pressure at the origin. Right: The same perturbations as the left panel, but here $\sigma \simeq -0.3495$, which satisfies the fourth boundary condition imposed by \citet{ryu87}, being that the perturbation to the pressure vanish at the origin. Here the velocity diverges as $\propto 1/\xi^2$ near the origin, and the energy is not globally conserved.}
    \label{fig:stsols}
\end{figure}

Notice that our solution for the perturbation to the pressure does not go to zero at the origin, which is the fourth boundary condition imposed by \citet{ryu87}. However, our boundary condition -- which demands that the flux of energy at the origin be zero and hence that the total energy be conserved -- is satisfied for any non-diverging pressure and velocity (specifically Equation \ref{bcst}). Our solutions, being written in terms of the well-behaved, unperturbed solutions, are clearly convergent near the origin, and Figure \ref{fig:stsols} shows this convergence explicitly for $n = 0$ and $\gamma = 1.5$. On the other hand, the solutions that satisfy the vanishing pressure boundary condition of \citet{ryu87}, which are shown in the right panel of this figure (and have a decaying Eigenmode of $\sigma \simeq -0.3495$), possess a perturbation to the velocity that diverges as $\propto 1/\xi^2$ near the origin. Therefore, the solutions found by \citet{ryu87} do not globally conserve energy owing to their finite energy flux at the origin.

\bibliographystyle{aasjournal}
\bibliography{refs}

\end{document}